\documentclass{aa}  
\usepackage{gensymb}
\usepackage{graphicx}
\usepackage{txfonts}
\begin{document}

   \title{Confusion of extragalactic sources in the far infrared: a baseline assessment of the performance of PRIMAger in intensity and polarization}
   
    \authorrunning{B\'ethermin et al.}
    \titlerunning{Confusion in the far infrared: assessment of PRIMAger performance}

   \author{Matthieu B\'ethermin\inst{1,2} \and Alberto D. Bolatto\inst{3} \and François Boulanger\inst{4} \and Charles M. Bradford\inst{5,6} \and Denis Burgarella\inst{2} \and Laure Ciesla\inst{2} \and James Donnellan\inst{7} \and Brandon S. Hensley\inst{5} \and Jason Glenn\inst{8} \and Guilaine Lagache\inst{2} \and Enrique Lopez-Rodriguez\inst{9} \and  Seb Oliver\inst{7} \and Alexandra Pope\inst{10} \and Marc Sauvage\inst{11}}
          
    \institute{Universit\'e de Strasbourg, CNRS, Observatoire astronomique de Strasbourg, UMR 7550, 67000 Strasbourg, France. \email{matthieu.bethermin@astro.unistra.fr} \and
    Aix Marseille Univ, CNRS, CNES, LAM, Marseille, France \and
    Department of Astronomy, University of Maryland, College Park, MD 20742, USA \and
    Laboratoire de Physique de l’\'Ecole Normale Sup\'erieure, Universit\'e Paris Science et Lettres, Centre National de la Recherche Scientifique, Sorbonne Universit\'e, Universit\'e de Paris, 75005, Paris, France \and
    Jet Propulsion Laboratory, California Institute of Technology, 4800 Oak Grove Drive, Pasadena, CA 91109, USA \and
    California Institute of Technology, 1200 E California Blvd, Pasadena, CA 91125, USA \and
    Astronomy Centre, Department of Physics and Astronomy, University of Sussex, Falmer, Brighton, BN1 9QH, UK \and
    NASA Goddard Space Flight Center, Greenbelt, MD 20771, USA \and
    Kavli Institute for Particle Astrophysics and Cosmology (KIPAC), Stanford University, Stanford, CA 94305, USA \and
    Department of Astronomy, University of Massachusetts, Amherst, MA 01003 \and
    AIM, CEA, CNRS, Université Paris-Saclay, Université de Paris, 91191 Gif-sur-Yvette, France
    }
          
    \date{Received XXX; accepted XXX}

% \abstract{}{}{}{}{}
% 5 {} token are mandatory
 
  \abstract{}
  {Because of their limited angular resolution, far-infrared telescopes are usually affected by confusion phenomenon. Since several galaxies can be located in the same instrumental beam, only the brightest objects emerge from the fluctuations caused by fainter sources. The probe far-infrared mission for astrophysics imager (PRIMAger) will observe the mid- and far-infrared (25--235\,$\mu$m) sky both in intensity and polarization. We aim to provide predictions of the confusion level and its consequences for future surveys.}
  {We produced simulated PRIMAger maps affected only by the confusion noise using the simulated infrared extragalactic sky (SIDES) semi-empirical simulation. We then estimated the confusion limit in these maps and extracted the sources using a basic blind extractor. By comparing the input galaxy catalog and the extracted source catalog, we derived various performance metrics as completeness, purity, and the accuracy of various measurements (e.g., the flux density in intensity and polarization or the polarization angle).}
  {In intensity maps, we predict that the confusion limit increases rapidly with increasing wavelength (from 21\,$\mu$Jy at 25\,$\mu$m to 46\,mJy at 235\,$\mu$m). The confusion limit in polarization maps is more than two orders of magnitude lower (from 0.03\,mJy at 96\,$\mu$m to 0.25\,mJy at 235\,$\mu$m). Both in intensity and polarization maps, the measured (polarized) flux density is dominated by the brightest galaxy in the beam, but other objects also contribute in intensity maps at longer wavelength ($\sim$30\,\% at 235\,$\mu$m). We also show that galaxy clustering has a mild impact on confusion in intensity maps (up to 25\,\%), while it is negligible in polarization maps. In intensity maps, a basic blind extraction will be sufficient to detect galaxies at the knee of the luminosity function up to z$\sim$3 and 10$^{11}$\,M$_\odot$ main-sequence galaxies up to z$\sim$5. In polarization for the most conservative sensitivity forecast (payload requirements), $\sim$200 galaxies can be detected up to z=1.5 in two 1\,500\,h surveys covering 1\,deg$^2$ and 10\,deg$^2$. For a conservative sensitivity estimate, we expect $\sim$8\,000 detections up to z=2.5 opening a totally new window on the high-z dust polarization. Finally, we show that intensity surveys at short wavelength and polarization surveys at long wavelength tend to reach confusion at similar depth. There is thus a strong synergy between them.}
   {}
%
%________________________________________________________________

\maketitle

\section{Introduction}

Far-infrared wavelengths are key to understand galaxy evolution across cosmic times. Studies of the Cosmic Infrared Background \citep[CIB, e.g.,][]{Hauser2001,Lagache2005,Dole2006, Berta2011,Bethermin2012b} revealed that more than half of the relic emission from galaxies and their host nuclei is located in the 8--1000\,$\mu$m range with a peak around 150\,$\mu$m. The UV photons emitted by young stars are absorbed by dust and their energy is re-emitted in the mid- and far-infrared. These wavelengths thus trace obscured star formation in the Universe.

While the CIB had been detected in the nineties \citep{Puget1996,Fixsen1998,Hauser1998}, characterizing all the individual galaxy populations producing it remains difficult. Since far-infrared radiation can only be observed from space and the stratosphere, the diameter of the telescope primary mirror is limited. The angular resolution ($\theta\approx\lambda/D_{tel}$, where $\lambda$ is the wavelength and $D_{tel}$ the telescope diameter) is thus severely limited by diffraction, and at the longest wavelengths it can be up to few tens of arcseconds. At this resolution, several high-z galaxies can be located in the same beam leading to background fluctuations, and only the brightest objects emerge above the fluctuations from unresolved faint galaxies. This phenomenon is called confusion \citep[e.g.,][]{Condon1974,Lagache2003,Dole2004}, and must be taken into account to design mid- and far-infrared telescopes and surveys.

With its actively-cooled 85\,cm mirror, the \textit{Spitzer} space telescope \citep{Werner2004} resolved more than 80\,\% of the CIB at 24\,$\mu$m into individual sources \citep[e.g.,][]{Papovich2004,Bethermin2011}, but only a small fraction ($\lesssim$10\,\%) emerged above the confusion around the peak of the CIB at 160\,$\mu$m \citep[e.g.,][]{Dole2004,Frayer2006a}. Thanks to its 3.5\,m mirror, the \textit{Herschel} space observatory \citep{Pilbratt2010} allowed us to resolve the majority of the CIB into individual sources at its peak \citep{Berta2011,Magnelli2013}, but not at longer wavelengths \citep{Oliver2010,Bethermin2012b}. The passive cooling of the mirror ($\sim$85\,K) caused a high background and limited the sensitivity. \textit{Herschel} could thus not reach the confusion limit below 100\,$\mu$m, and the sizes of the deep field were limited between 100 and 200\,$\mu$m \citep{Lutz2011}. For both \textit{Spitzer} and \textit{Herschel}, sub-confusion flux density regimes were probed using statistical approaches as stacking of galaxy population known from shorter wavelengths \citep[e.g.,][]{Dole2006,Bethermin2012b} or P(D) analysis, the analysis of the 1-point distribution of intensity in the maps \citep[e.g.,][]{Glenn2010,Berta2011}, or using source extractors relying on priors from shorter wavelengths \citep[e.g.,][]{Magnelli2009,Roseboom2010,Hurley2017}.

The PRobe far-Infrared Mission for Astrophysics (PRIMA) project uses a 1.8\,m space-based telescope  cryogenically-cooled to 4.5\,K with new generation detectors that take full advantage of the low thermal background. One of the two payload instruments is the PRIMA imaging camera (PRIMAger, \citealt{Burgarella_PRIMAger_AAS,Meixner_PRIMAger_AAS}).
PRIMAger has two main bands. The first one is an hyperspectral band (PRIMAger Hyperspectral Imaging: PHI) which provides imaging with a linear variable filter at a spectral resolution R$\sim$10 over 25 to 80\,$\mu$m. The second band (PRIMAger Polarization Imaging: PPI) provides 4 broad band filters between 91 and 235\,$\mu$m, sensitive to polarization. PRIMAger will operate with 100\,mK cooled kinetic inductance detectors which allows for an incomparable improvement of sensitivity in far infrared.

An observatory like PRIMA will cover a wide range of science topics such as, but not limited to, origins of planetary atmospheres, evolution of galaxies, and build-up of dust and metals through cosmic time \citep{Moullet2024}. In addition, it will offer for the first time spaceborne high-sensitivity far-infrared polarimetric capabilities in the far infrared. So far no high-z source has been polarimetrically detected at these wavelengths, and only one strongly-lensed starburst galaxy at z=2.6 has been published in the sub-millimeter \citep{Geach2023}. In addition, \citet{Chen2024} reported in a preprint a second sub-millimeter object at z=5.6 exhibiting kiloparsec-scale ordered magnetic fields while we were revising this paper.

Since high-z galaxies will not be spatially resolved, we will only detect the integrated polarization, if the magnetic fields driving the dust polarization in the various regions of a galaxy are ordered and their polarized flux densities add up at least partially. Else, in the disordered case, the signals from the various regions cancel out, leading to a very small integrated polarization. The integrated polarization fraction can vary with various physical parameters such as, e.g., the intrinsic dust polarization, the geometry of the galaxy, or the depolarization caused by turbulence (see, e.g., Sect.\,6.3 of \citealt{Andre2019}). Pioneering studies in the local Universe showed that two main mechanisms lead to organized polarization patterns and thus significant integrated polarized fractions in star-forming galaxies: organized magnetic fields in disk galaxies \citep{Lopez-Rodriguez2022} and starburst-driven outflows \citep{Lopez-Rodriguez2023}. In addition, active galactic nuclei (AGN) can also exhibit high polarization fractions in the far infrared \citep[e.g.,][]{Lopez-Rodriguez2018,Marin2020}, and could also lead to polarized outflows.

The goal of this paper is to assess the impact of the confusion on PRIMAger performance both in intensity and polarization based on the simulated infrared dusty extragalactic sky (SIDES, \citealt{Bethermin2022}) tools, and demonstrate the feasibility of high-z polarization surveys. 
%We focus on performances expected from blind source extractors, while prior-based extractors will be discussed in a companion paper (Donnellan et al. in prep.).
%

It is important to realize that confusion arising from simple source extractions, as assessed here, is not an ultimate limit for well designed surveys and instrumentation. In this paper, we will focus on the "classical" confusion limit for basic blind source extractors. Super-resolution techniques can certainly break through the classical confusion limit determined in these calculations to extract accurate SED information from fainter sources.  More importantly, prior information derived from catalogs at un-confused wavelengths is very effective at improving the flux extraction to much fainter limits. PRIMAger's hyperspectral architecture is especially designed to take advantage of this type of technique to break through confusion.
In this paper we focus on performances expected from basic source extractors, while prior-based extractors, which show improvements by factors of several beyond the confusion noise established here, will be discussed in a companion paper \citep{Donnellan2024}.

In Sect.\,\ref{sect:simulation}, we introduce the SIDES simulation and describe how it was adapted to perform PRIMA forecasts. In particular, we describe the extension of SIDES to polarization in Sect.\,\ref{sect:simpolar}. We then describe in Sect.\,\ref{sect:methods} the methods used to extract sources from the confusion-driven simulations and to assess the expected performances. We then present our results in intensity in Sect.\,\ref{sect:intensity} and in polarization in Sect.\,\ref{sect:polar}. Finally, we discuss the impact of confusion on future PRIMAger surveys and the expected number of detections in polarization in Sect.\,\ref{sect:surveys}. We conclude in Sect.\,\ref{sect:conclusion}.

In this paper, we will use the terminology "in intensity" to described all the quantities associated to standard photometric surveys and usually derived from specific intensity maps (e.g., flux density of point sources). The term "in polarization" will corresponds to quantities extracted from the polarization maps (Sect.\,\ref{sect:simpolar}) such as polarized flux density or the polarized fraction of the flux density (shortened hereafter by polarization fraction).

%%%%%%%%%%%%%%%%%%%%%%%%% SIMULATIONS %%%%%%%%%%%%%%%%%%%%%%

\section{Description of our simulation}
\label{sect:simulation}

Confusion depends on both the intrinsic nature of the sources being observed and the telescope and instrument providing the data. We discuss these two aspects of our simulation in turn.

\subsection{The SIDES simulation}

The confusion phenomenon is highly connected to the flux density distribution of galaxies \citep[e.g.,][]{Condon1974, Dole2004} and mildly by their spatial distribution as we will show in this study. To produce accurate forecasts of the confusion limit in the far infrared, we thus need a realistic model of the statistical source properties at these wavelengths. 

The simulated infrared dusty extragalactic sky (SIDES, \citealt{Bethermin2017}\footnote{The material associated to SIDES can be found at \url{https://data.lam.fr/sides/home}.}) is a semi-empirical model populating dark-matter halos from numerical simulations using recent observed physical relations. In this paper, we use the 2\,deg$^2$ version of SIDES. It connects the halo mass to the stellar mass using a sub-halo abundance matching technique \citep[e.g.,][]{Behroozi2013}. A fraction of the galaxies are drawn to be star-forming based on their stellar mass and redshift, and emits in the far-infrared. Their star formation rate (SFR) is then drawn based on the evolution of the main-sequence of star-forming galaxies, which is the relation between SFR and the stellar mass evolving with redshift \citep[e.g,][]{Schreiber2015}. The observed scatter around this relation is taken into account by SIDES, and a population of high-SFR outliers is labeled as starbursts. Different spectral energy distributions (SEDs) are then attributed to galaxies depending on whether they are starbursts or not. These SEDs evolve with redshift following the observations of warmer dust at higher redshift \citep[e.g.,][]{Bethermin2015a}. A temperature scatter on the SED templates is also included in the simulation. The AGN contribution is not included in the model, but \citet{Bethermin2012c} showed that it has a small impact on the galaxy number counts and thus the confusion noise (see Eq.\,\ref{eq:confusion} for the link between number counts and confusion noise).

This model reproduces successfully a large set of observables. The source number counts from the mid-infrared to the millimeter are very well reproduced after taking into account the resolution effects leading to the blending of some galaxies (\citealt{Bethermin2017}, see also \citealt{Bing2023} for recent results in the millimeter). The simulation produces the correct redshift distributions and number counts in redshift slices. This capability of reproducing the galaxy flux density and redshift distributions over a large set of wavelengths suggests that both the SED and redshift distribution of galaxies are realistic, and confirms the relevance of our model to derive confusion limits. In addition, statistical measurements suggest that faint sources below the detection limits are also properly modeled. For instance, the  histogram of pixel intensities in \textit{Herschel}/SPIRE maps \citep[250--500\,$\mu$m,][]{Glenn2010}, also called P(D), is well reproduced after taking into account the clustering \citep{Bethermin2017}. However, some flux density and wavelength ranges targeted by PRIMAger were never observed before (e.g., there is a lack of deep observations between 24 and 70\,$\mu$m), and we have to rely on the capability of our model to extrapolate correctly in these ranges.

Finally, the CIB anisotropies (e.g., \citet{Planck_CIB2013,Viero2013}), which is the integrated background from dust emission of galaxies at all redshifts, are also correctly recovered by SIDES, including cross-power spectra between different wavelengths \citep{Bethermin2017,Gkogkou2023}. The CIB anisotropies at small scale are dominated by the shot noise from galaxies below the detection threshold, while the signal at large scale is caused by the galaxy clustering. This agreement enhances our confidence that our model characterizes both the clustering and flux distribution of faint sources. The faithfully reproduced cross-power spectra indicate that  the galaxy colors are also reasonable.

\begin{figure*}
\centering
\begin{tabular}{ccc}
\includegraphics[width=5.95cm]{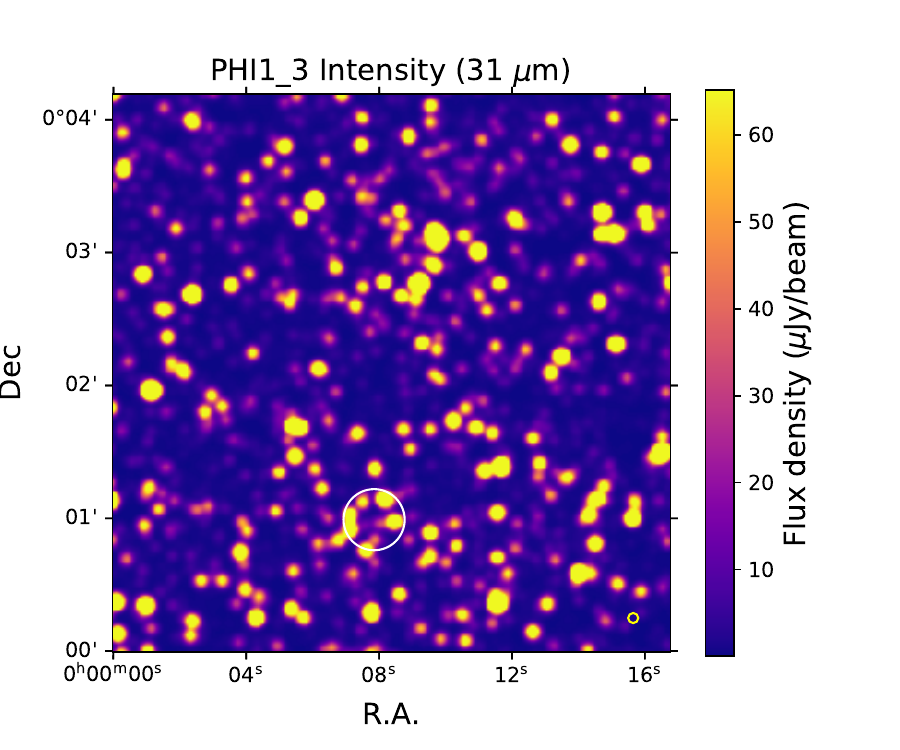} & \includegraphics[width=5.9cm]{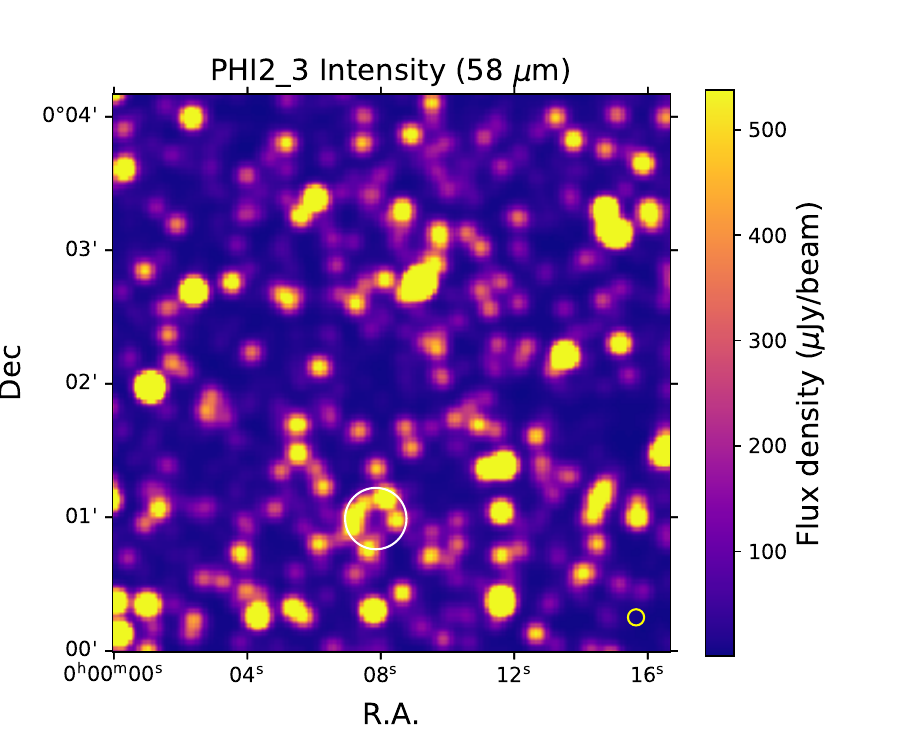} & \includegraphics[width=5.9cm]{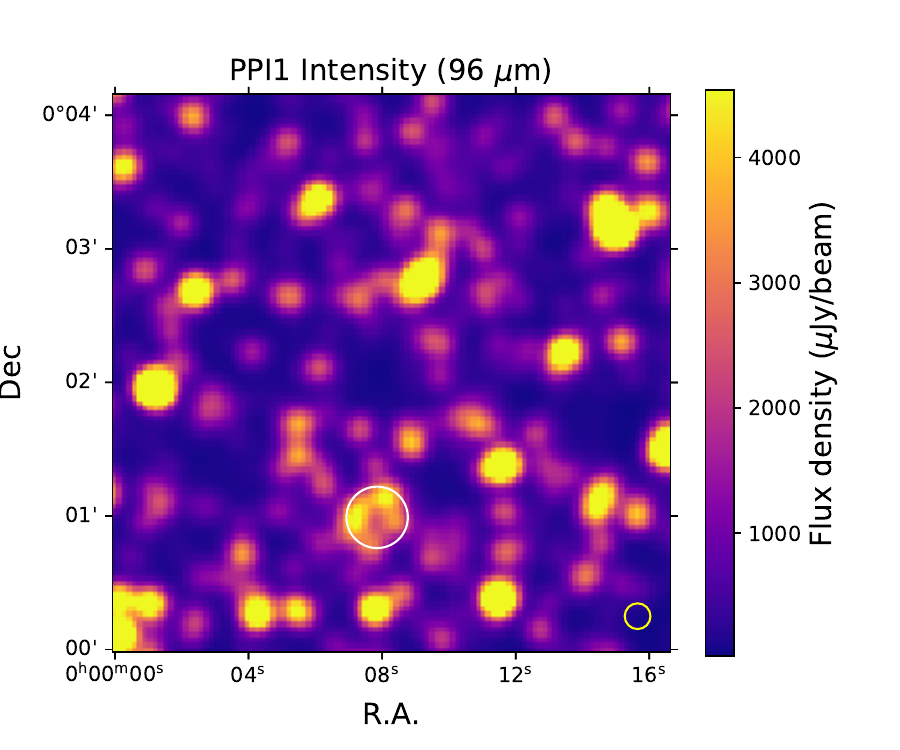}\\
\includegraphics[width=5.9cm]{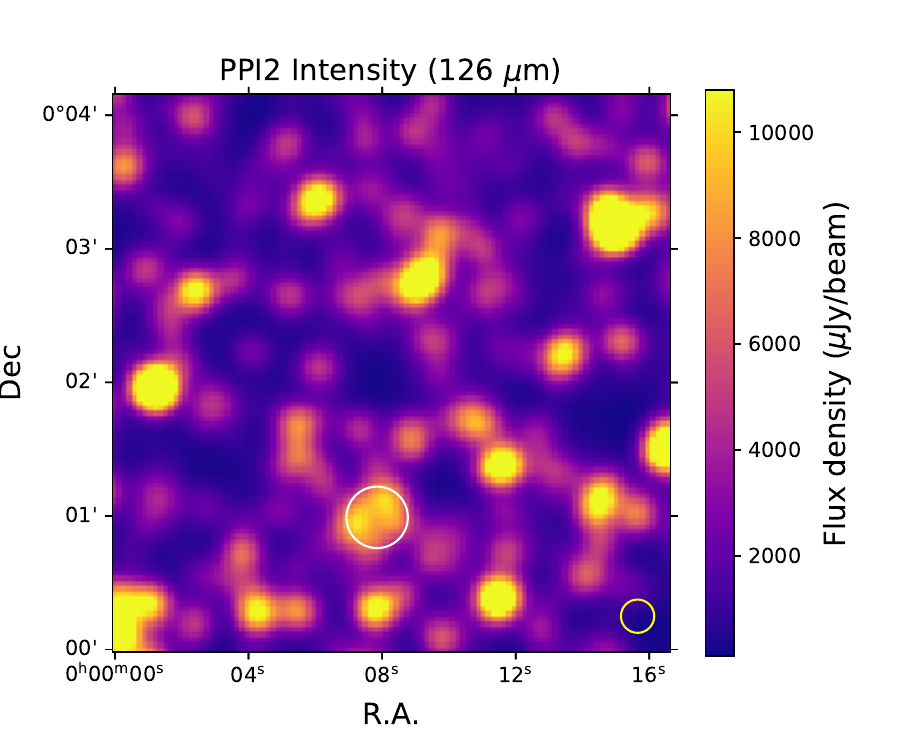} & \includegraphics[width=5.9cm]{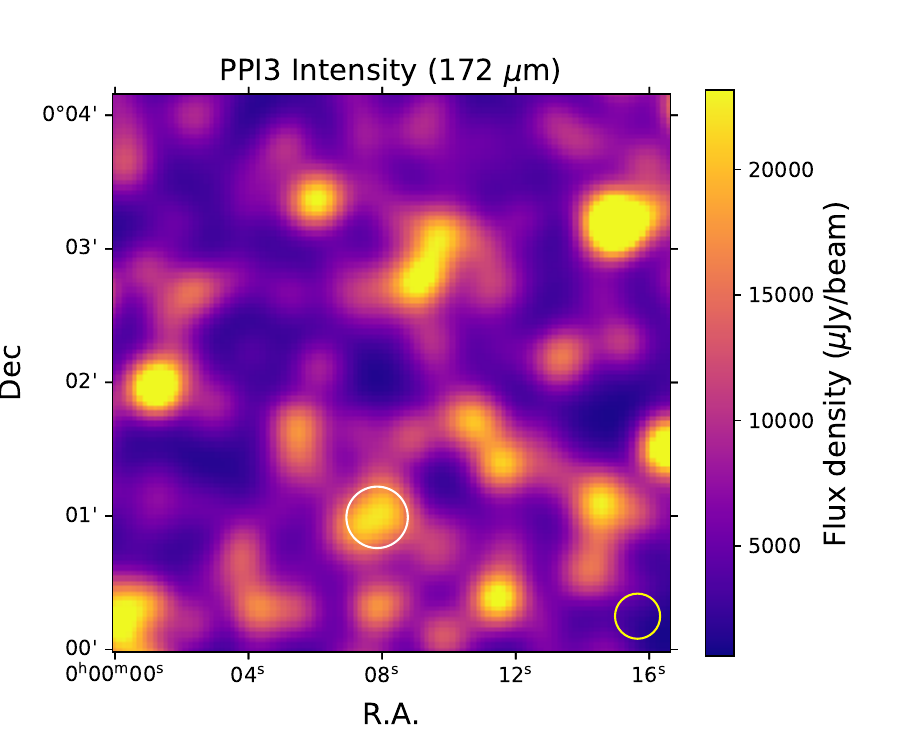} & \includegraphics[width=5.9cm]{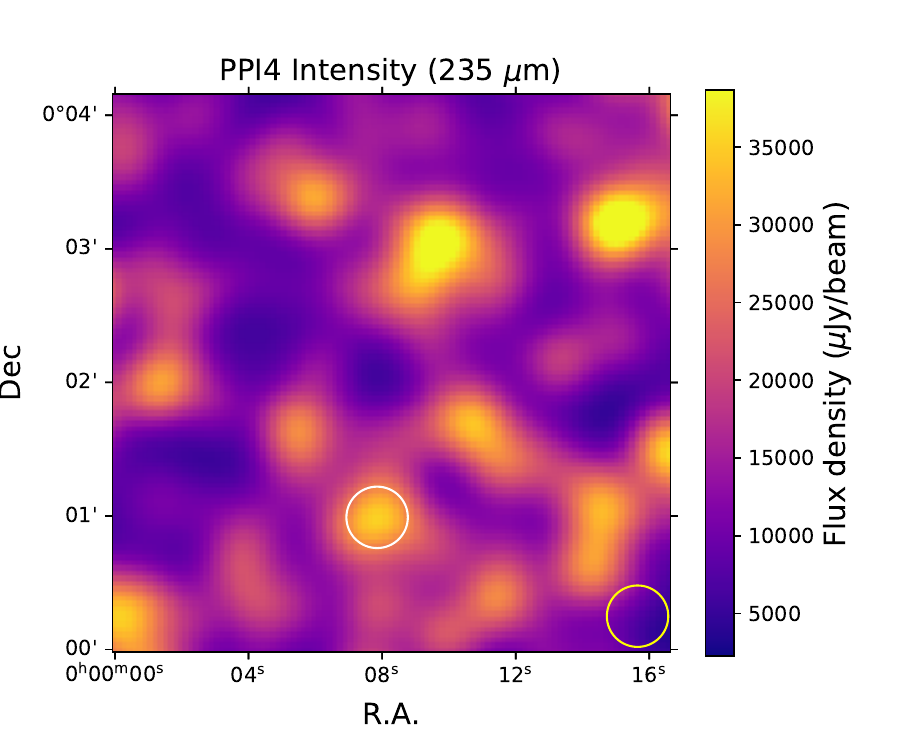}\\
\includegraphics[width=5.9cm]{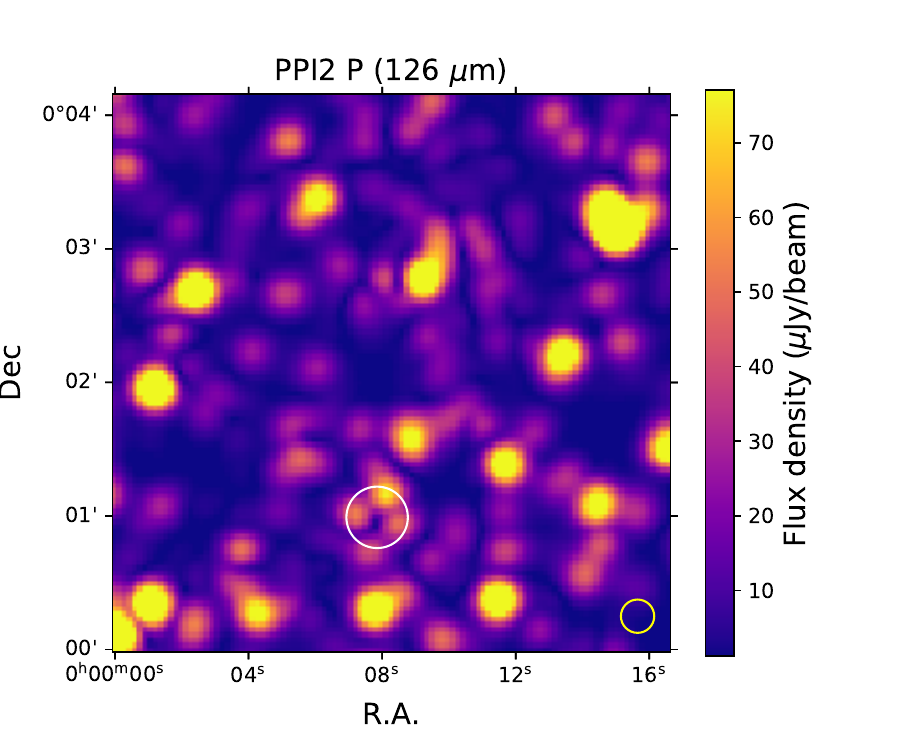} & \includegraphics[width=5.9cm]{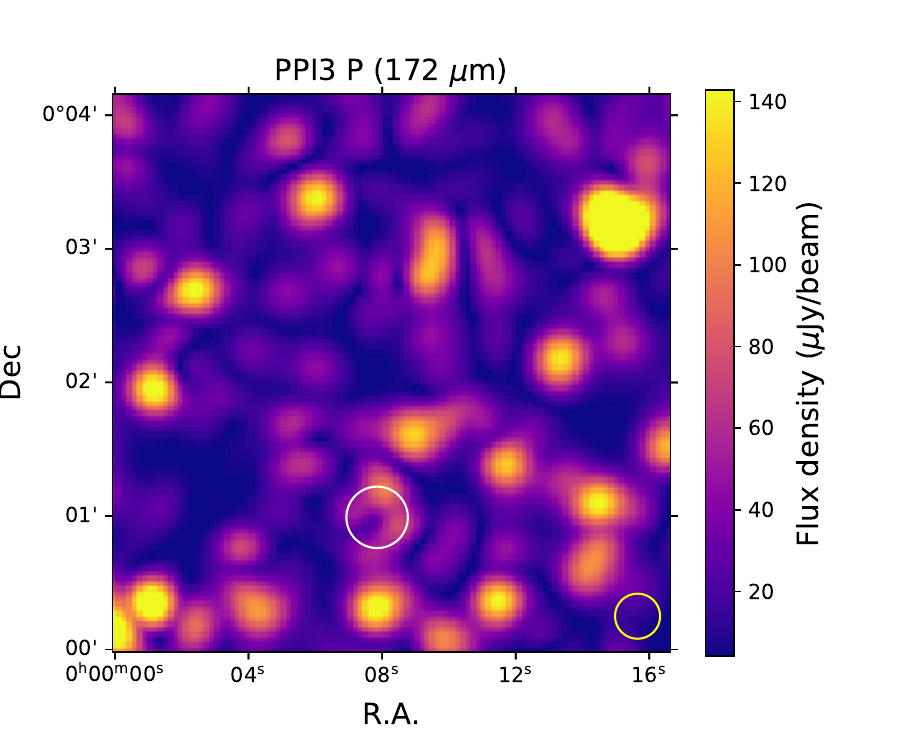} & \includegraphics[width=5.9cm]{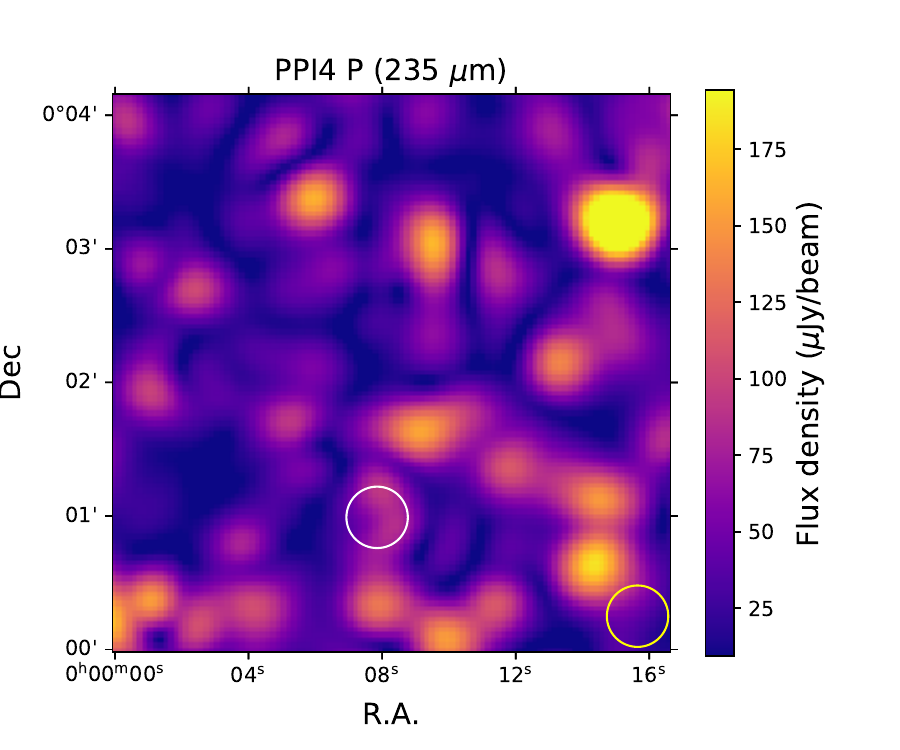}\\
\includegraphics[width=5.9cm]{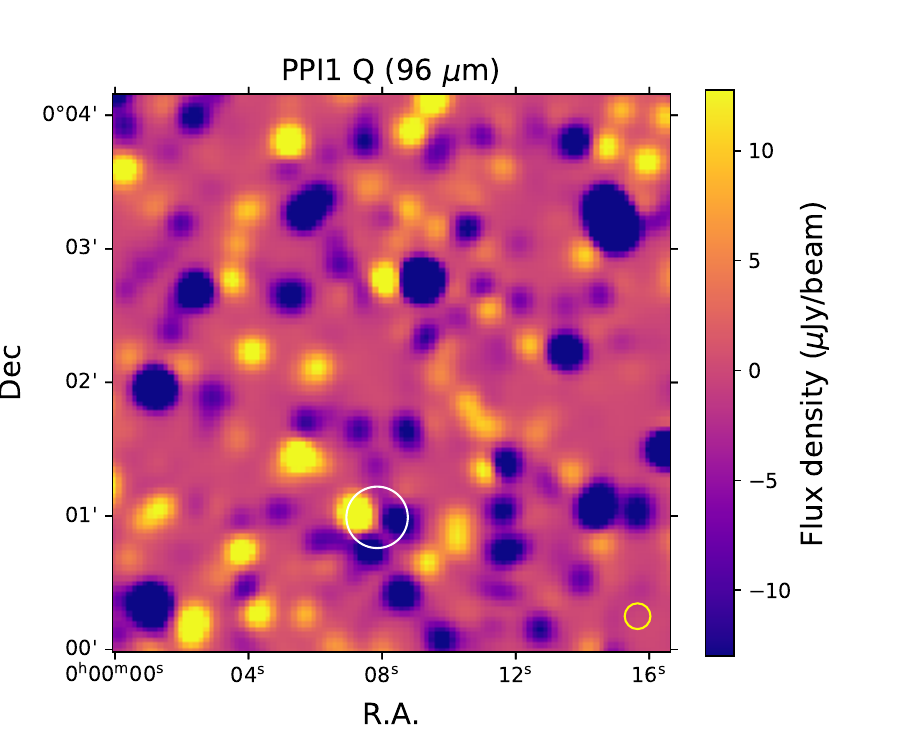} & \includegraphics[width=5.9cm]{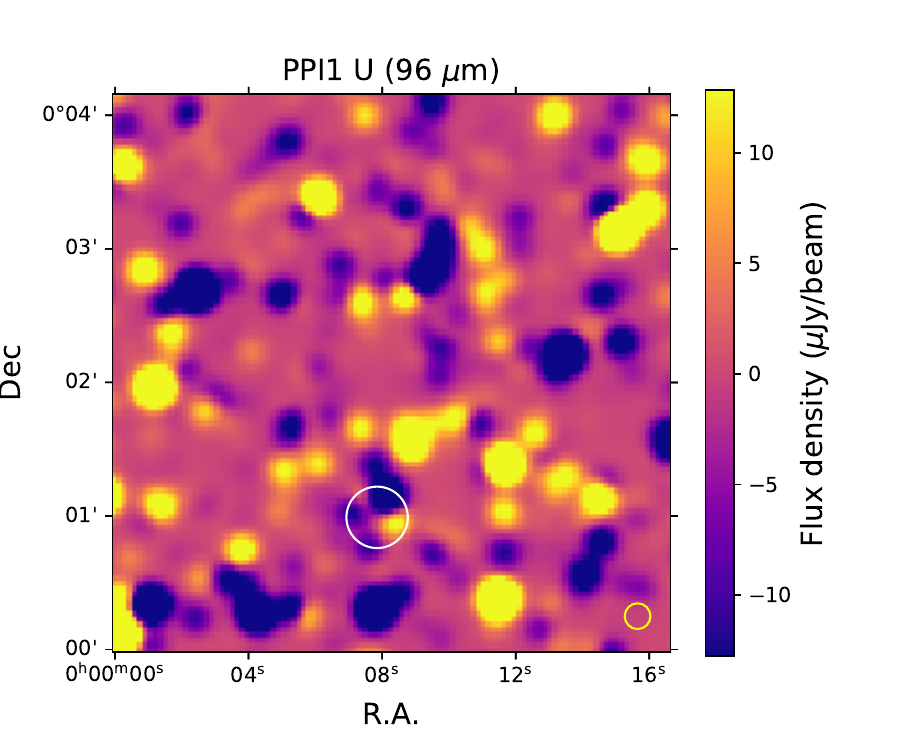} & \includegraphics[width=5.9cm]{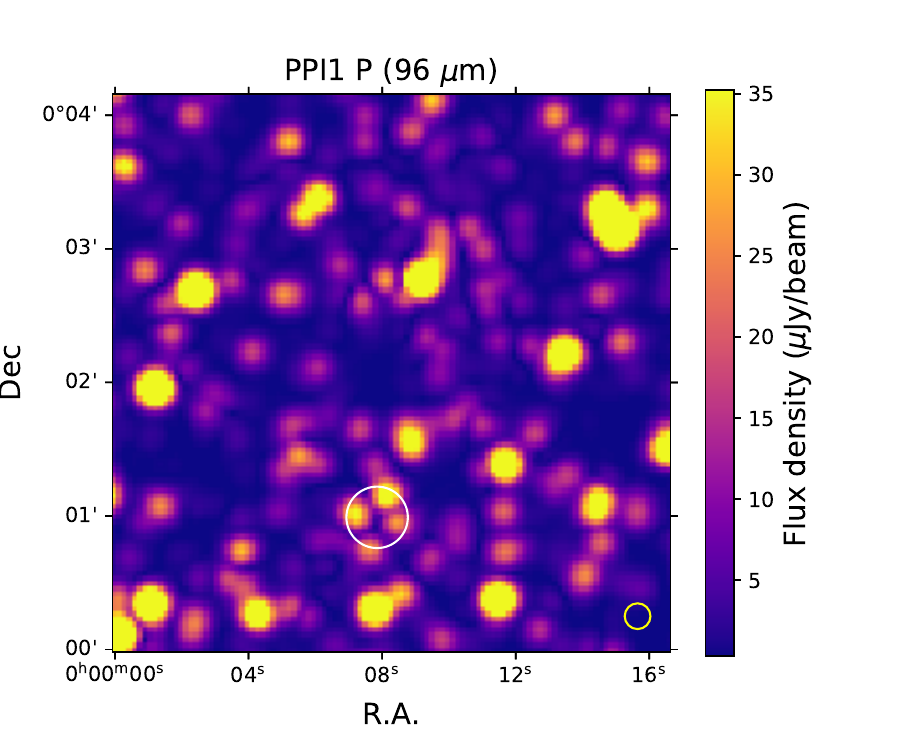}\\
\end{tabular}
\caption{\label{fig:maps} Cutouts of our simulated PRIMAger noiseless maps produced by SIDES. The first two rows present the intensity maps of the various bands. The third row show the polarized flux ($P=\sqrt{Q^2+U^2}$) maps of the PPI2, PPI3, and PPI4 bands, which can be compared with intensity maps in the same bands in the second row (see discussion in Sect.\,\ref{sect:simpolar}). The fourth row contains the Q, U, and P maps in PPI1 band, which illustrates how the Q and U maps combined into the P map. The PPI4 source indicated with a white circle is discussed in Sect.\,\ref{sect:simpolar}. The instrumental beam size is indicated using a yellow circle in the bottom-right corner.}

\end{figure*}

\begin{table}
\caption{\label{tab:bands}Summary of the properties of the various PRIMAger filters used in our analysis.}
\centering
\begin{tabular}{lcccc}
\hline
\hline
Band& Central  & Filter & Beam & With  \\
name & wavelength & width & FWHM & polarization? \\
 & [$\mu$m] & [$\mu$m] & [\arcsec] & \\
\hline
PHI1\_1 & 25.0 & 2.5 & 4.1 & No \\
PHI1\_2 & 27.8 & 2.8 & 4.3 & No \\
PHI1\_3 & 30.9 & 3.1 & 4.6 & No \\
PHI1\_4 & 34.3 & 3.4 & 4.9 & No \\
PHI1\_5 & 38.1 & 3.8 & 5.2 & No \\
PHI1\_6 & 42.6 & 4.3 & 5.7 & No \\
PHI2\_1 & 47.4 & 4.7 & 6.2 & No \\
PHI2\_2 & 52.3 & 5.2 & 6.7 & No \\
PHI2\_3 & 58.1 & 5.8 & 7.3 & No \\
PHI2\_4 & 64.5 & 6.5 & 8.0 & No \\
PHI2\_5 & 71.7 & 7.2 & 8.8 & No \\
PHI2\_6 & 79.7 & 8.0 & 9.7 & No \\
\hline
PPI1 & 96.3 & 23.0 & 11.6 & Yes \\
PPI2 & 126 & 31.0 & 15.0 & Yes \\
PPI3 & 172 & 43.0 & 20.3 & Yes \\
PPI4 & 235 & 61.0 & 27.6 & Yes \\
\hline
\end{tabular}
\tablefoot{As explained in Sect.\,\ref{sect:simmaps}, the PHI1 and PHI2 bands are linear-variable filters and are represented in our analysis by 6 representative sub-filters (e.g., PHI1\_X for the Xth representative filter of band PHI1).}
\end{table}

\subsection{Simulated PRIMAger maps}

\label{sect:simmaps}

The confusion limit is the faintest flux density at which we can extract sources reliably in the limit of zero instrumental noise. In this paper, the instrumental noise refers both to the noise coming from the instrument itself and the photon noise from the various diffuse astrophysical backgrounds. To estimate the confusion limit, we produced simulated PRIMAger maps without instrumental noise using SIDES. Our simulations do not contain Galactic cirrus, which can potentially impact the photometry but are usually very faint in fields chosen for the deep surveys. However, in polarization, Galactic cirrus are expected to have a typically 5 times higher polarization fraction than unresolved galaxies (5\,\% versus 1\,\%, see \citealt{Planck_polar} and Sect.\,\ref{sect:simpolar}). The survey footprints will thus have to be chosen very carefully to minimize the cirrus contamination in polarization.

We assume a Gaussian beam with a full width at half maximum (FWHM) as listed in Table\,\ref{tab:bands}. These model beams are only determined by the optics and do not take into account the impact of the pointing accuracy, the scanning strategy, nor the future map-making pipeline process. The impact of these effects are discussed in Appendix\,\ref{sect:instrpix_impact}.

The flux densities in the PHI1 and PHI2 bands are derived assuming rectangular filters with spectral resolution $R = \lambda / \Delta \lambda = 10$. These bands employ linearly variable filters which, for simplicity, we decided to represent them with six effective filters spanning the wavelength range of each band (PHI1\_1 to PHI1\_6 for the PHI1 band and PHI2\_1 to PHI2\_6 for the PHI2 band). This simplification has little consequence for confusion since it is caused by galaxies at various redshifts, and the polycyclic aromatic hydrocarbon (PAH) spectral features are thus smoothed. It would be consequential in other circumstances, such as redshift estimation from PAH emission bands. For PPI bands, we also assume rectangular filters, but with the currently planned widths that are listed in Table\,\ref{tab:bands}. 

We produce simulated maps without instrumental noise using the integrated map maker of SIDES ({\verb+make_maps.py+}, \citealt{Bethermin2022}). To ensure a sufficient sampling of the beam while keeping the data volume of the maps reasonable, we chose pixel sizes of 0.8\arcsec\  in the PHI1 band, 1.3\arcsec\ in PHI2 band, and 2.3\arcsec\ in the PPI bands. This allows us to have at least 5 pixels per FWHM.

The cutouts (1/20 of the field width, 4.24\arcmin) of the resulting maps are presented in Fig.\,\ref{fig:maps} (first two rows for the intensity maps). For simplicity, only one representative filter per band is shown for PHI. As expected, because of the higher resolution at shorter wavelengths, the number of clearly defined (unconfused) sources is much higher in PPI1 than in PPI4.

\subsection{Adding integrated polarization to SIDES}

\label{sect:simpolar}

The PPI bands will measure the linear polarization. To study the impact of confusion on polarization measurements, we added a simple polarization model to SIDES motivated by observations of the local Universe. 

We consider only the integrated polarization coming from the entire galaxy, because PRIMAger will not spatially resolve distant galaxies ($z \gtrsim 0.1$). As the plane-of-the-sky magnetic field orientations vary across the galaxy, the integrated dust polarization fraction from the entire galaxy is measured to be lower than those from individual lines of sight. Observations of a sample of local galaxies with the SOFIA telescope \citep[the SALSA survey,][]{Lopez-Rodriguez2022} demonstrated that the average integrated polarization fraction is $\sim$1\,\%. A similar value ($p=0.6\pm0.1\,\%$) has been measured in a lensed high-z galaxy \citep{Geach2023}. Since the physics of the dust polarization is very complex and necessitates physical parameters not included in SIDES, we chose a semi-empirical probabilistic approach anchored to the local Universe observations. In addition, more physical galaxy evolution models tend to struggle to reproduce far-infrared observables in intensity and their capability to produce dust polarization has never been tested.

%explain the polar fraction in SIDES
In SIDES, we draw the galaxy integrated polarization fractions ($p$) from a Gaussian model of the distribution observed in the local Universe by \citet{Lopez-Rodriguez2022} centered on 1\,\% and with a standard deviation of 0.3\,\%. The method used to derive these values is presented in Appendix\,\ref{sect:polar_dependence}. Since the 3-$\sigma$ confidence interval of the central value of the polarized fraction goes from 0.7\,\% to 1.3\,\%, we will also discuss scenarios with these extreme values to illustrate the uncertainties on our forecasts. For simplicity, we assume that the relative scatter ($\sigma / \mu$) on the polarized fraction is constant (0.3) in all the scenarios to be able to apply a simple rescaling to the polarized maps. In the current version of the model, we do not consider a dependence of the polarization with wavelength or the presence of a starburst in the galaxy. \citet{Lopez-Rodriguez2023} showed that starburst outflows can produce a specific wavelength-dependent signature. However, SIDES does not contain a model for outflow, and we could not calibrate an empirical dependence on these parameters using the SALSA sample since we did not find any statistically-significant effect on the integrated polarization (see Appendix\,\ref{sect:polar_dependence}). 

%https://ui.adsabs.harvard.edu/abs/2015MNRAS.450.2195S/abstract
%https://ui.adsabs.harvard.edu/abs/2018MNRAS.481.4753C/abstract

%explain the draw of the angle
As in \cite{Lagache2020}, we neglect the intrinsic alignments between the integrated polarization angles of galaxies. This is primarily motivated by the small ($<$5\,\%) probability of a spiral galaxy to be aligned with its neighbors found both in observations \citep{Singh2015} and in simulations \citep{Codis2018}. We thus drew randomly the polarization angle $\alpha$ from a uniform distribution between 0 and $\pi$. The various integrated Stokes parameter ($Q$, $U$) and the polarized flux density ($P$) are then derived for each simulated galaxy using:
\begin{eqnarray} 
\label{eq:polar}
\notag
Q &=& p I \cos{2 \alpha},\\
U &=& p I \sin{2 \alpha},\\
\notag
P &=& \sqrt{Q^2+U^2} = p I,
\end{eqnarray} 
where $I$ is the intensity.

The $Q$ and $U$ maps are built using the method described in Sect.\,\ref{sect:simmaps} using the flux densities in $Q$ and $U$ instead of $I$. Contrary to $I$, $Q$ and $U$ can be negative, and the flux of the sources in the same beam do not add up systematically. It is thus important to compute the $Q$ and $U$ maps before generating the observed $P$ map by combining them quadratically. The polarized maps are presented in Fig.\,\ref{fig:maps} (last two rows). The comparison between the intensity and polarized maps (second and third rows) demonstrates immediately that the polarized maps are not rescaled versions of the intensity maps. While the flux of blended sources add up in intensity map to produce extended bright blobs, this is not the case in polarization. If the polarization angles are not aligned, the flux of two sources can potentially lead to depolarization. A good example can be found around the coordinates (00h00m08s, +00$\degree$01'00"). It is highlighted with a white circle in Fig.\,\ref{fig:maps}. The source is relatively bright in intensity in the PPI4 band, but barely visible in polarization in the same band. The PPI1 $Q$, $U$, and $P$ maps (last row), which have a better resolution than the PPI4 map, help to understand the origin of this effect. The PPI1 $P$ map exhibits four components in the beam. In the $Q$ map, the eastern component has a positive signal, while the southern and western ones are negative. The northern component has no significant $Q$ signal. In the $U$ map, the northern component has a strong negative signal, the southern and eastern ones have a weaker negative signal, and the western component has a positive signal. When the beam is larger, the four components merge partially canceling each other $Q$ and $U$ signal. This leads to a small polarized flux density $P$.

%%%%%%%%%%%%%%%%%%%%%%%%% METHODS %%%%%%%%%%%%%%%%%%%%%%

\section{Source extraction and determination of our confusion metrics}

\label{sect:methods}

\subsection{Our basic source extractor}

\label{sect:extractor}

In this paper, our goal is to determine the baseline performance that can be expected from PRIMAger at the confusion limit, i.e. when the instrumental noise is negligible.  For this baseline we purposefully chose a basic source detection and photometry method, making use of the standard {\verb+photutils+} package \citep{photutils}. This method is not intended to be optimal, and instead provide a robust estimate of the performance expected with a basic blind extraction algorithm.

We expect significantly  better performance for a more advanced photometry method (XID+, \citealt{Hurley2017}) relying on priors and this is presented in \citet{Donnellan2024}. 

Point source detection and photometry methods typically employ an image filter. It is common for this to be optimized for the detection of isolated sources in the presence of uncorrelated instrumental noise, in which case the image is convolved with the PSF.
Since we work at the instrumental-noise-free limit and confusion is our main concern this filter is not optimal.  Indeed, filtering broadens the effective beam and increases the confusion noise. Optimal filters are discussed in \citet{Donnellan2024}. Here, for simplicity, we choose to apply no filtering.

The sources are detected by searching for the brightest pixel above a given threshold (choice discussed in Sect.\,\ref{sect:conflim_def}) within a local region using the {\tt find\_peaks} algorithm. The local region is defined to be 5$\times$5\,pixel square corresponding roughly to the beam half-light radius. Our maps are in units of Jy/beam, and the flux of the sources is estimated by recording the value of the central pixel in the background subtracted map. Finally, we determined the sub-pixel centroid of the sources using the {\verb+centroid_sources+} algorithm using the same $5\times5$ region size used in  detection.

\begin{figure}
\centering
\includegraphics[width=9cm]{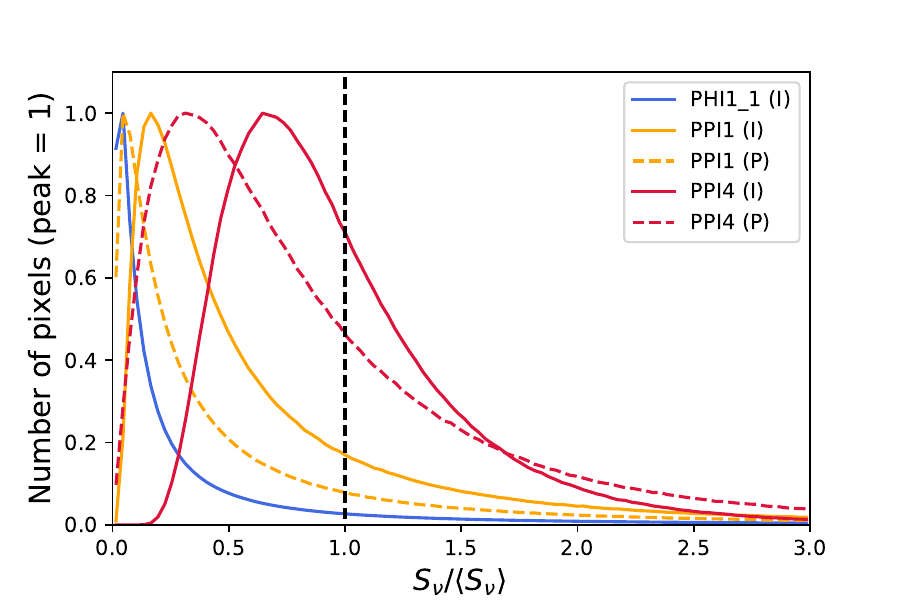}
\caption{\label{fig:map_histos} Histogram of pixel flux densities of our simulated PRIMAger maps in various bands (PHI1\_1 in blue, PPI1 in orange, and PPI4 in red) in intensity $I$ (solid lines) and polarization $P$ (dashed lines). The x-axis is normalized by mean of the map, while the y-axis is normalized to unity at the peak. The vertical dashed line correspond to the mean of the map.}
\end{figure}

%\subsection{Background, extraction threshold, and basic confusion limit estimate}
\subsection{Background}

\label{sect:extr_setting}

To detect sources and perform photometry it is crucial to evaluate the background. 

This is not trivial for data from far-infrared observatories, since they are  not usually absolute photometers. Real maps often have a zero mean enforced in order to filter out instrumental and celestial backgrounds and foregrounds. In addition, because of confusion, no region is free from sources to be suitable for estimating the background. To illustrate this, we show the flux density distribution from our PRIMAger simulated maps in Fig.\,\ref{fig:map_histos}. These simulated maps have a true zero, which corresponds to the absence of any galaxy emission. However, diffuse background (e.g., cosmic microwave background) and foregrounds (e.g., zodiacal light) are not included in the simulation and nor do they include instrumental backgrounds. In practice, we cannot use this zero for the photometry, since it cannot be measured in real data. Furthermore the faint sources provide an unresolved background that we should remove for unbiased photometry.

We thus chose to use the mode of the distribution to define the zero point. In Fig.\,\ref{fig:map_histos}, the x-axis is normalized by the overall mean of the map. The mode is well below the mean (dashed vertical line). At short wavelength (PHI1 in blue), where the beam is small, the mode is very close to the true zero, since most beams contains only very faint sources. At long wavelengths (PPI4 in red), few pixels are close to zero and the mode is closer to the mean, since each beam contains a significant number of sources creating a background. Similar behavior is observed in polarization.

\begin{table}
\centering
\caption{\label{tab:conf_levels} Summary of the 5$\sigma_{\rm conf}$ classical confusion limits (Sect.\,\ref{sect:conflim_def}) obtained with our minimal source extractor in intensity (I) and polarization (P), together with the 50\,\% and 80\,\% completeness flux densities (Sect.\,\ref{sect:puco_est}) and the source surface density.}
\begin{tabular}{lcccccc}
\hline
\hline
Band & Type & Central & 5-$\sigma$ & 50\,\% & 80\,\% & Surface \\
name &  & wavelength & limit & comp. & comp. & density \\
 &  & $\mu$m & mJy & mJy & mJy & deg$^{-2}$\\
\hline 
PHI1\_1 & I & 25.0 & 0.021 & 0.02 & 0.046 & 43962 \\
PHI1\_2 & I & 27.8 & 0.027 & 0.027 & 0.055 & 39602 \\
PHI1\_3 & I & 30.9 & 0.037 & 0.037 & 0.07 & 35070 \\
PHI1\_4 & I & 34.3 & 0.051 & 0.051 & 0.091 & 30680 \\
PHI1\_5 & I & 38.1 & 0.072 & 0.071 & 0.11 & 26357 \\
PHI1\_6 & I & 42.6 & 0.11 & 0.1 & 0.16 & 22080 \\
PHI2\_1 & I & 47.4 & 0.16 & 0.16 & 0.25 & 18089 \\
PHI2\_2 & I & 52.3 & 0.25 & 0.25 & 0.35 & 15281 \\
PHI2\_3 & I & 58.1 & 0.4 & 0.39 & 0.54 & 12722 \\
PHI2\_4 & I & 64.5 & 0.66 & 0.64 & 0.82 & 10498 \\
PHI2\_5 & I & 71.7 & 1.1 & 1.1 & 1.4 & 8545 \\
PHI2\_6 & I & 79.7 & 1.9 & 1.8 & 2.1 & 6890 \\
\hline
PPI1 & I & 96.3 & 4.6 & 4.4 & 5.1 & 4346 \\
PPI2 & I & 126.0 & 12 & 11 & 13 & 2227 \\
PPI3 & I & 172.0 & 28 & 27 & 30 & 908 \\
PPI4 & I & 235.0 & 46 & 42 & 49 & 306 \\
\hline
PPI1 & P & 96.3 & 0.03 & 0.032 & 0.041 & 5898 \\
PPI2 & P & 126.0 & 0.078 & 0.082 & 0.096 & 3326 \\
PPI3 & P & 172.0 & 0.17 & 0.18 & 0.21 & 1582 \\
PPI4 & P & 235.0 & 0.25 & 0.27 & 0.32 & 697 \\
\hline
\end{tabular}
\tablefoot{The polarized flux density is assumed to be in average 1\,\% of the photometric flux density of a galaxy.}
\end{table}

\subsection{Classical confusion limit estimates and source extraction threshold}

\label{sect:conflim_def}

In the absence of clustering, the variance of the background fluctuations coming from sources below the detection limit ($\sigma_{\rm conf}$) can be computed using \citep{Condon1974,Lagache2000}:
\begin{equation}
\label{eq:confusion}
\sigma_{\rm conf}^2 = \int \int b^2 d \Omega \times \int_0^{S_{\rm lim}} S^2 \frac{dN}{dS} dS,
\end{equation}
where $b$ is the beam function, $S_{\rm lim}$ is the flux limit above which sources can be detected, $\frac{dN}{dS}$ the number of galaxies per flux interval and per solid angle (usually called differential number counts). This equation is implicit, since $S_{\rm lim}$ is usually defined to be $5\sigma_{\rm conf}$ in the confusion limited case and needs to be solved e.g. by iteration. The limit is essential as without the integral would be divergent. However, the choice of where to place the limit is inherently subjective and the choice of $5\sigma$ is a convention without a strong rationale\footnote{The commonly presented rationale is that sources above the limit can be identified and removed and so would not contribute to the residual fluctuations, but a $5\sigma$ threshold will correspond to different detection probabilities depending on the underlying source counts}.

Equation \ref{eq:confusion} is no longer valid if we take into account galaxy clustering \citep{Lagache2020}. \citet[][Sect.\,5.1 and Fig.\,12]{Bethermin2017} showed that the clustering tends to broaden the histogram of pixel flux densities i.e. increase the fluctuations. This is expected since bright sources tend to bunch up together leading to stronger positive fluctuations, while low-density area tends to be even emptier. Since the analytical computation would be very complex, we decided to use an iterative map-based method inspired by Eq.\,\ref{eq:confusion}. We compute the initial standard deviation of the map ($\sigma_{\rm map,0}$), and then recomputed it iteratively after masking the pixels 5\,$\sigma_{\rm map, k}$ above background (see Sect.\,\ref{sect:extr_setting}), where $\sigma_{\rm map, k}$ is the standard deviation at the k-th iteration. After a few tens of iterations, $\sigma_{\rm map}$ converges to the confusion noise $\sigma_{\rm conf}$. The results are summarized in Table\,\ref{tab:conf_levels}.

To extract sources from the simulated maps using the method described in Sect.\,\ref{sect:extractor}, we set the detection threshold to $5\sigma_{\rm conf}$ after subtracting the mode. The photometry is also performed on the mode-subtracted map. The surface density of the detected sources goes from 43\,962 sources per deg$^2$ in the PHI1\_1 band to 306 sources per deg$^2$ in the PPI4 band. All the values are provided in Table\,\ref{tab:conf_levels}. The surface density of sources above the confusion limit is slightly higher in polarization than in intensity in the same band (see discussion in Sect.\,\ref{sect:puco_P}). The confusion metrics expected for a different mean polarization fraction $\mu_p$ can be computed applying a ($\mu_p$/1\,\%) scaling factor (except for the surface densities, see Appendix\,\ref{sect:other_polfrac}).

\subsection{Matching of the input and output catalogs}

\label{sect:match}

Having extracted the sources on the simulated maps, we searched for the source counterparts in the simulated galaxy catalog. Since there can be numerous simulated galaxies in the beam, we chose to define the brightest galaxy in the half-light radius of the beam as the main counterpart. In single-dish far-infrared and submillimeter data, the flux of a source can come from multiple component \citep[e.g.][]{Karim2013,Hayward2013,Scudder2016,Bethermin2017,Bing2023}. In our approach this multiplicity will become apparent as the observed flux density will be larger than the associated counterpart from the input catalog. 

We found no systematic offset between the position of the brightest source and the observed centroid. The peak of the distribution of radial separations is less than the half-light radius by a factor of at least 3. At this cutoff radius, the histogram has a value less than 2\,\% of the peak. This shows that the exact choice of the search radius for the brightest counterpart should have a negligible impact on the final results.

The results are presented in Sect.\,\ref{sect:photo_acc_I} for the intensity and Sect.\,\ref{sect:photo_acc_P} for the polarization. 

\subsection{Completeness and purity estimates}

\label{sect:puco_est}

The completeness as a function of the flux density (polarized or not, in this section we use the term flux density to discuss both the intensity and polarization cases) is a key characterization of the source detection performance. A classical definition of the completeness is the fraction of sources at a given flux density in the input catalog (in our analysis, the SIDES simulated galaxy catalog) that are recovered in the output catalog produced by the source extractor. 
However, since several simulated galaxies can be found in the beam of a single source, recovery can be ambiguous. In this paper, we will use two definitions of a recovered source. In the definition A, we consider that a galaxy from the simulated catalog is recovered if it is located in the half-light radius of any source extracted from the associated simulated map. However, in this case, the completeness does not tend to zero at low flux density. At first order (no clustering), it converges to the fraction of the map encircled in the half-light radii of the extracted sources (up to 7\,\%). To avoid this problem, we introduce a definition B, where the galaxy must satisfy the additional condition of being the brightest source in half-light radius. The results are discussed in Sect.\,\ref{sect:comp_I} and \ref{sect:puco_P}. 

The definition of the purity is also non trivial. The purity is the fraction of ``true'' sources in a sample of detected sources, with the remained being artifacts caused by noise. In our case, there is no instrumental noise, and the density of galaxy in SIDES is so high that every beam contains several simulated galaxies. To declare a detection as ``true'' if there is a simulation counterpart in the beam would thus not be meaningful. Since our goal is to show that the flux of bright individual galaxies can be measured despite the confusion, we chose to consider an extracted source as ``true'' if the brightest counterpart is at least half of the measured flux density (definition A). At long wavelength, the recovered flux density can be systematically overestimated due to the multiplicity described above (see also Sect.\,\ref{sect:photo_acc_I}). We thus computed a second estimate of the purity (definition B) after correcting for this systematic bias, scaling by the median flux density ratio between the brightest galaxy in the beam and the extracted source.

%%%%%%%%%%%%%%%%%%%%%%%%% INTENSITY %%%%%%%%%%%%%%%%%%%%%%

\begin{figure*}
\centering
\begin{tabular}{cc}
\includegraphics[width=8cm]{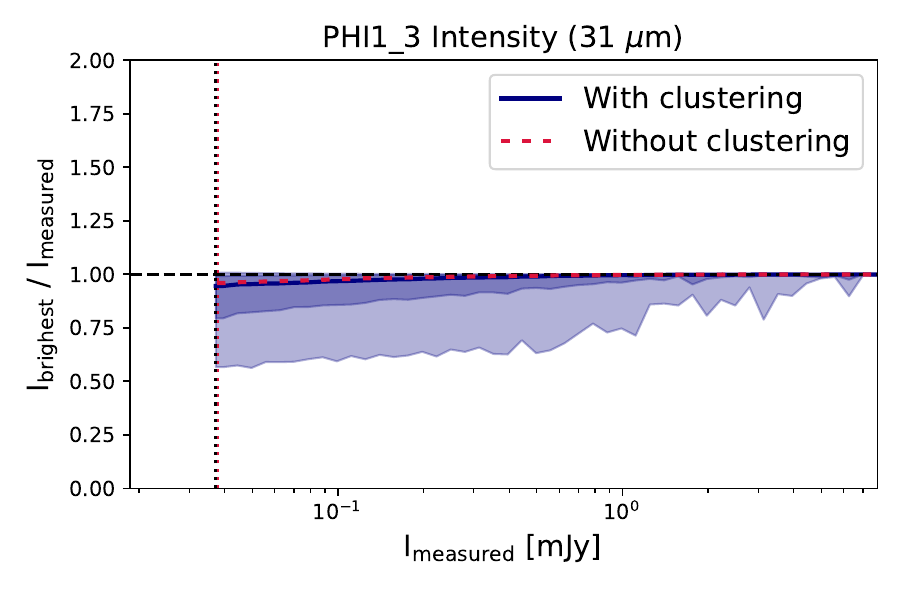} & \includegraphics[width=8cm]{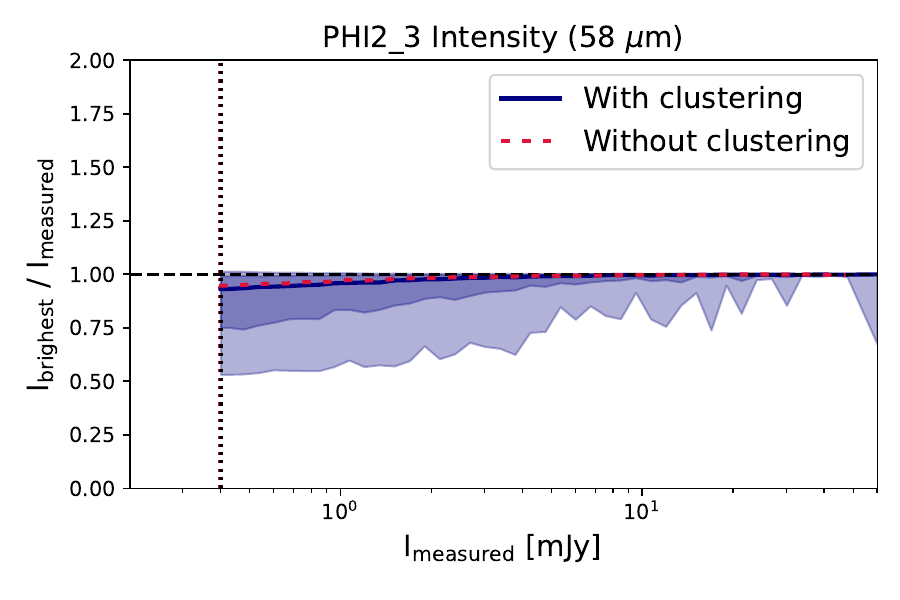}\\
\includegraphics[width=8cm]{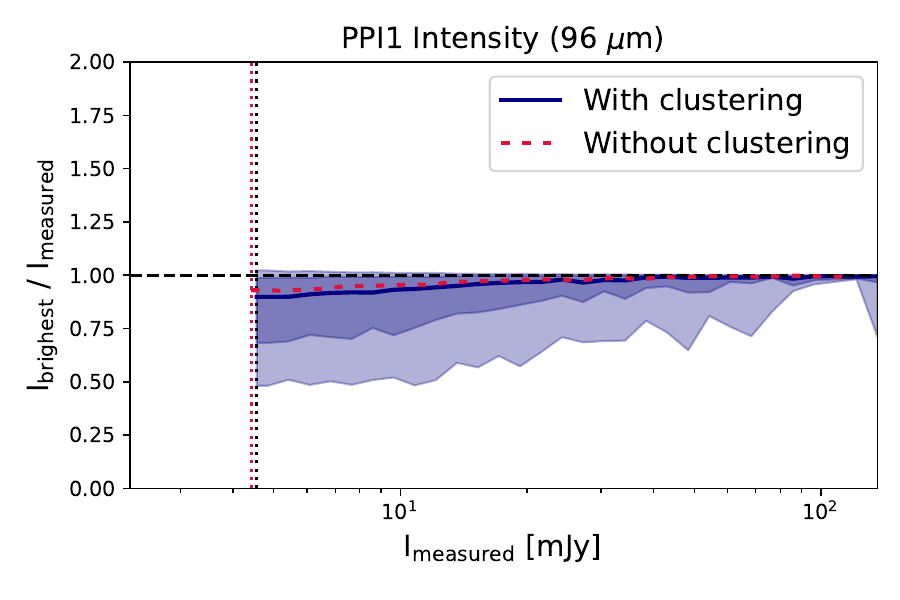} & \includegraphics[width=8cm]{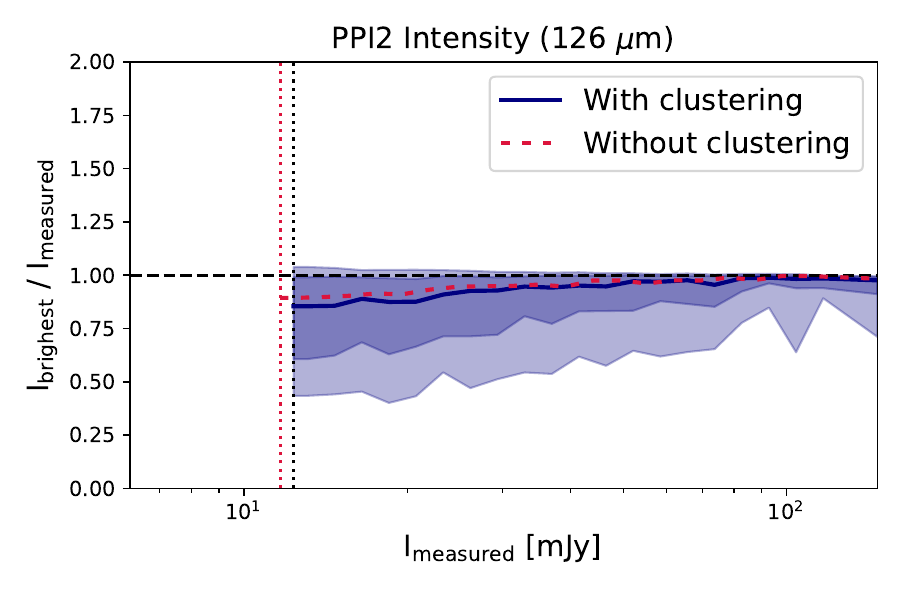}\\
 \includegraphics[width=8cm]{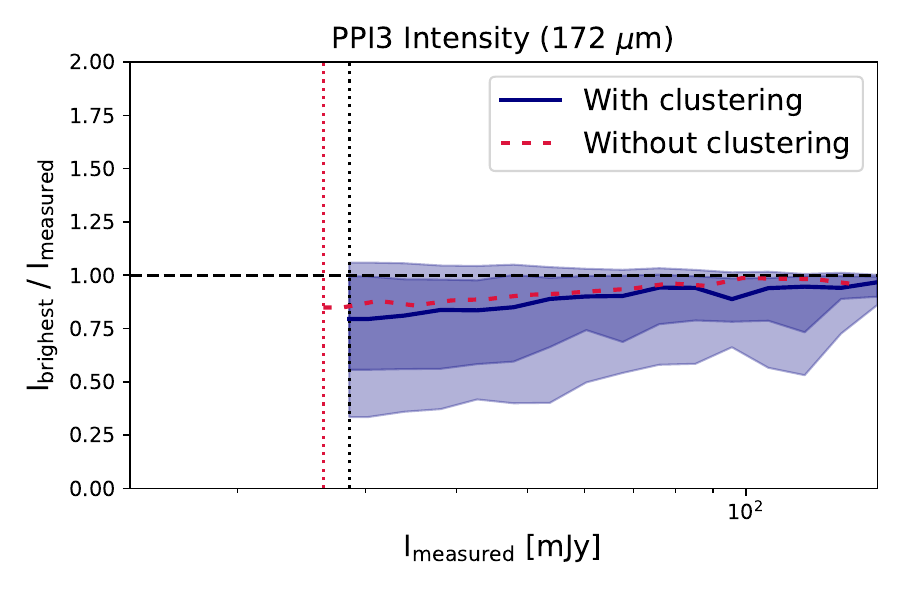} & \includegraphics[width=8cm]{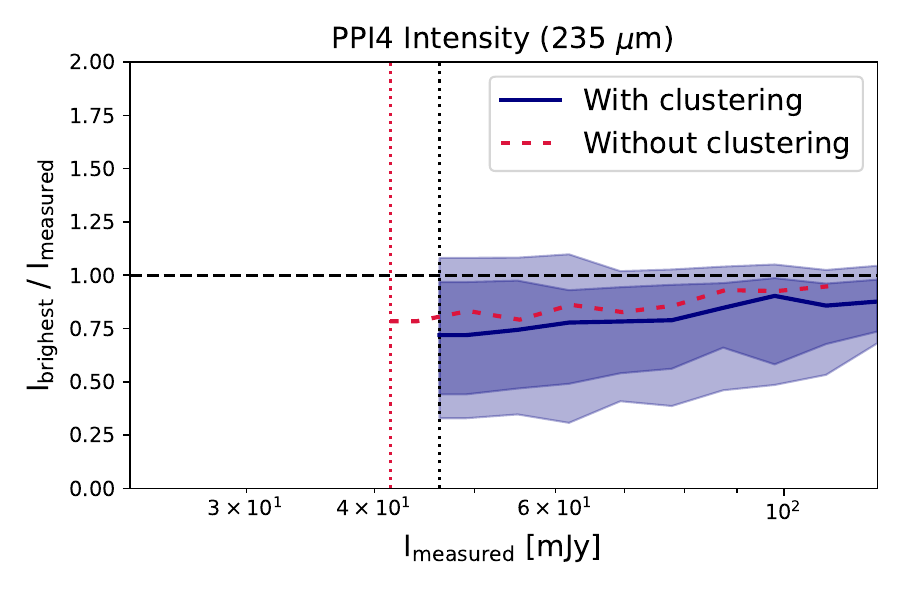} \\
\end{tabular}
\caption{\label{fig:photacc_I} Ratio between the flux density of the brightest simulated galaxy in the beam (see Sect.\,\ref{sect:match}) and the measured flux density in the simulated map as a function of the measured flux density. The various panels corresponds to the various PRIMAger bands in intensity (see title above the panel). For PHI bands, we show only the third representative filters. The dark blue solid line is the median value. The dark and light blue areas represent the 16--84\,\% and 2.3--97.7\,\% ranges, respectively, which are equivalent to 1\,$\sigma$ and 2\,$\sigma$ in the Gaussian case. 
The horizontal dashed line is the one-to-one ratio. The black and red vertical dotted lines are the classical confusion limit estimated in Sect\,\ref{sect:conflim_def} used as detection threshold to produce the output catalog with and without clustering, respectively. The red dashed line represents the median flux density ratio in absence of clustering.}
\end{figure*}

\begin{figure*}
\centering
\begin{tabular}{cc}
\includegraphics[width=8cm]{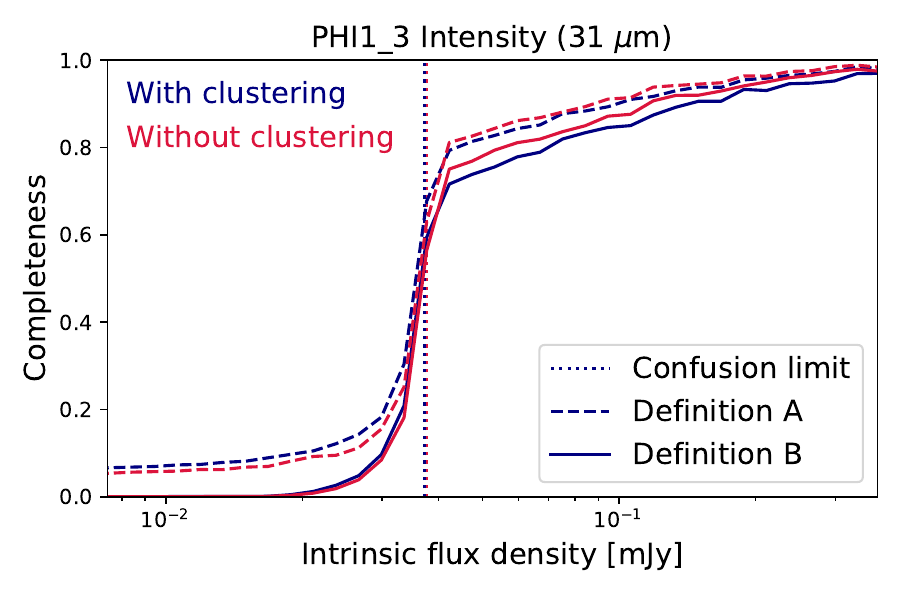} & \includegraphics[width=8cm]{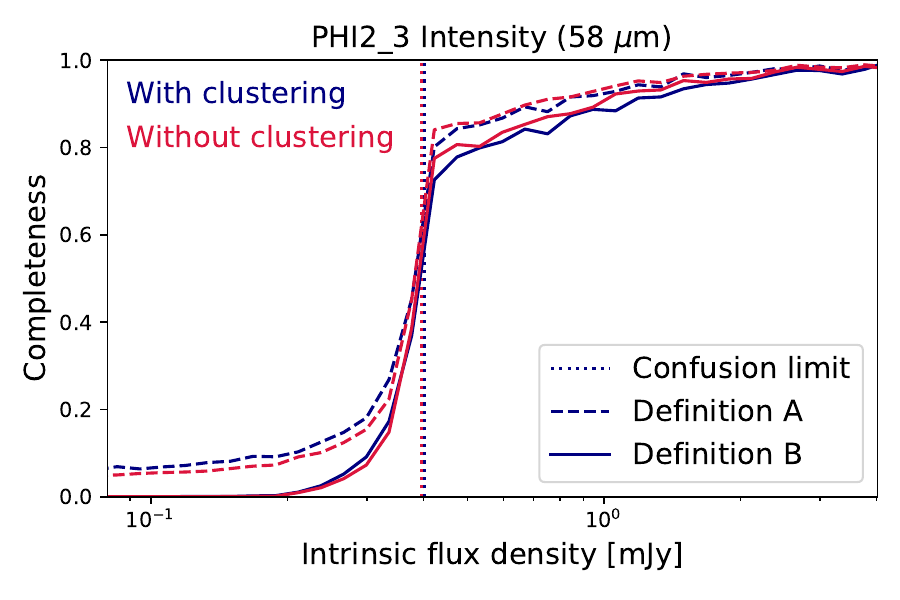} \\
\includegraphics[width=8cm]{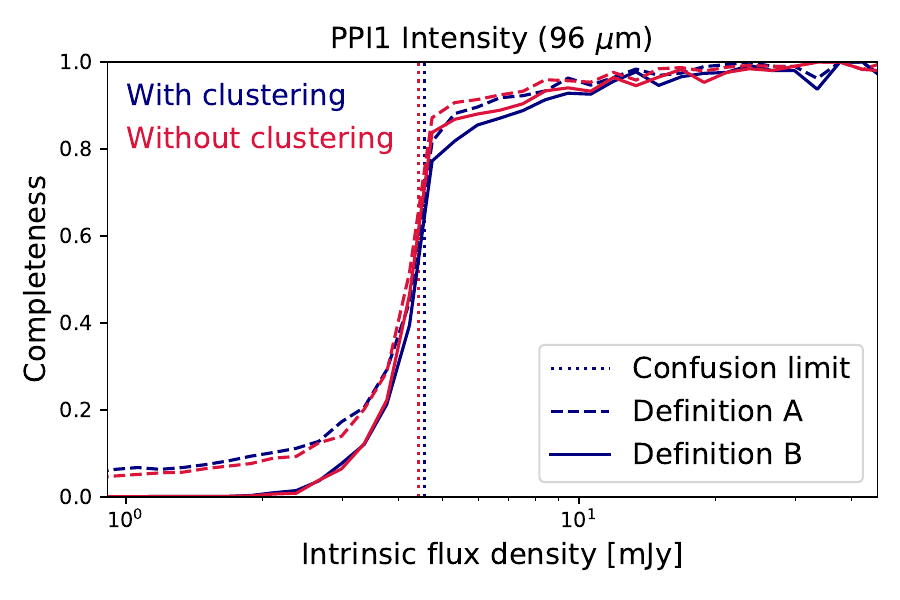} & \includegraphics[width=8cm]{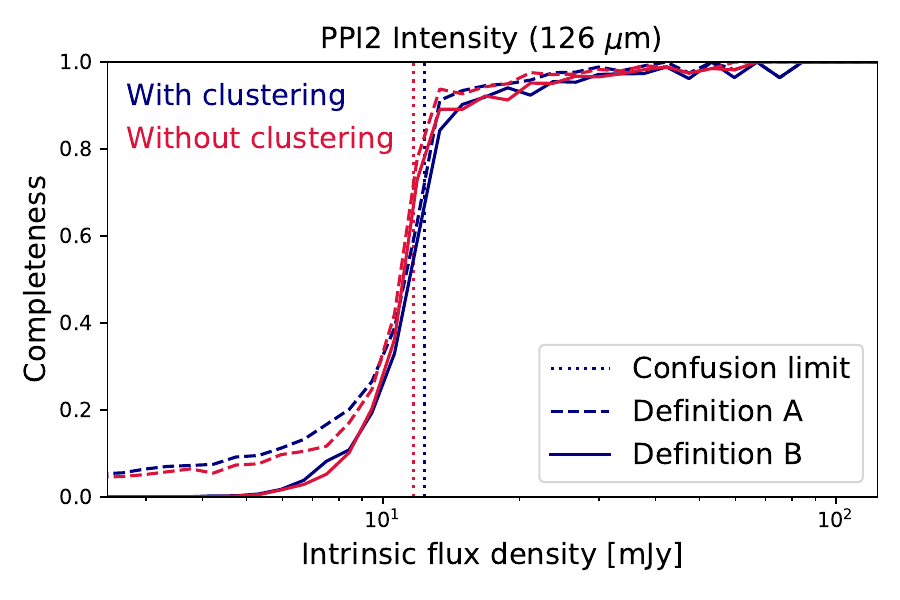} \\
\includegraphics[width=8cm]{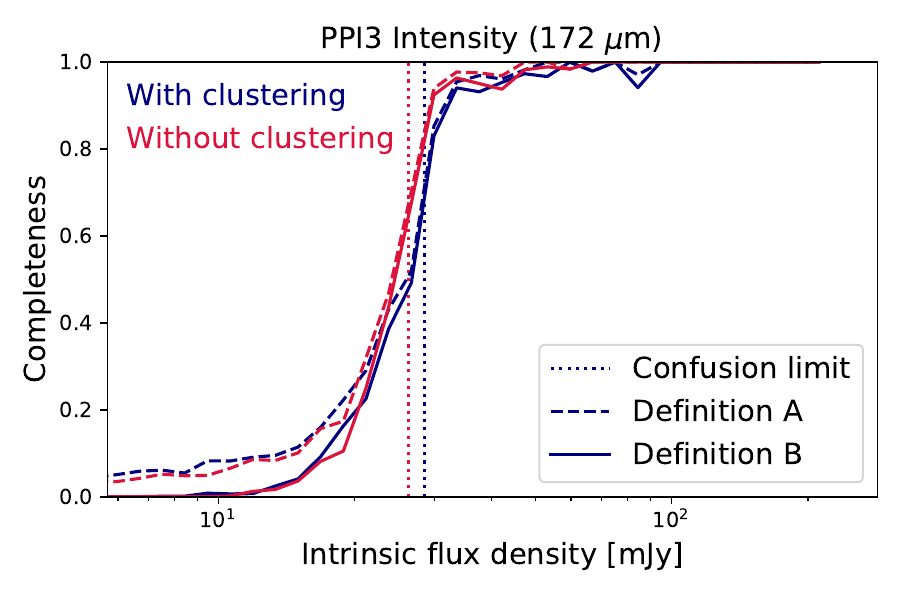} & \includegraphics[width=8cm]{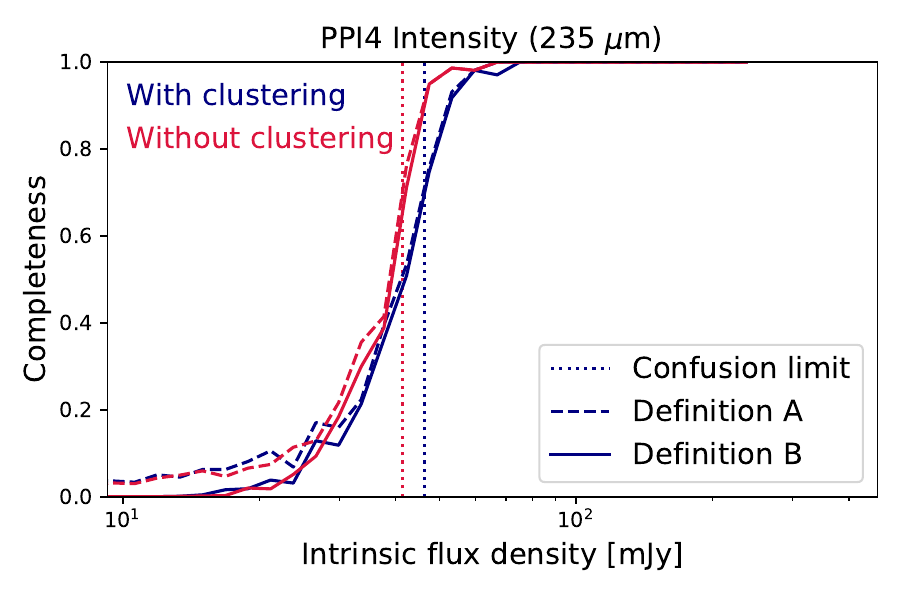} \\
\end{tabular}
\caption{\label{fig:comp_perband_I} Completeness as a function of the intrinsic flux density of a galaxy in intensity in the input simulated catalog. The six panels correspond to the same bands as in Fig.\,\ref{fig:photacc_I}. The red and blue lines show the results with and without clustering, respectively. The dotted and solid lines correspond to the definition A (a detected galaxy is in the beam of an extracted source) and B (a galaxy is detected only if it is the brightest in the beam of the extracted source) of the completeness described in Sect.\,\ref{sect:puco_est}. The vertical dotted line is classical confusion limit computed in Sect\,\ref{sect:conflim_def}.}
\end{figure*}

\begin{figure}
\centering
\includegraphics[width=8cm]{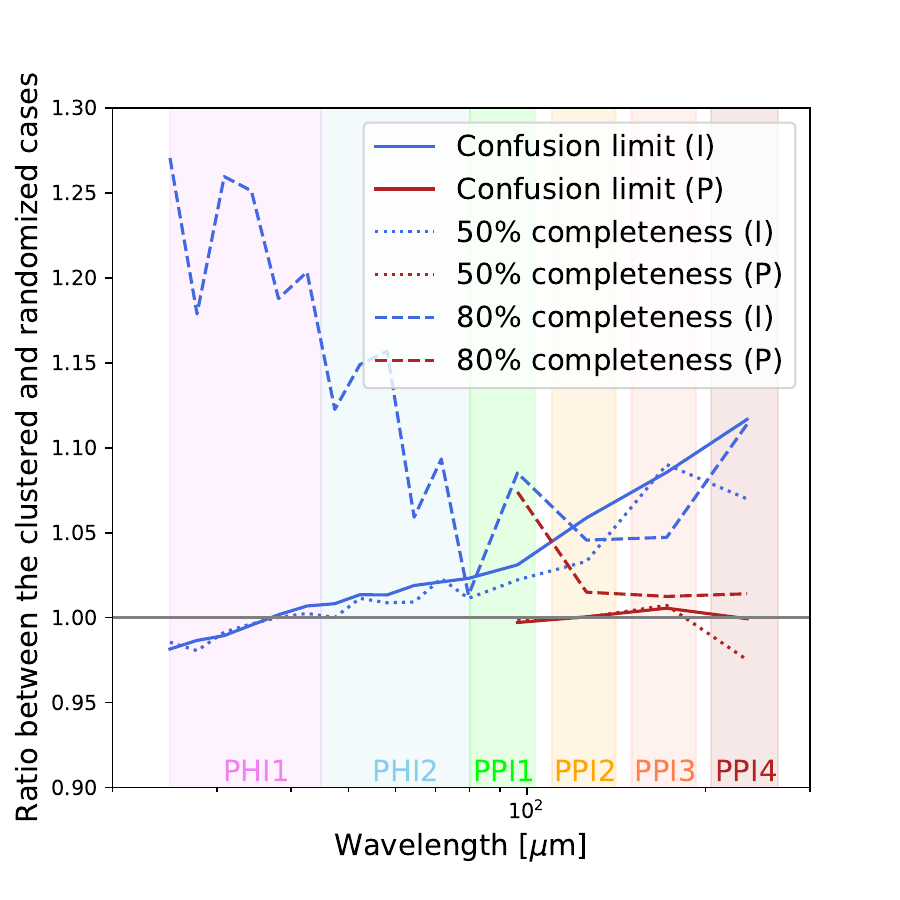}
\caption{\label{fig:comp_clustering} Ratio between values found using the simulation with (original SIDES) and without (randomized positions) clustering for the following performance metrics: 5$\sigma$ classical confusion limit (solid line), 50\,\% completeness flux density (dotted line), and 80\,\% completeness flux density (dashed line). The intensity is in blue and the polarization in dark red. The intensity is discussed in Sect.\,\ref{sect:comp_I} and the polarization in Sect.\,\ref{sect:puco_P}.}
\end{figure}

\begin{figure}
\centering
\includegraphics[width=8cm]{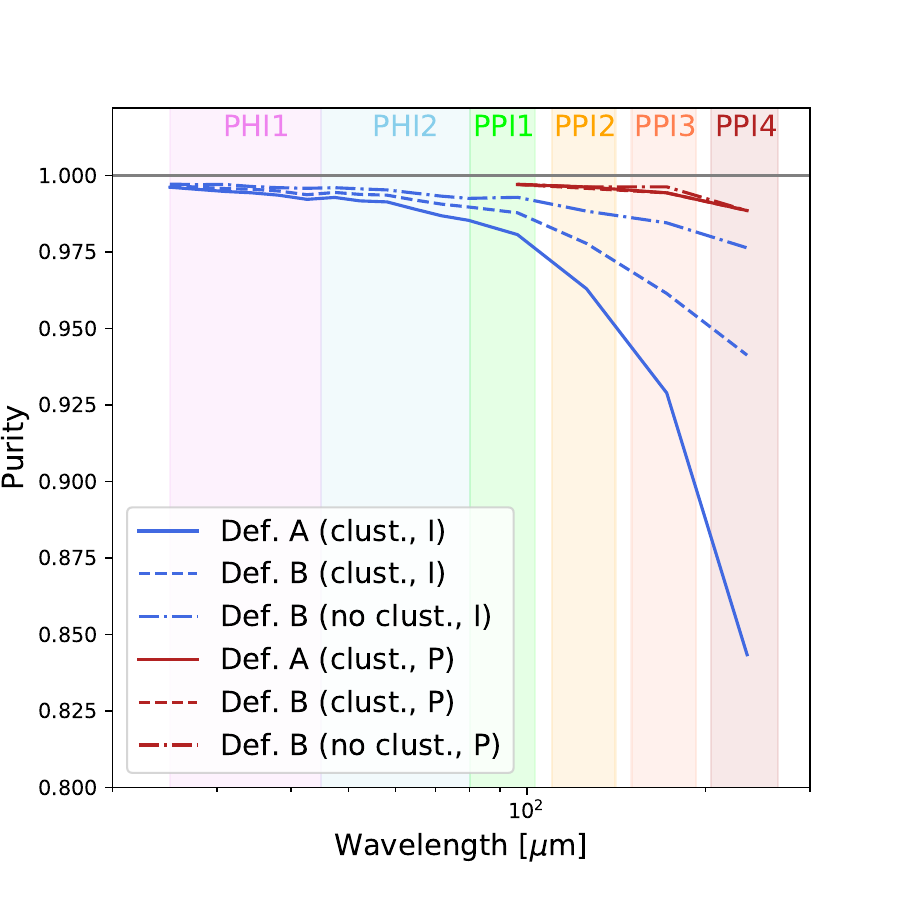}
\caption{\label{fig:purity_sythesis} Purity of the sample extracted from the simulated map above the $5\sigma_{conf}$ threshold in intensity (blue) and polarization (dark red). The solid lines correspond to the definition A of purity (see Sect.\,\ref{sect:puco_est}), where a source is considered "true" if the brightest galaxy in the beam is at least half of the measured flux. The dashed and dot-dash lines are corresponding to the definition B with and without clustering, respectively. In this second definition, we lower the minimal flux density to consider a source to be "true" by the flux density excess factor measured in Sect.\,\ref{sect:photo_acc_I}. We discuss the intensity in Sect.\,\ref{sect:purity_I} and the polarization in Sect.\,\ref{sect:puco_P}.}
\end{figure}

\begin{figure*}
\centering
\begin{tabular}{cccc}
\includegraphics[width=6cm]{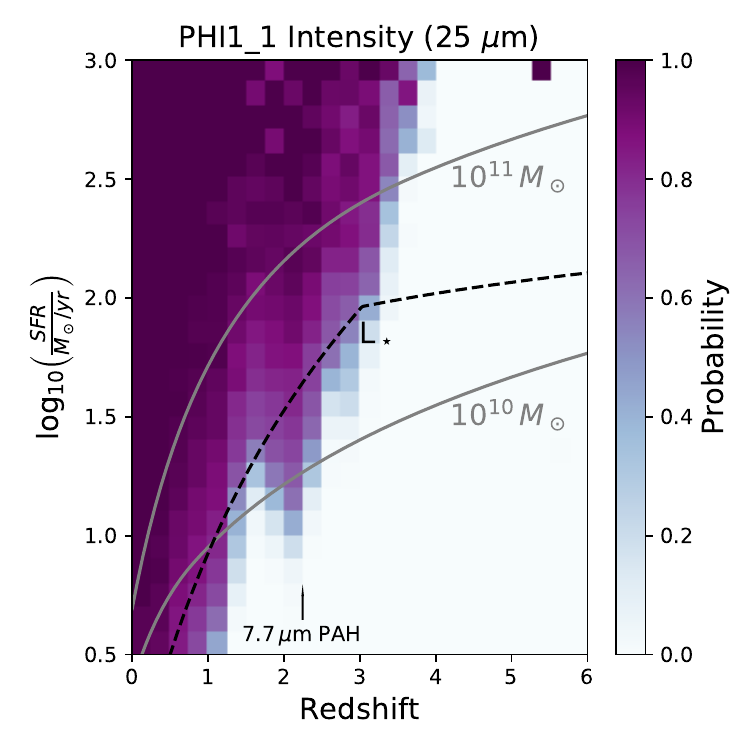} & \includegraphics[width=6cm]{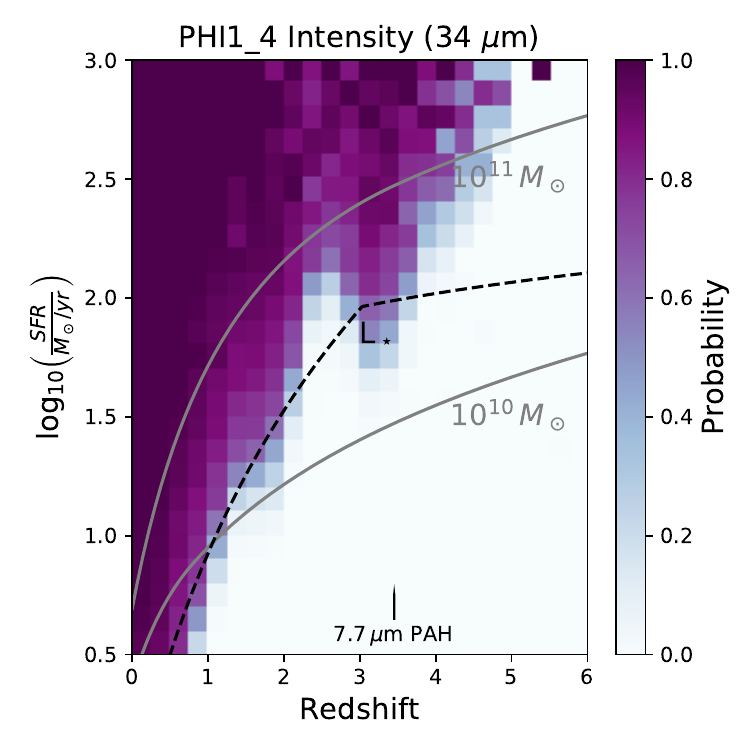} & \includegraphics[width=6cm]{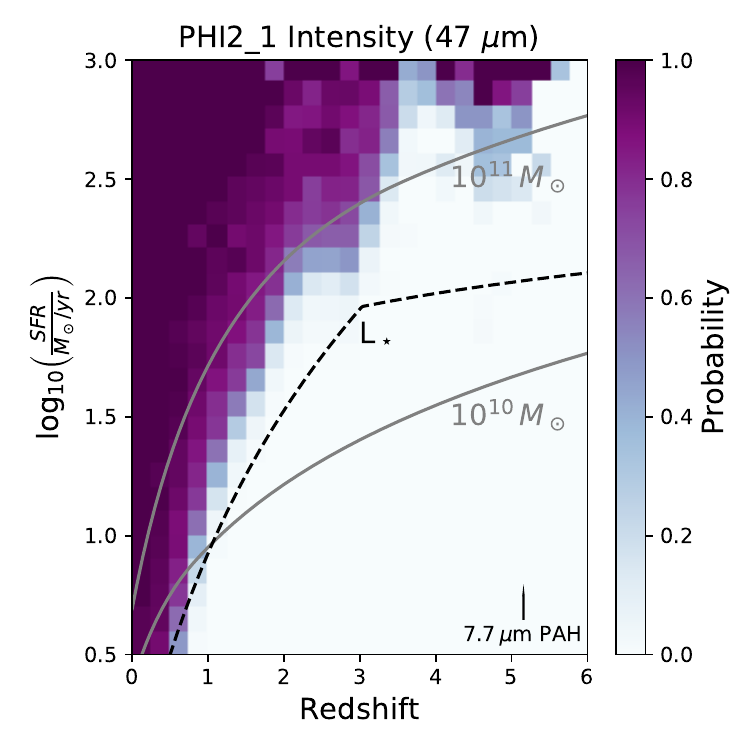}\\
\includegraphics[width=6cm]{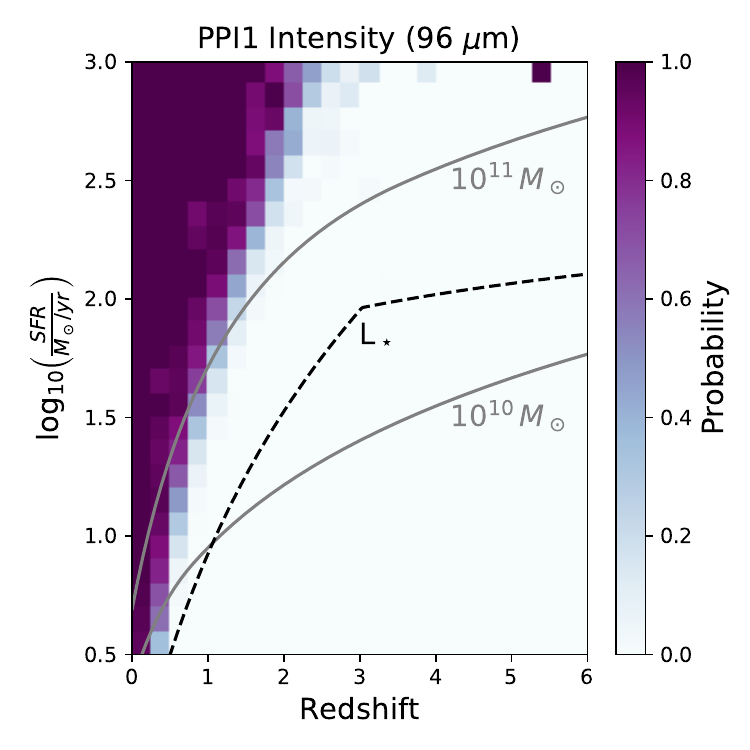} & \includegraphics[width=6cm]{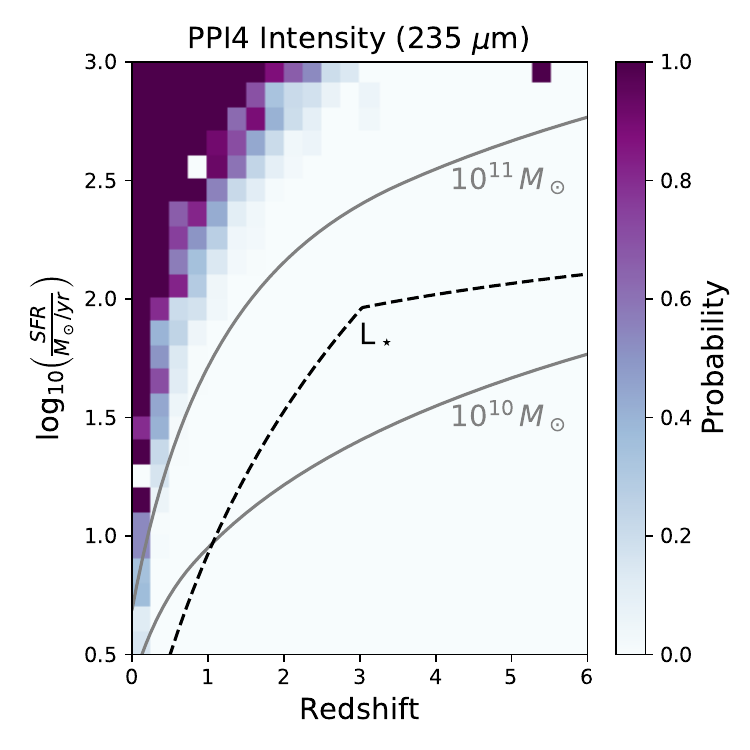} & \includegraphics[width=6cm]{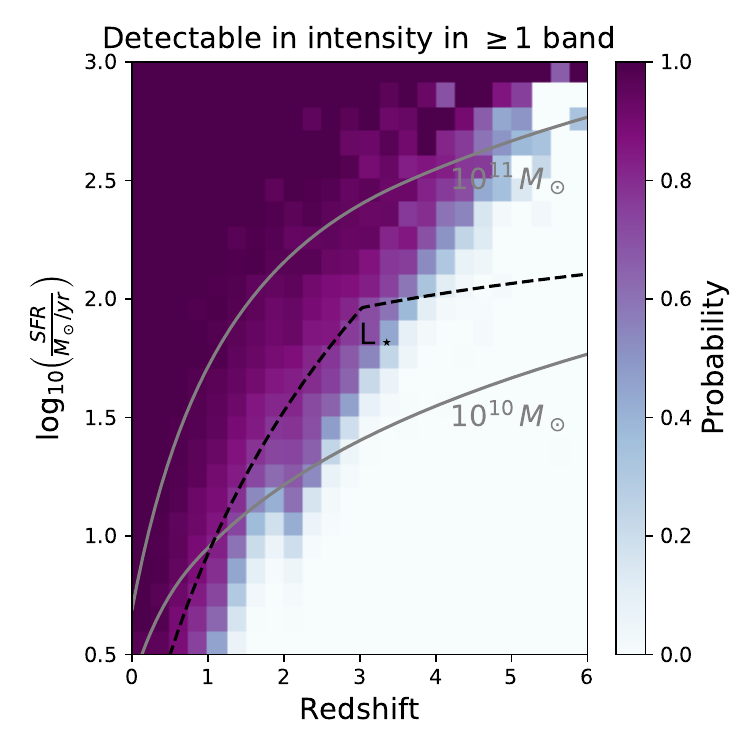} \\
\end{tabular}
\caption{\label{fig:SFRz_I} Probability (color coded) to detect a galaxy with our basic blind source extractor in an intensity map affected only by confusion as a function of its position in the SFR-$z$ plane. The lower right panel is the probability to detect the source in at least one band, while the other panels are for a selection of single bands. The two gray tracks show the position of a galaxy exactly on the main-sequence relation \citep{Schreiber2015} for various stellar masses. The black dashed line shows the evolution of the knee of the infrared luminosity function L$_\star$ measured by \citet{Traina2024}.} 
\end{figure*}

\begin{figure*}
\centering
\begin{tabular}{ccc}
\includegraphics[width=6cm]{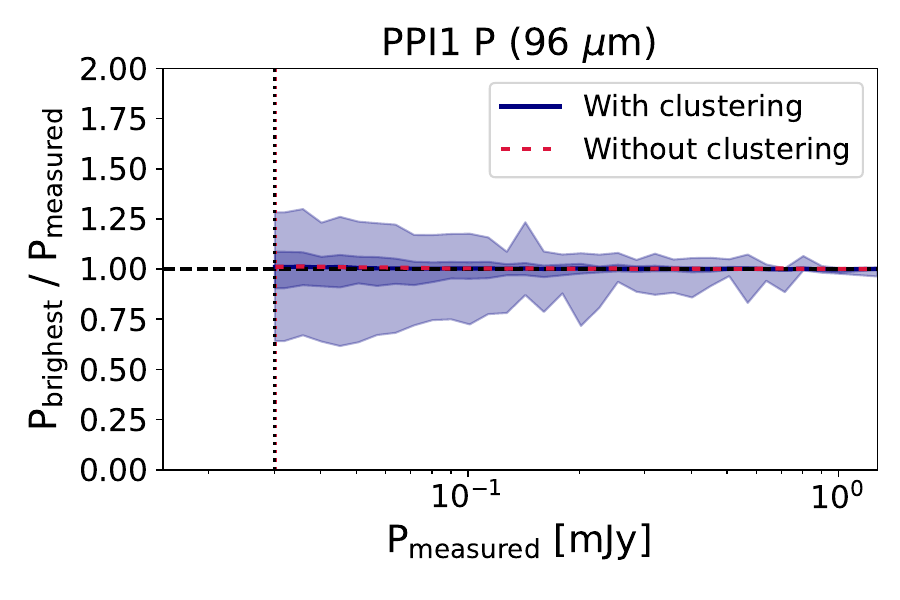} & \includegraphics[width=6cm]{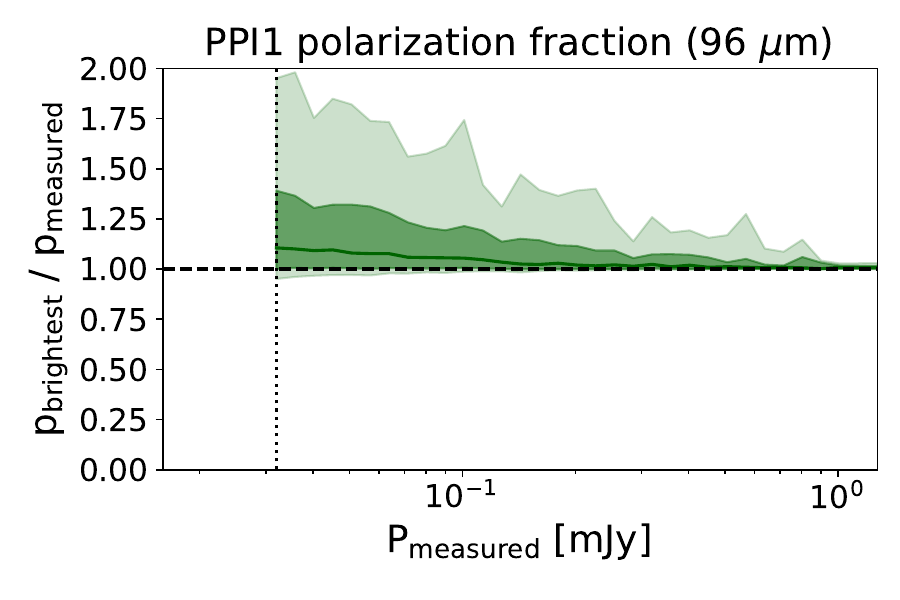} & \includegraphics[width=6cm]{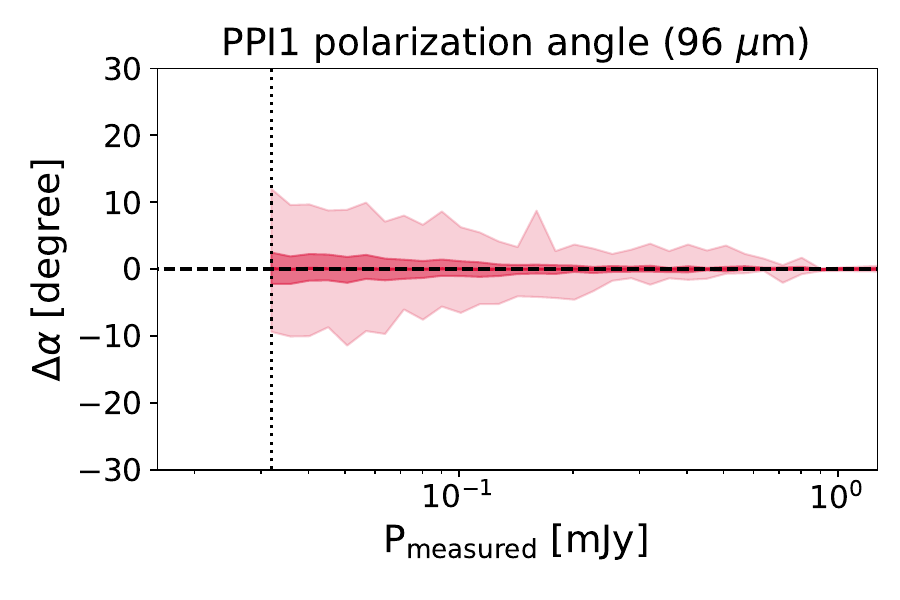}  \\
\includegraphics[width=6cm]{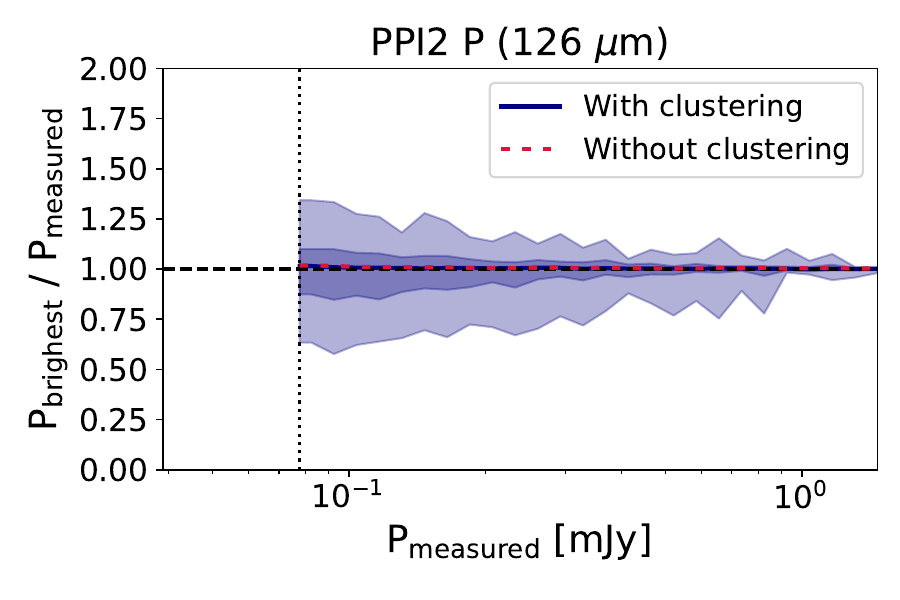} & \includegraphics[width=6cm]{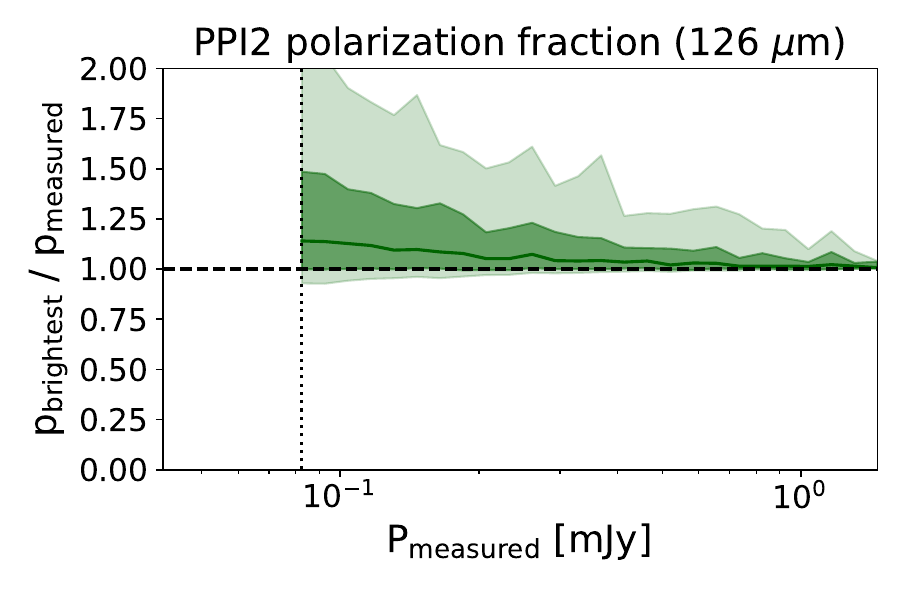} & \includegraphics[width=6cm]{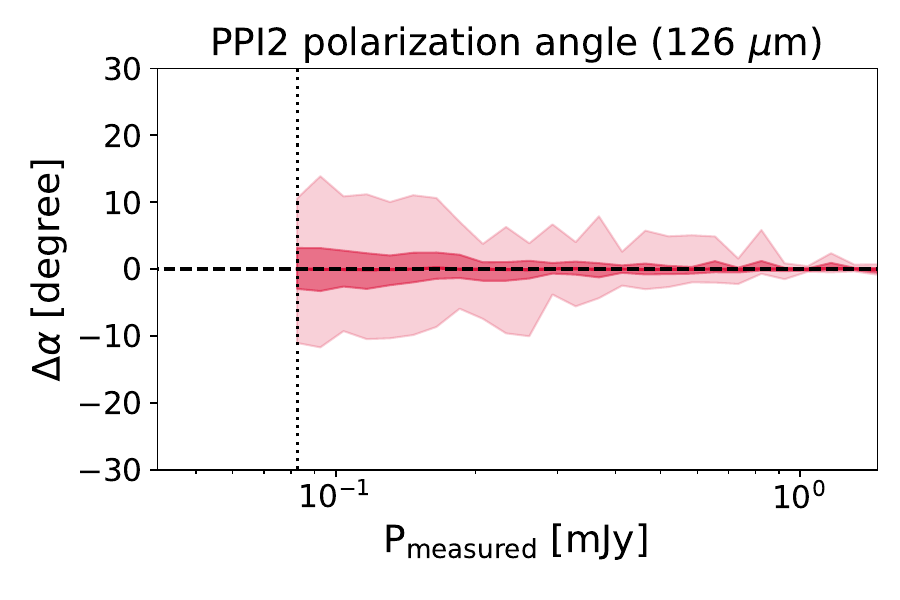}  \\
\includegraphics[width=6cm]{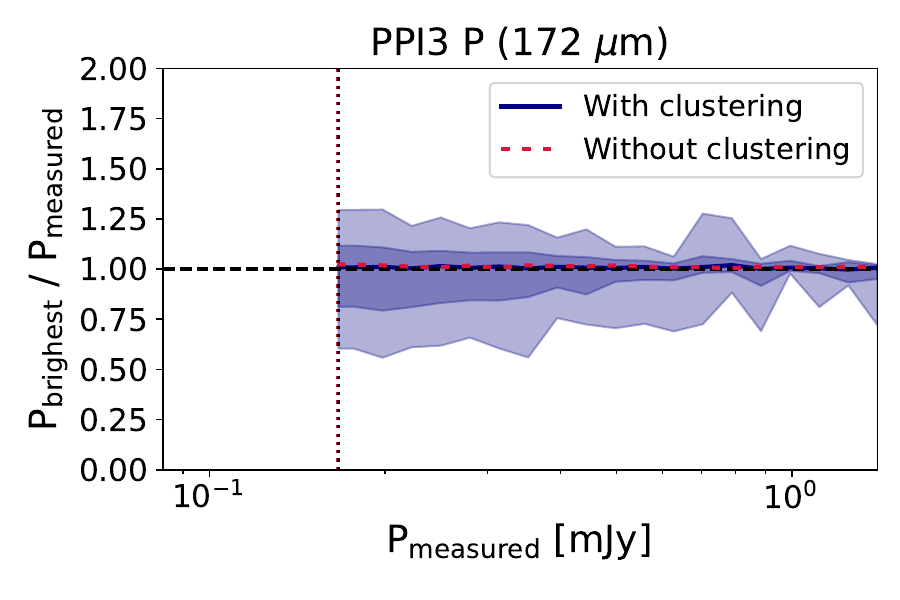} & \includegraphics[width=6cm]{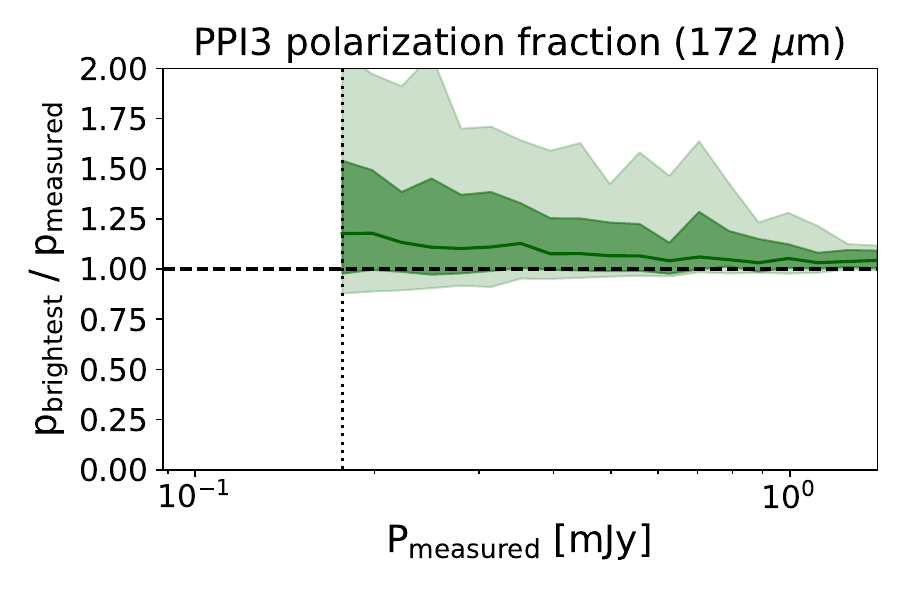} & \includegraphics[width=6cm]{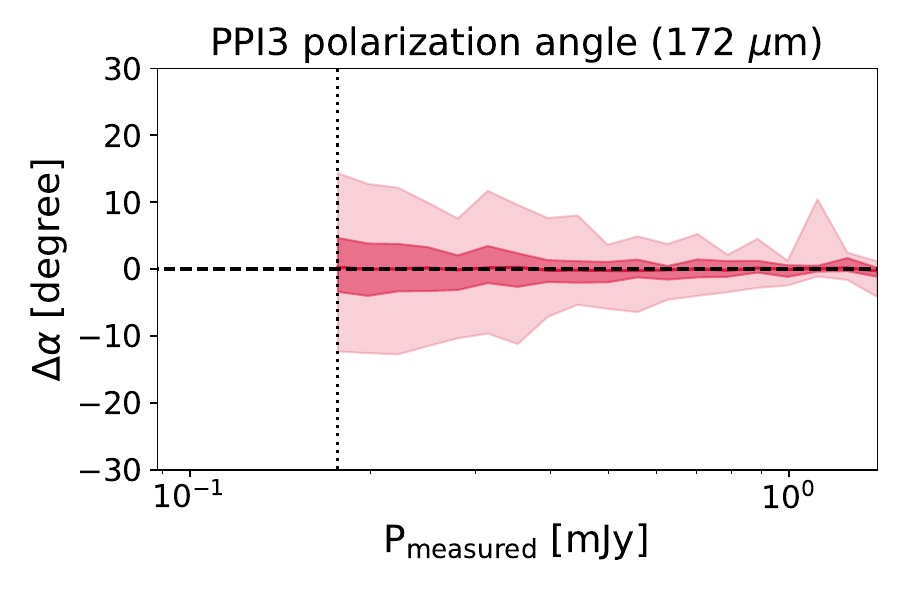}  \\
\includegraphics[width=6cm]{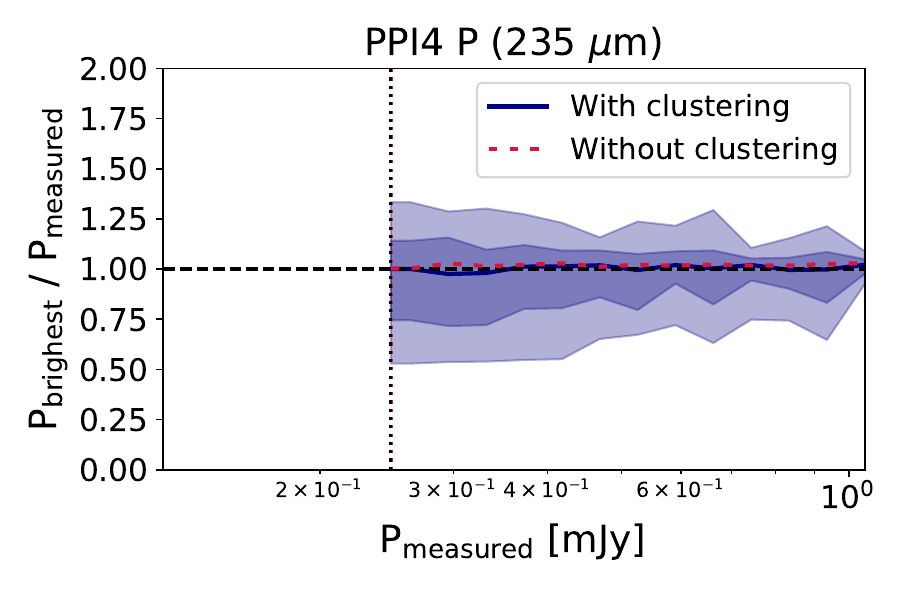} & \includegraphics[width=6cm]{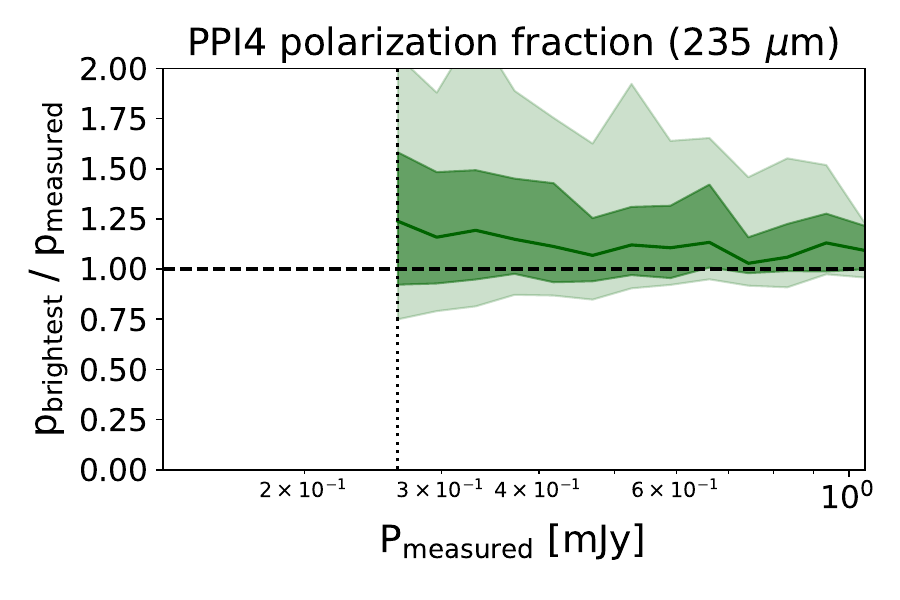} & \includegraphics[width=6cm]{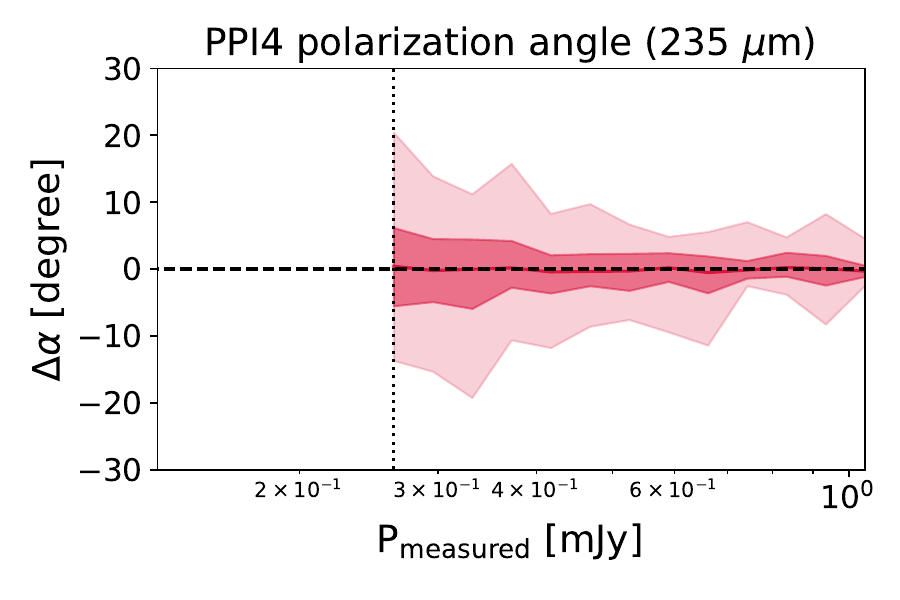}  \\
\end{tabular}
\caption{\label{fig:acc_P} Left panels: ratio between the polarized flux density $P$ of the brightest simulated galaxy in the beam and the measured one in the $P$ map. The symbols are the same as in Fig.\,\ref{fig:photacc_I}. Central panels: same thing for the polarized fraction $p$. Right panels: difference between the polarization angle $\alpha$ of the brightest source in the beam (see Sect.\,\ref{sect:simpolar}) and the measured angle. The rows correspond to the various polarized band from PPI1 to PPI4.}
\end{figure*}

\begin{figure}
\centering
\includegraphics[width=8cm]{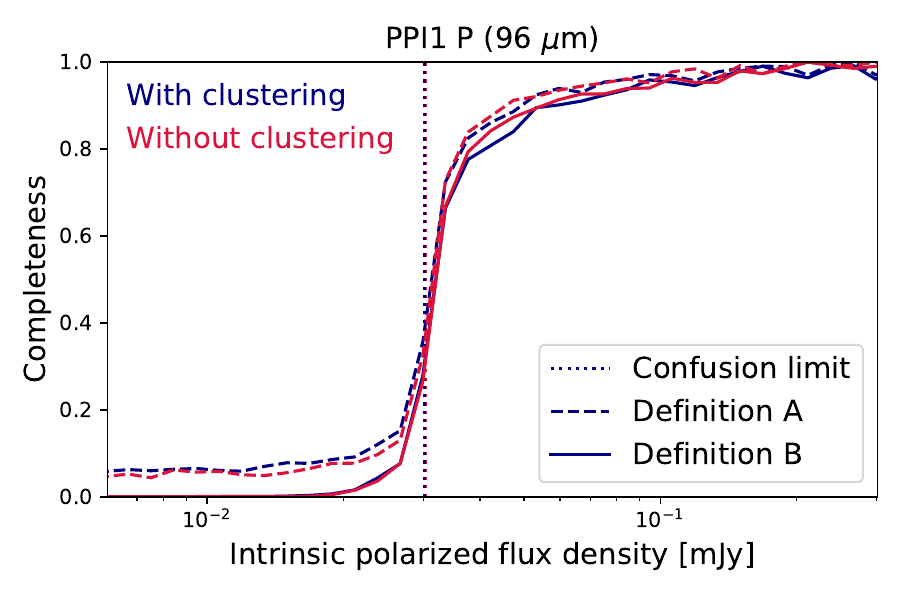}
\includegraphics[width=8cm]{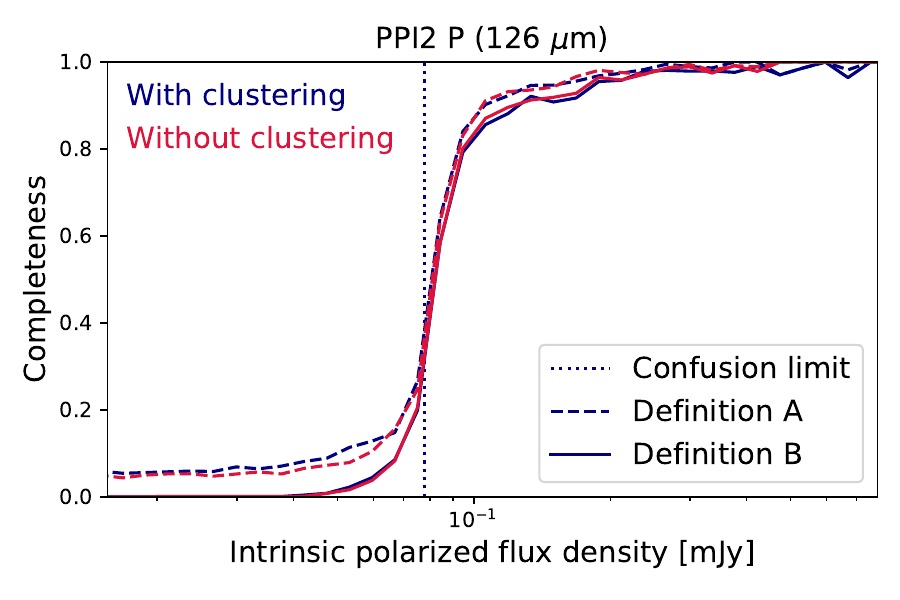}
\includegraphics[width=8cm]{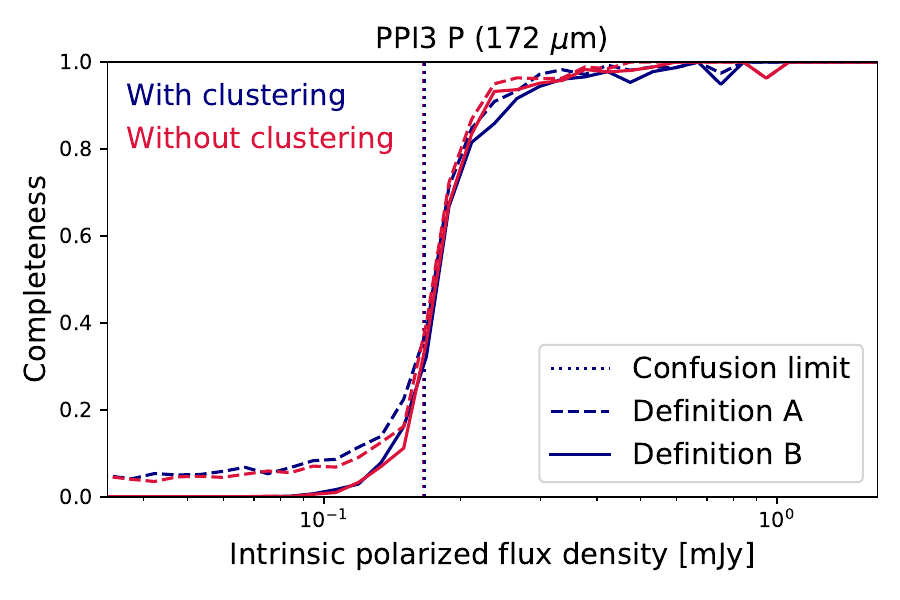}
\includegraphics[width=8cm]{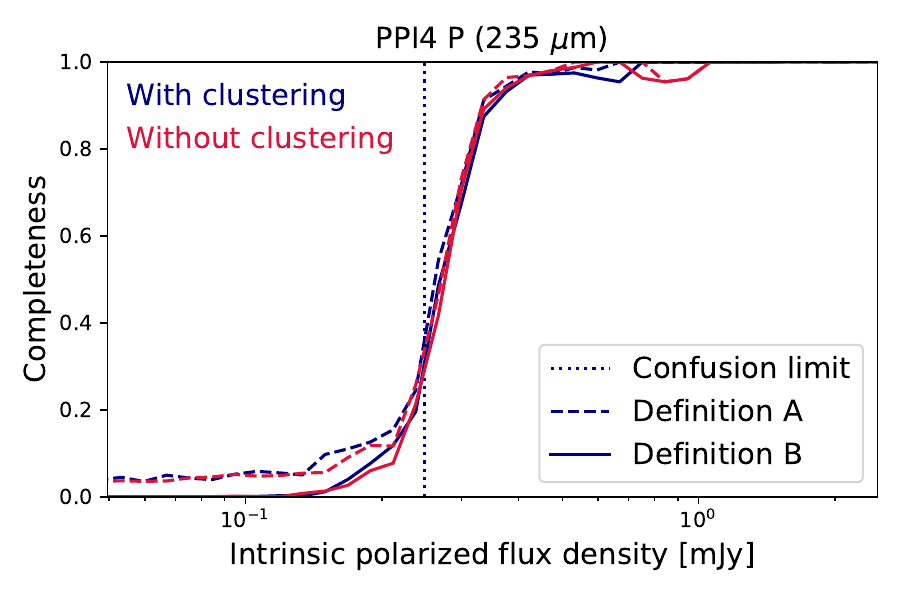}
\caption{\label{fig:comp_perband_P} Same figure as Fig.\,\ref{fig:comp_perband_I}, but for the four polarized bands.}
\end{figure}

\begin{figure*}
\centering
\begin{tabular}{ccc}
\includegraphics[width=6cm]{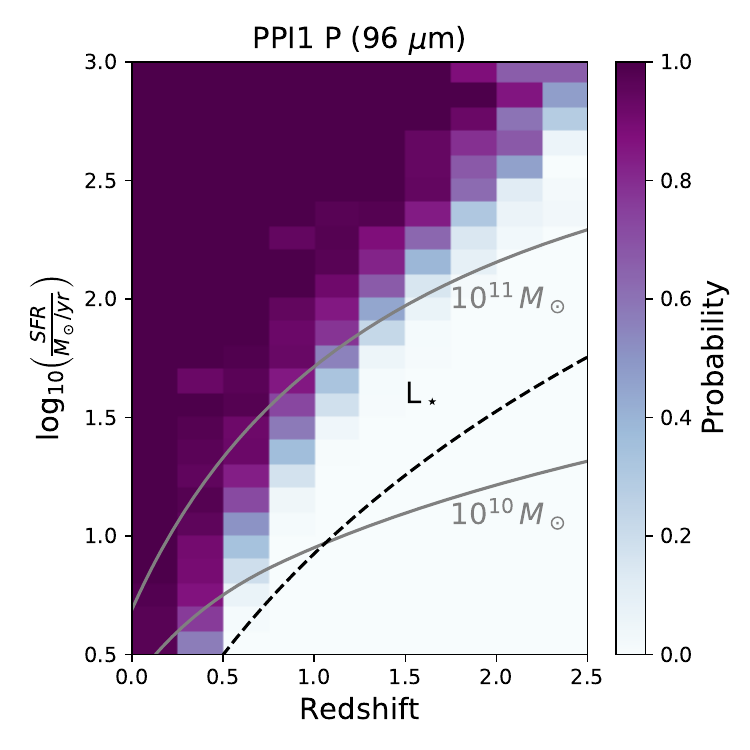} & \includegraphics[width=6cm]{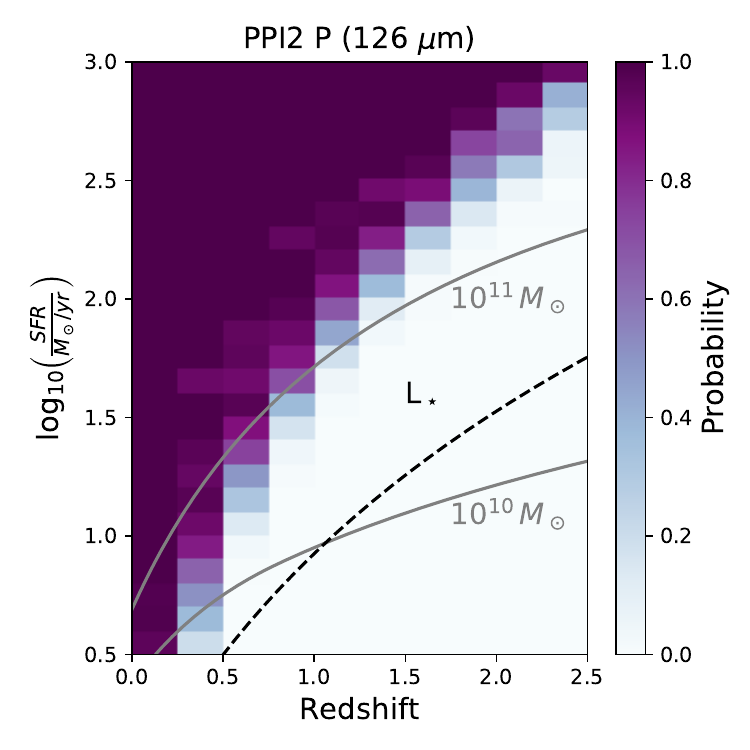} & \includegraphics[width=6cm]{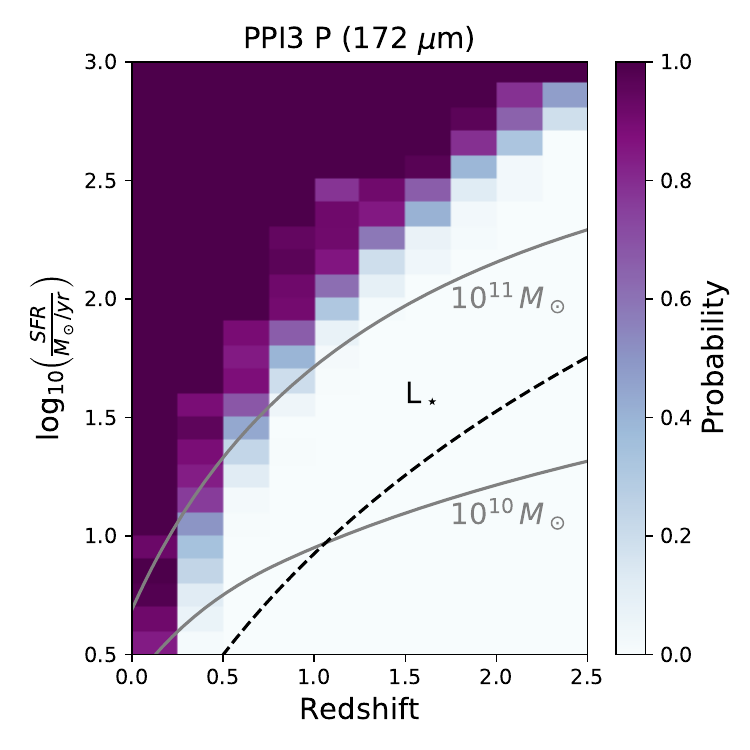}\\
\end{tabular}
\begin{tabular}{cc}
\includegraphics[width=6cm]{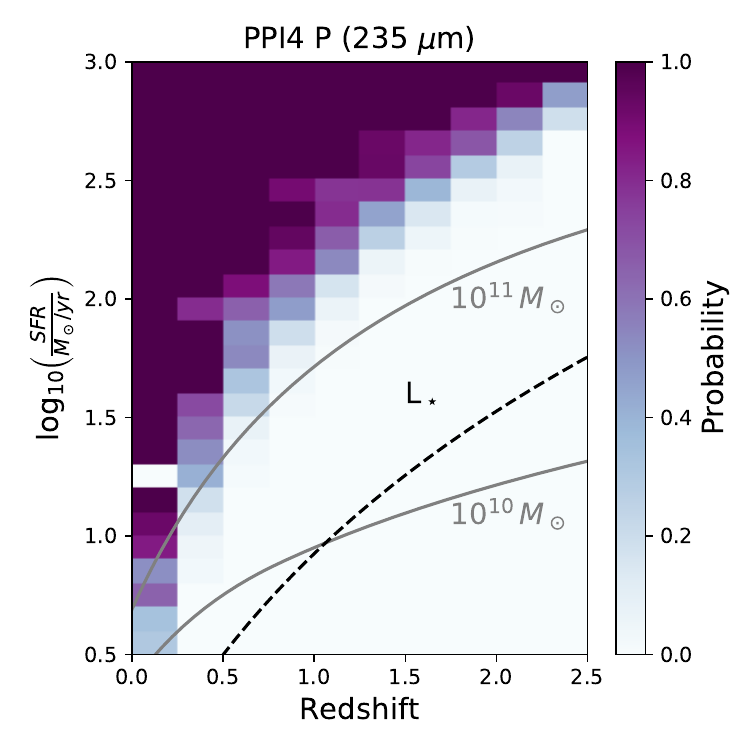} & \includegraphics[width=6cm]{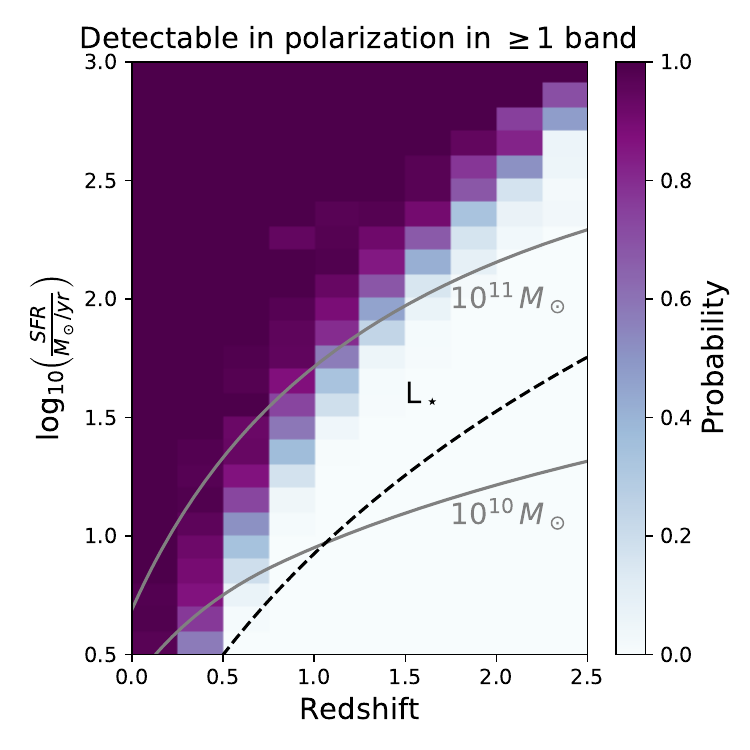}\\
 \end{tabular}
\caption{\label{fig:SFRz_P} Same figure as Fig.\,\ref{fig:SFRz_I} but for polarization.}
\end{figure*}

\section{Confusion in intensity}

\label{sect:intensity}

In this section, we focus on the impact of confusion on intensity data.

\subsection{Photometric accuracy}

\label{sect:photo_acc_I} 

A key question in photometry of sources with confused data is whether the measured flux densities is dominated by a single bright galaxy or comes from several objects. Modeling can take into account blending effects before  comparing predictions to data \citep[e.g.,][]{Bing2023}, however, most conventional astronomy relies on photometry of individual objects. We compared the flux density of the brightest source in the beam  to the measured flux in the map using the matching algorithm presented in Sect.\,\ref{sect:match}.

In Fig.\,\ref{fig:photacc_I}, we show the ratio between the flux density of the brightest galaxy in the beam and the measurement in the map. This is the inverse of the classical output versus input ratio used to illustrate the flux boosting, but our choice has the advantage to provide immediately the relative contribution of the brightest galaxy. In PHI bands, the median ratio is above 0.91 at the classical confusion limit and converges rapidly to unity at higher flux density. We also studied the distributions around the median using the 16--84\,\% and 2.3--97.7\,\% ranges (corresponding to 1 and 2$\sigma$ for Gaussian distributions). We note that the distribution is highly asymmetrical. Only rare outliers have an underestimated flux density (ratio$>$1), while the lowest 2.3\,\% can have a flux density underestimated by half in these bands. This is a consequence of the large tail of positive outliers in the histogram of pixel flux densities (see Fig.\,\ref{fig:map_histos}). These overestimated flux densities are mainly caused by the blending of two sources with a similar flux density. Advanced deblending algorithms are usually very efficient to mitigate this effect \citep{Donnellan2024}. 

In PPI bands, the contribution from other sources in the beam becomes more significant with larger beams at longer wavelength. The median flux density ratio at the classical confusion limit decreases with increasing wavelength from 0.90 to 0.72. This effect has already been discussed in the case of the \textit{Herschel}/SPIRE instrument by \citet{Scudder2016} and \citet{Bethermin2017}, and is mainly caused by sources at other redshifts while a small  contribution ($\lesssim$5\,\%) comes from physically-related sources. The dispersion of the ratio around the median value also becomes more symmetrical with increasing wavelength. This is expected since the histogram of pixel flux densities becomes more symmetrical at longer wavelength (see Fig.\,\ref{fig:map_histos}). This is consequence of the central limit theorem as the approximately Poisson distribution of source fluxes becomes more Gaussian with a larger number of source per beam.

We investigated if a different choice of background could impact the measured flux density excess using the extreme example of the PPI4 band. As discussed in Sect.\,\ref{sect:extr_setting}, the mode is the most natural choice in a noiseless case, but a higher background could reduce the excess. The median flux density excess is 15.5\,mJy. Choosing a higher background as the median or the mean, would have led to an excess of 12.9 and 10.4\,mJy, respectively. It does not change qualitatively the results.

We also estimated the flux uncertainties using half of the 84--16\,\% interval (corresponding to $1\sigma$ for a Gaussian distribution). At the classical confusion limit, in PHI bands, the relative flux  uncertainties range from 8\,\% in band PHI1\_1 to 15\,\% in band PHI2\_6, and are better than the 20\,\% expected from our 5\,$\sigma$ construction of the extraction threshold. This is because the distribution is highly non-Gaussian and the clipped variance is more sensitive to outliers. These relative uncertainties below 20\,\% confirm that our estimate of classical 5$\sigma$ confusion level is conservative in these bands, and deeper catalogs can be obtained with more advanced extraction methods (see \citealt{Donnellan2024}). In PPI bands, the performance  degrades with increasing wavelength (15\,\% in PPI1, 19\,\% in PPI2, 22\,\% in PPI3, 26\,\% in PPI4). The accuracy in PPI3 and PPI4 bands is slightly worse than the 20\% expected. In these bands, blind-extracted catalogs will be challenging to use in intensity, and their interpretation will either require complex statistical corrections of the fluxes or incorporating the effects of angular resolution through statistical models \citep[e.g.,][]{Hayward2013,Cowley2015,Bethermin2017,Bing2023}. In contrast, techniques such as prior-based source extraction will enable accurate measures of the fluxes (e.g., \citealt{Donnellan2024}).

Finally, we investigated the impact of clustering on the flux density bias by comparison with results following randomization of the source positions (without clustering). This illustrated in Fig.\,\ref{fig:photacc_I} by the red dashed line. The effect is almost negligible in PHI bands. In PPI bands and in absence of clustering, the brightest galaxy in the beam of sources at the classical confusion limit contributes to 3\,\%, 4\,\%, 6\,\%, and 8\,\% more than in the clustering case. The clustering does not explain fully the effect discussed previously, but it reinforces it. 

\subsection{Completeness}

\label{sect:comp_I}

The probability that a survey will be able to detect sources as a function of flux density, called completeness, is also an important performance criterion. In Fig.\,\ref{fig:comp_perband_I}, we present the completeness obtained using our minimal extractor in simulated noiseless PRIMAger maps. We show the two definitions of completeness introduced in Sect.\,\ref{sect:puco_est} for both the clustered and randomized cases.

At low flux densities, the completeness converges to a non zero value with the definition A. As discussed in Sect.\,\ref{sect:match} and \ref{sect:puco_est}, this is caused by the faint galaxies in the beam of a brighter object being considered as detected. We do not observe this behavior for the definition B, where only the brightest galaxy in the beam is considered to be detected. 

Both definitions reach 50\,\% close to the classical 5$\sigma$ confusion limit used as threshold by our source extractor (see Sect.\,\ref{sect:conflim_def}). In the case of Gaussian noise (or any symmetrical noise), we would expect to have exactly 50\,\% at the extraction threshold, since only half of the sources will be on a positive noise realization. However, in the PPI3 and PPI4 bands, the completeness is slightly larger ($\sim$60\,\%). This could be that in these bands the flux is not coming from a single object and the flux is boosted by the neighbors (Sect.\,\ref{sect:photo_acc_I} and Fig.\,\ref{fig:photacc_I}).

At short wavelength, the completeness curve increases only slowly above the 5-$\sigma$ classical confusion limit (one full flux density decade from 80\,\% to 95\,\% in PHI1), especially for definition B of the completeness. This suggests that the sources close to bright sources tend to be missed, since this second definition does not consider a faint galaxy as detected when in the beam of a brighter one. At long wavelength, the transition is much sharper and converge rapidly to unity. There is also not much dependence on the definition. 

Because the definition A at faint flux density does not converge to zero and the difference between the two definitions remains small around the confusion limit, we will use only the definition B of the completeness in the rest of this paper. In Table\,\ref{tab:conf_levels}, we summarize the classical confusion limits and the 50\,\% and 80\,\% completeness levels found in the various bands.

Finally, we investigated the wavelength dependence of the clustering impact on completeness. In Fig.\,\ref{fig:comp_clustering}, we compare the value of the completeness determined in our clustered simulation and after randomizing the positions (no clustering). The ratio between the clustered and the random case increases with increasing wavelength for both the 5$\sigma$ classical confusion limit and the 50\,\% completeness flux density, and reaches $\sim$10\,\% in the PPI4 band. This is expected, since the clustering tends to broaden the pixel flux histogram \citep[e.g.,][]{Bethermin2017}. The behavior in the short-wavelength side of the PHI1 band, where the clustered case has a lower classical confusion limit, is less intuitive. This is a small effect ($\sim$2\,\%), and could be due to the strong blending of the bright sources in the clustered case decreasing the density in the rest of the field. Finally, the 80\,\% completeness flux density has a more complex U-shaped trend with wavelength with a minimum between band PHI2 and PPI1. The rise above band PPI1 has a similar explanation as for the other quantities.  A possible explanation for the strong impact ($\sim$25\,\%) of clustering at short wavelength is that the bright sources tend to cluster with each other and a small fraction of objects well above the global classical confusion limit are missed since they are in the vicinity of brighter objects. At longer wavelength, the flux density ratio between the brightest and faintest detectable sources is smaller, reducing the impact of this effect.

\subsection{Purity}

\label{sect:purity_I}

The third criterion to evaluate the quality of the catalogs is the purity. Surveys usually aim for 80 to 95\,\% depending on whether the scientific goal is a pure statistical measurement or building a clean sample for detailed follow-up studies. As discussed in Sect.\,\ref{sect:puco_est}, the definition of purity in confusion-limited data is not trivial. In Fig.\,\ref{fig:purity_sythesis}, we show the purity as a function of wavelength for our two definitions. 

In PHI bands, the purity is always excellent ($>$98\,\%) whatever the definition, even though it slightly decreases with increasing wavelength. This suggests that we were  conservative in our choice of extraction threshold. We could thus expect to go deeper for statistical studies in confusion-limited data using more aggressive source extraction algorithms.

In PPI bands, the purity degrades rapidly with increasing wavelength. In the case of definition A (source considered "true" if the brightest counterpart produces more than half of the measured flux density), it drops  to 84\,\% in PPI4. This is mainly because the median ratio between the flux density of the brightest galaxy in the beam and the measured flux is below unity (see Sect.\,\ref{sect:photo_acc_I}). If we use the definition B of the purity for which we correct the measured flux densities by the median ratio, the purity rises to 94\,\%. This demonstrates that it is the main reason of the lower purity at longer wavelength. In the case of real data, this average correction could be calibrated using artificial sources injections in the data or using end-to-end simulations. Finally, if we use the definition B in absence of clustering, the result increases to 97.6\,\% and is close to the performance reached in PHI bands. The clustering thus has also a mild impact on the degradation of PPI-band purity.

\subsection{Detection probability in the SFR-$z$ plane in intensity}

\label{sect:SFRz_I}

In the previous sections, we characterized the classical confusion limit only in term of flux density. However, to understand its impact on the observatory science, it is essential to consider the impact on  intrinsic physical properties. We thus computed the probability to detect a galaxy above the classical confusion limit as a function of SFR and redshift. The border between the regions of low and high probability is blurred, since our model has a diversity of SEDs and our simulation produces a completeness curve which have a continuous transition from 0 to 1 (see Sect.\,\ref{sect:comp_I}). The results are presented in Fig.\,\ref{fig:SFRz_I} together with the tracks corresponding to galaxies of various stellar masses following the main sequence relation of \citet{Schreiber2015} and the evolution of the knee of the infrared luminosity function \citep[L$_\star$,][]{Traina2024}.

For the shortest wavelength (PHI1\_1 band centered on 25\,$\mu$m, upper left corner of Fig.\,\ref{fig:SFRz_I}), the 10$^{10}$\,M$_\odot$ and 10$^{11}$\,M$_\odot$ main-sequence galaxies are recovered up to $z\sim$2.5 and $z\sim$3.5, respectively. The L$_\star$ galaxies are slightly above the detection border up to z$\sim$3 and undetected above. Overall, the border between detections and non detections moves towards higher SFR with increasing redshift. However, we can identify some specific features. At $z\lesssim$2, we are probing the 10\,$\mu$m rest-frame dip in SED between the various PAH bands, and galaxies are harder to detect (higher SFR limit). At $z\gtrsim$2, the 7.7\,$\mu$m PAH band enters the representative filters making the galaxies easier to detect (lower SFR limit). Around $z=3$, the typical SFR at which galaxies are detected increases sharply, and only some rare outliers can be detected. This is the consequence of the absence of strong dusty features below 6\,$\mu$m rest-frame in the SED of star-forming galaxies.

At longer wavelength in PHI bands, at $z<$1.5, the SFR sensitivity decreases with increasing wavelength. At $z>1.5$, the PAH features boost the SFR sensitivity in some specific redshift range (e.g., $z\sim$3.4 in PHI1\_4 at 34\,$\mu$m and $z\sim$5.1 in PHI2\_1 at 47\,$\mu$m). In PPI bands, the PAH corresponds to very high redshifts and the border between detection and non detection area is less complex.

Finally, we combined all the bands to derive the probability to detect galaxies in at least one band (lower right corner of Fig.\,\ref{fig:SFRz_I}). This illustrates the parameter space, which could be probed by PRIMAger above the classical confusion limit. A galaxy exactly on the main-sequence relation and with a stellar mass of 10$^{10}$\,M$_\odot$ and 10$^{11}$\,M$_\odot$ can be detected up to $z\sim$2.5, and $z\sim$5, respectively. The L$_\star$ galaxies are detected up to z$\sim$3.5. The border between detection and non detection is almost featureless, since the PAH slides through the various representative sub-filters. This illustrates how hyperspectral imaging can help to deal with confusion. However, the dip at $z\sim$1.5 seen in PHI1\_1 is still present, since there is no shorter wavelength to observe this redshift range around 7.7\,$\mu$m rest-frame. \\

%%%%%%%%%%%%%%%%%%%%%%%%% POLAR %%%%%%%%%%%%%%%%%%%%%%%%%

\section{Confusion in polarization}

\label{sect:polar}

In this section, we discuss the impact of confusion on polarization data. We do not take into account the instrumental noise in this section. If we assume a different mean polarized fraction $\mu_p$, both the confusion noise and the source polarized flux density scales as ($\mu_p$/1\,\%).  Consequently, the x-axis of Fig.\,\ref{fig:acc_P} and \ref{fig:comp_perband_P} and the values in Table\,\ref{tab:conf_foreground} must be shifted by this factor, while Fig.\,\ref{fig:SFRz_P} is unchanged.

\subsection{Photometric accuracy}

\label{sect:photo_acc_P} 

Since the polarized flux density $P$ is the quadratic combination of $Q$ and $U$ (Eq.\,\ref{eq:polar}), the pixel values of the $P$ map are always positive, while $Q$ and $U$ pixels can be either positive or negative (with a zero mean in absence of alignment between galaxies). However, contrary to the intensity maps where flux densities always add up, two bright sources at the same position and with the same polarized flux density $P$ can, in principle, lead to a null $P$ flux intensity map if their polarization angles $\alpha$ differs by $\pi$/2, since the sum of their $Q$ and $U$ values will be zeros. As illustrated by Fig.\,\ref{fig:map_histos}, the mode of the $P$-map histogram is strictly positive, and even in polarization it is important to define carefully the background.

In Fig.\,\ref{fig:acc_P} (left panels), we show the ratio between the brightest galaxy polarized flux density $P$ in the beam and the measured value in the simulated map. In contrast with the intensity maps (Sect.\,\ref{sect:photo_acc_I} and Fig.\,\ref{fig:photacc_I}), we do not observe any bias in the median flux ratio. This is likely to be because the contribution of several sources in the beam is not fully additive if their polarization angles are not aligned. However, similarly to the intensity, we still observe an increase of the half width of the 1\,$\sigma$ confidence region from 9\,\% to 20\,\% from PPI1 to PPI4, but overall the dispersion is slightly lower than it is for intensity.

We can thus recover the polarized flux density of sources just above the classical confusion limit with a good accuracy, while it is not the case in intensity (see Sect.\,\ref{sect:photo_acc_I}). However, the polarized flux density is much weaker than the intensity, and detecting it will require much deeper data. In addition, as discussed in Sect.\,\ref{sect:simpolar}, we did not include galaxy alignments, which could produce a small polarized flux density excess similarly to what happens in intensity.

\subsection{Recovering polarized angles and polarized fraction}

The polarized fraction $p$ ($P/I$) is another useful quantity to characterize distant unresolved galaxies. We derived $p$ for each galaxy detected in the $P$ map,  extracting the value of $I$ at the same position in the intensity map. In Fig.\,\ref{fig:acc_P} (central panels), we show the polarized fraction ratio between the brightest galaxy in the beam (highest $P$) and the measurement in the simulated map. At low polarized flux, the intrinsic polarized fraction of the brightest galaxy is significantly larger than the measured one. This is a natural consequence of the negligible bias found for the polarized flux density measurements ($P$, Sect.\,\ref{sect:photo_acc_P}) and the significant bias found in intensity ($I$, Sect.\,\ref{sect:photo_acc_I} and Fig.\,\ref{fig:photacc_I}), since $p_{\rm brightest}/p_{\rm measured} = (P_{\rm brightest}/P_{\rm measured}) \times (I_{\rm measured}/I_{\rm brightest})$ with the first factor being close to one and the second being significantly above  (i.e., the inverse of the quantity shown in Fig.\,\ref{fig:photacc_I}).

We also tested our ability to recover the polarization angle $\alpha$. We measure it from the $Q$ and $U$ values found at the position of sources detected in the $P$ maps:
\begin{equation}
\alpha = \begin{cases}
\frac{1}{2} \arctan(\frac{U}{Q}) & \text{ if } Q \ge 0, \\
\frac{1}{2} \arctan(\frac{U}{Q}) + \frac{\pi}{2} & \text{ if } Q < 0.
\end{cases}
\end{equation}
We then compute the difference between the intrinsic polarization angle of the brightest galaxy in the beam and the measured angle ($\Delta \alpha$). Since $\Delta \alpha$ is defined modulo $\pi$ for a polarization angle, we shifted all the values between $-\pi/2$ and $+\pi/2$. The results are presented in Fig.\,\ref{fig:acc_P} (right panels). We do not identify any significant bias. Just above the classical confusion limit, the precision remains high and the half width of the 16--84\,\% region is 2, 3, 4, and 6\,deg in PPI1, PPI2, PPI3, and PPI4, respectively. However, the region equivalent to 2\,$\sigma$ is more than two times broader (11, 11, 13, and 17\,deg, respectively), highlighting that the impact of the confusion noise on angle measurements is non Gaussian.

\subsection{Purity and completeness}

\label{sect:puco_P} 

The purity of the samples extracted from polarization maps is excellent ($>$98\,\%, see Fig.\,\ref{fig:purity_sythesis}), and is barely affected by the choice of definition of the clustering. This is not surprising as the low level of flux boosting by the neighbors on $P$ means definitions A and B consider very similar matches. Finally, the clustering is also not expected to have a strong impact, since the polarized flux density does not add up as it does for the intensity data.

The completeness curves as a function of the intrinsic galaxy $P$ have a rather similar shape to those found for $I$, but the transition between low and high completeness appears at lower flux densities (see Fig.\,\ref{fig:comp_perband_P}). As shown in Table\,\ref{tab:conf_levels} summarizing the completeness in both intensity and polarization, the polarized flux density limits are up to a factor of 1.8 lower than the product of the limits in intensity by the mean polarization fraction $\mu_p$. This is again probably caused by the the non-additivity of the polarized flux density inside a beam in polarization mitigating slightly the blending problems. Consequently, the surface density of sources above the classical confusion limit in a given band is higher in polarization than in intensity (see Table\,\ref{tab:conf_levels}).

Finally, we find that the impact of clustering on the 50\,\% completeness polarized flux density and the classical confusion limit is negligible (see Fig.\,\ref{fig:comp_clustering}, red lines). This is the consequence of the confusion in polarization being driven by chance polarization alignments rather than the local source density. There is a small impact ($<$10\,\%) on the 80\,\% completeness polarized flux density, which could have the same cause as the effect seen at short wavelength in intensity (see discussion in Sect.\,\ref{sect:comp_I}).

\subsection{Polarized detection probability in the SFR-$z$ plane above the classical confusion limit}

\label{sect:SFRz_P}

We also studied the probability of recovering a source above the classical confusion limit using the same method as described in Sect.\,\ref{sect:SFRz_I} for the intensity. The results are shown in Fig.\,\ref{fig:SFRz_P}.

In the PPI bands, only z$<$2.5 galaxies emerge from the confusion, similarly to intensity. The PPI1 band probes 10$^{10}$\,M$_\odot$, 10$^{11}$\,M$_\odot$, and L$_\star$ galaxies up to z$\sim$0.5, z$\sim$1.5, and z$\sim$0.5, respectively. Above $z=2.25$, the probability of detection remains small even for the most strongly star-forming galaxies. Although they are less sensitive at low redshift, the PPI2 and PPI3 bands are slightly better at catching these extreme sources, since they observe them closer to their peak of emission. Consequently, the probability to detect a source in at least one band is very similar to the probability to detect it in the PPI1 band with the exception of the the tail at z$>$2.25 and SFR$\sim$1000\,M$_\odot$/yr.

\begin{figure}
\centering
\includegraphics[width=8cm]{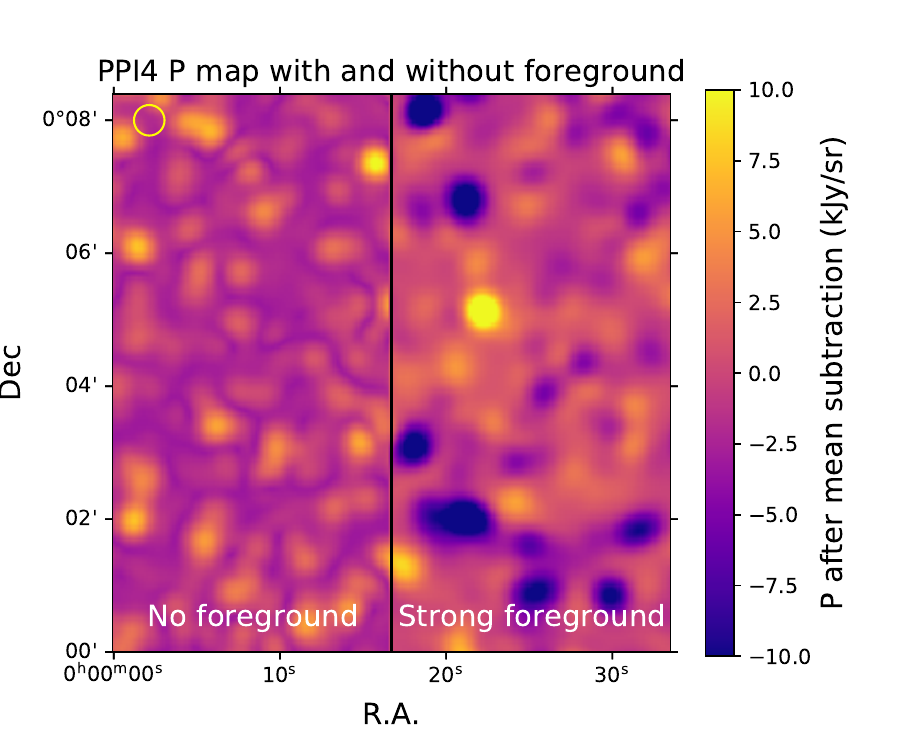}
\caption{\label{fig:foreground_effect} Comparison between the PPI4 P maps in absence (only CIB, left side) and in presence (right side) of a strong foreground (1000\,MJy/sr, see Sect.\,\ref{sect:color_foreground}) illustrating the different behavior of the CIB confusion noise depending on the foreground strength. Since P is much higher on the right side, we subtracted the mean of each side to obtain a better visualization. The yellow circle in the top-left corner shows the instrumental beam size.}
\end{figure}

\begin{table*}
\caption{\label{tab:conf_foreground} Estimated fluctuations caused by the CIB in polarized surface brightness density maps in absence and in presence of a strong foreground (see Sect\,\ref{sect:color_foreground}), minimal foreground surface brightness to obtain a 10\,\% precision on the foreground polarized color (confusion-noise only), and correlation coefficient of the polarized CIB signal between two polarized bands at this minimal surface brightness.}
\centering
\begin{tabular}{lcccc}
\hline
\hline
 & PPI1 & PPI2 & PPI3 & PPI4 \\
 \hline
Polarized CIB 1$\sigma$ fluctuations at native resolution in kJy/sr & 1.7 & 2.6 & 3.0 & 2.4  \\
1$\sigma$ confusion noise at native resolution in kJy/sr (strong foreground case) & 1.8 & 2.9 & 3.4 & 2.9 \\
Polarized CIB 1$\sigma$ fluctuations at PPI4 angular resolution in kJy/sr & 1.2 & 1.8 & 2.3 & 2.4 \\
1$\sigma$ confusion noise at PPI4 angular resolution in kJy/sr (strong foreground case) & 1.4 & 2.2 & 2.8 & 2.9 \\
Polarized surface brightness limit to obtain a 10\,\% uncertainty on the color with PPI4 in kJy/sr & 11 & 19 & 20 & --  \\
Correlation coefficient between a band and PPI4 at the polarized surface brightness limit & 0.79 & 0.87 & 0.96 & -- \\
\hline
\end{tabular}
\end{table*}

\subsection{Impact of high-z galaxy confusion on measurements of Galactic and low-z diffuse emission in polarization}

\label{sect:color_foreground}

The confusion noise is not only a problem for studying high-redshift galaxies. The fluctuations of the polarized CIB can impact both diffuse foreground and background measurements. The case of the cosmic microwave background has already been extensively discussed by \citet{Lagache2020}. While the contribution of astrophysical components in intensity is additive, the polarization is a vectorial quantity and leads to a more complex combination of the various components. In intensity, the confusion noise from the CIB only is sufficient to estimate its impact on the foreground science (e.g., diffuse Galactic emission and nearby or spatially-resolved galaxies). In contrast, as we will show in this section, the impact of the CIB in polarization depends on the foreground polarized surface brightness.

To estimate the fluctuations caused by background sources, we convert our simulated $Q_{\rm CIB}$ and $U_{\rm CIB}$ maps to MJy/sr and co-added them with a constant polarized foreground. For simplicity, we assume that this foreground is oriented on the $Q$ direction and denote this constant foreground value as $Q_f$ (by construction $U_f$=0\,MJy/sr). The value of the $P$ map combining the two components is thus:
\begin{equation}
P_{\rm tot} = \sqrt{ (Q_{\rm CIB} + Q_f)^2 + U_{ \rm CIB}^2}
\end{equation}
If $Q_f \ll Q_{\rm CIB}$, the results are similar to the case discussed in Sect.\,\ref{sect:simpolar} (except that the units are different). If $Q_f \gg Q_{\rm CIB}$, polarized CIB can be seen as a perturbation of the strong foreground signal:
\begin{equation}
\frac{\partial P_{\rm tot}}{\partial Q_{\rm CIB}} = \frac{Q_{\rm CIB} + Q_f}{P_{\rm tot}} \approx 1 \textrm{ and }
\frac{\partial P_{\rm tot}}{\partial U_{\rm CIB}} = \frac{U_{\rm CIB}}{P_{\rm tot}} \ll 1.
\end{equation}
We can thus see that the impact on the $P_{\rm tot}$ map depends on whether the CIB vector is aligned to the foreground or not.  If they are in the same direction the CIB component in the $Q$ direction will thus add or remove polarized flux density compared to the foreground alone. In contrast, the orthogonal component ($U$ in our construction) has no first-order impact on $P_{\rm tot}$.

The impact of this asymmetry generated by the strong foreground is illustrated Fig.\,\ref{fig:foreground_effect}. While the pure CIB map has mainly positive fluctuations, the sum of the CIB and the foreground exhibits both positive fluctuations (CIB and foreground polarization in the same direction) and negative fluctuations (orthogonal direction) at the position of the bright sources. We can also see that the fluctuations around the mean are larger in presence of a strong foreground.

We derive the classical confusion limit in the presence of the CIB and a foreground using a similar method as in Sect.\,\ref{sect:conflim_def}. However, since negative sources can appear when the foreground is included, we iteratively mask all the 5$\sigma$ outliers instead of only the positive ones. In Table\,\ref{tab:conf_foreground}, we tabulate the 1$\sigma$ fluctuations generated by the CIB in absence and in presence of a strong foreground. For the case of the strong foreground, we adopt $Q_f = 1000$\,kJy/sr at each wavelength, which is more than 5 orders of magnitude above CIB fluctuations. The values obtained for a fainter foreground would be between these two extreme cases. Finally, since some science cases will need color maps with a matched resolution, we also derive the same quantity after degrading the beam size to the PPI4 resolution.

The fluctuations measured in presence of a strong foreground are up to 20\,\% higher than in the CIB-only case. Hence, this is a small but non-negligible effect. If we had masked only the positive 5$\sigma$ outliers, the CIB fluctuations would have been up to 50\,\% higher in the strong foreground case, but unchanged in the pure CIB case since there are no strong negative fluctuations. Our table also shows that CIB fluctuations at PPI4 resolution are lower than at native resolution. In polarized surface brightness units, the signal from the constant foreground does not vary with the beam size, while a larger beam contains more sources and reduces the stochastic fluctuations.

Finally, we explored the impact of CIB on foreground color measurements. We use the ``astrodust'' model of dust emission and polarization \citep{Hensley2023} and have assumed that the dust is heated by a radiation field appropriate for diffuse atomic gas, to derive nominal input values of the PPI1/PPI4, PPI2/PPI4, and PPI3/PPI4 foreground colors of 0.52, 1.02, and 1.25, respectively. We varied the foreground polarized surface brightness fixing the input color, and derived the relative uncertainty on the measured foreground color produced by CIB fluctuations. We then interpolated between these values to determine the polarized surface brightness sensitivity limit corresponding to a 10\,\% uncertainty (see Table\,\ref{tab:conf_foreground}). The foreground polarized surface brightness limits to reach a 10\,\% precision on the foreground color are lower than 10 times the 1$\sigma$ CIB fluctuations, which is the limit expected based only on the numerator part of the color computation. However, as shown in the last row of Table\,\ref{tab:conf_foreground}, the confusion noise is highly correlated. Positive and negative fluctuations of the CIB are thus expected to impact both bands in a similar way, reducing their impact on the ratio. The correlation between bands thus mitigates the confusion noise in such analyses.

We thus showed that the confusion noise from the polarized CIB depends on the properties of the foreground and is also strongly correlated between bands. Our work provides first estimates of the impact of CIB to study Galactic emission and nearby galaxies in polarization. More complex simulations including full foreground models will be key to prepare these science cases.

%%%%%% DISCUSS THE CONSEQUENCES ON SURVEYS! %%%%%%%%%%%%%%%%%%%%%%

\begin{figure*}
\centering
\begin{tabular}{cc}
\includegraphics[width=8cm]{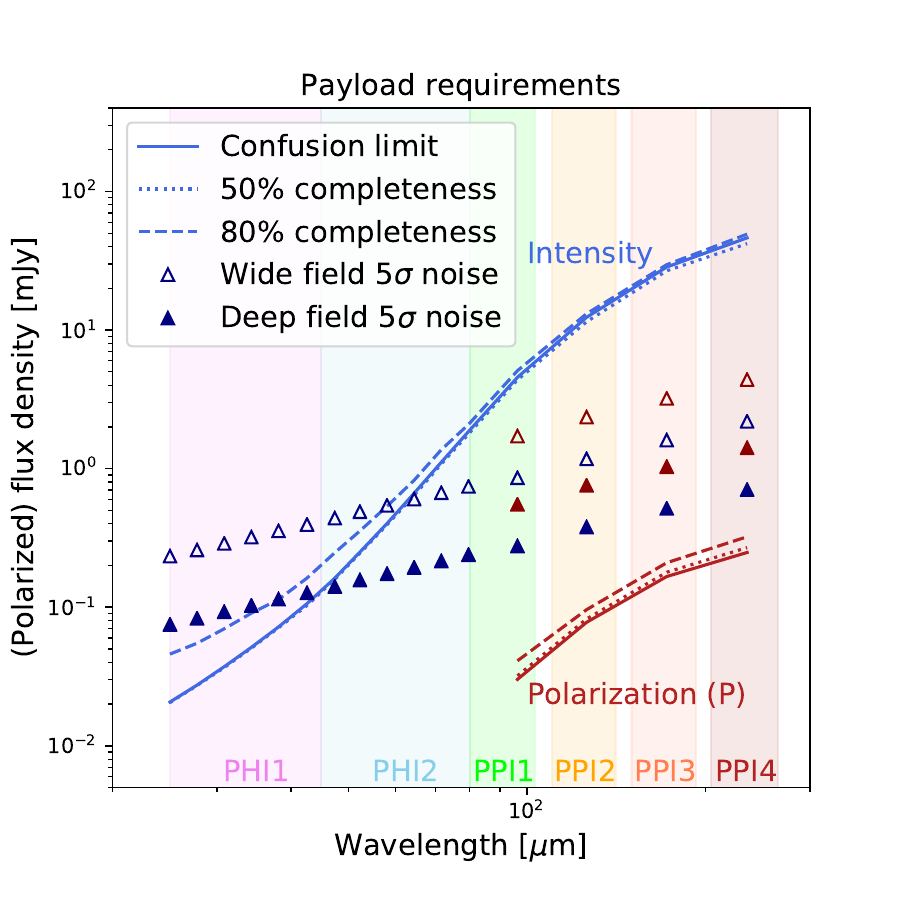} & \includegraphics[width=8cm]{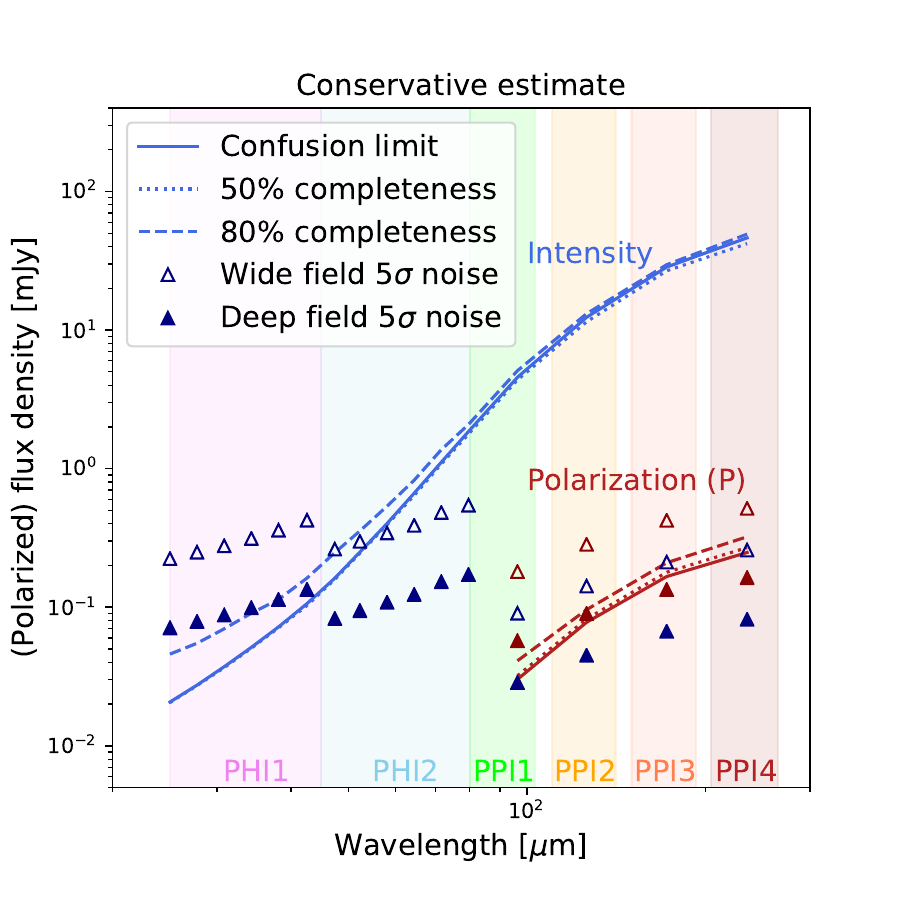}\\
\end{tabular}
\caption{\label{fig:comp_sythesis} Summary of the maximal depth reachable at the classical confusion limit as a function of wavelength and comparison with the expected PRIMAger instrumental depth. The left panel shows the survey depth for the payload required sensitivity, and the right panel corresponds to the conservative estimated sensitivities predicted by the instrumental teams. The open and filled upwards triangles correspond to the 5$\sigma$ instrumental sensitivity in the wide and deep surveys respectively. The solid, dotted, and dashed lines are the classical confusion limit, 50\,\%, and 80\,\% completeness flux densities, respectively. The blue symbols correspond to quantities derived from intensity maps (discussed in Sect.\,\ref{sect:disc_I}) and the red from polarization maps (see Sect.\,\ref{sect:disc_P}). Note that the flux density of a given galaxy is a factor of $\sim$100 lower in polarization than in intensity, since the mean polarization fraction is 1\,\%.}
\end{figure*}

\begin{table}
\caption{\label{tab:Ndet_polar}Summary of the number of expected detections (N$_{\rm det}$) above the polarized flux density limit P$_{\rm lim}$ (quadratic combination of the 5$\sigma$ confusion and 5$\sigma$ instrumental noise, see Sect.\,\ref{sect:disc_P}), their mean redshift, and their mean SFR for a deep and a wide 1500\,h PRIMAger survey assuming the payload required sensitivity and the conservative estimated sensitivity.}
\centering
\begin{tabular}{lcccc}
\hline
\hline
Band & P$_{\rm lim}$ & N$_{\rm det}$ & Mean $z$ & Mean SFR \\
& $\mu$Jy & & & M$_\odot$/yr \\
%%%copy below
\hline
\multicolumn{5}{c}{Deep field (1500h, 1deg$^2$)}\\
\multicolumn{5}{c}{with payload required sensitivity}\\
\hline
PPI1 & 553 & 86 & 0.27 & 71\\
PPI2 & 760 & 72 & 0.28 & 84\\
PPI3 & 1045 & 37 & 0.27 & 94\\
PPI4 & 1432 & 5 & 0.26 & 153\\
\hline
\multicolumn{5}{c}{Wide field (1500h, 10deg$^2$)}\\
\multicolumn{5}{c}{with payload required sensitivity}\\
\hline
PPI1 & 1718 & 100 & 0.12 & 28\\
PPI2 & 2353 & 75 & 0.10 & 22\\
PPI3 & 3215 & 25 & 0.07 & 12\\
PPI4 & 4394 & 5 & 0.03 & 2\\
\hline
\multicolumn{5}{c}{Deep field (1500h, 1deg$^2$)}\\
\multicolumn{5}{c}{with conservative estimated sensitivity}\\
\hline
PPI1 & 65 & 2546 & 0.59 & 61\\
PPI2 & 119 & 1940 & 0.62 & 78\\
PPI3 & 214 & 1050 & 0.68 & 119\\
PPI4 & 297 & 474 & 0.86 & 202\\
\hline
\multicolumn{5}{c}{Wide field (1500h, 10deg$^2$)}\\
\multicolumn{5}{c}{with conservative estimated sensitivity}\\
\hline
PPI1 & 183 & 5510 & 0.45 & 79\\
PPI2 & 294 & 4295 & 0.47 & 98\\
PPI3 & 455 & 2375 & 0.54 & 145\\
PPI4 & 573 & 880 & 0.63 & 215\\
\hline
%copy above
\end{tabular}
\end{table}

\begin{figure*}
\centering
\begin{tabular}{cc}
\includegraphics[width=7cm]{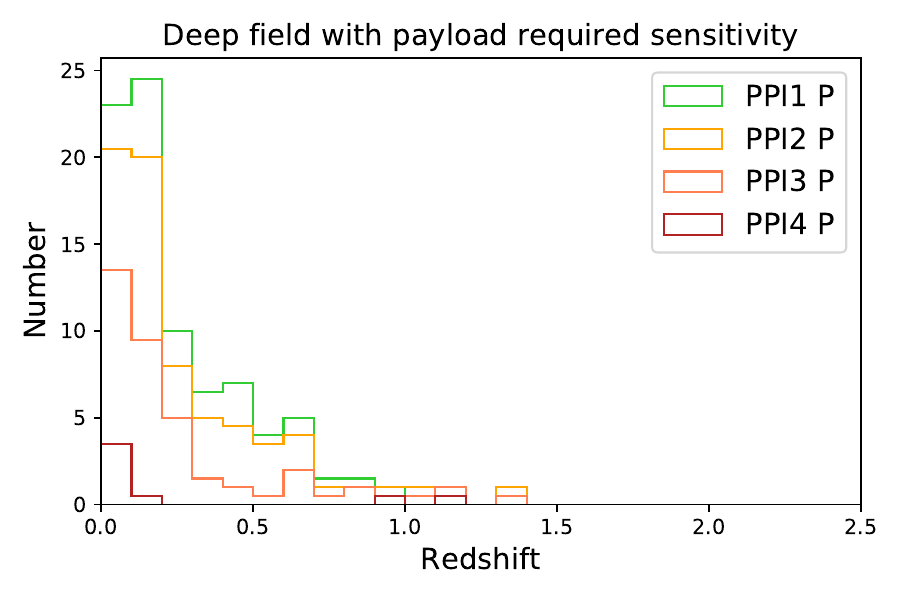} & \includegraphics[width=7cm]{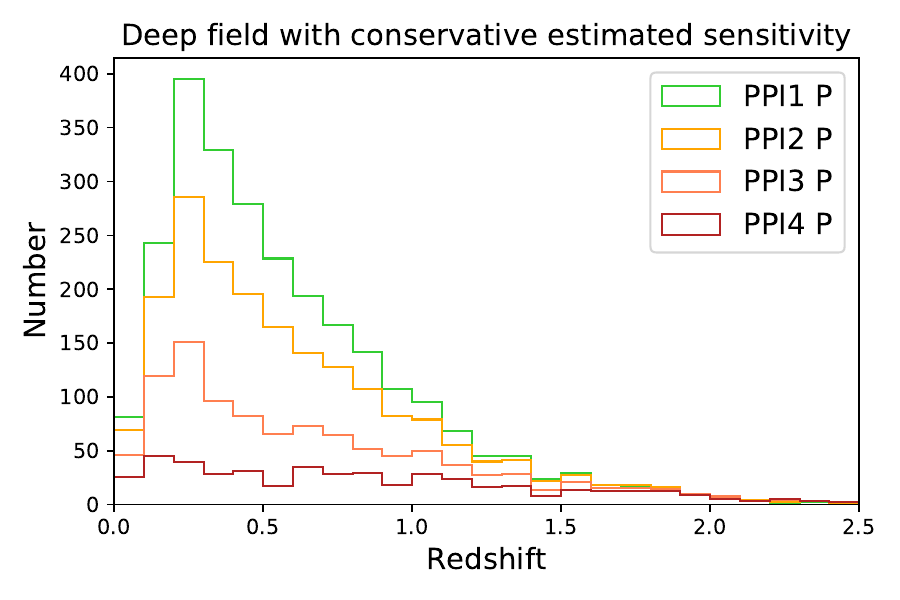}\\
\includegraphics[width=7cm]{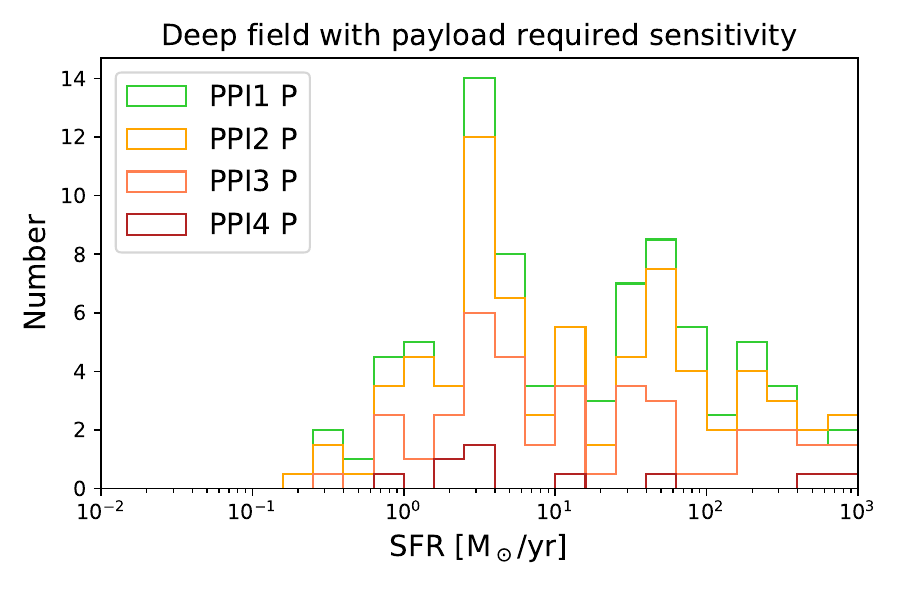} & \includegraphics[width=7cm]{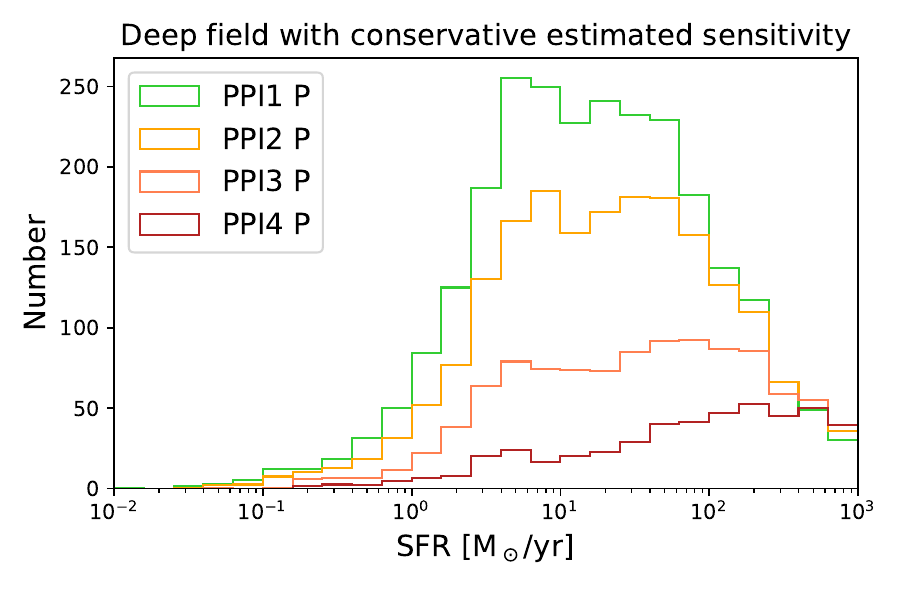}\\
\includegraphics[width=7cm]{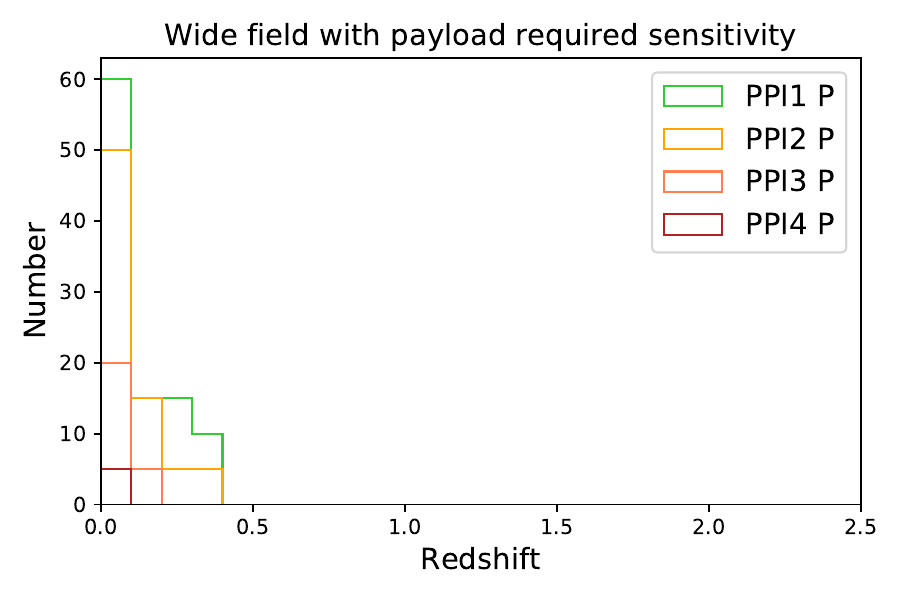} & \includegraphics[width=7cm]{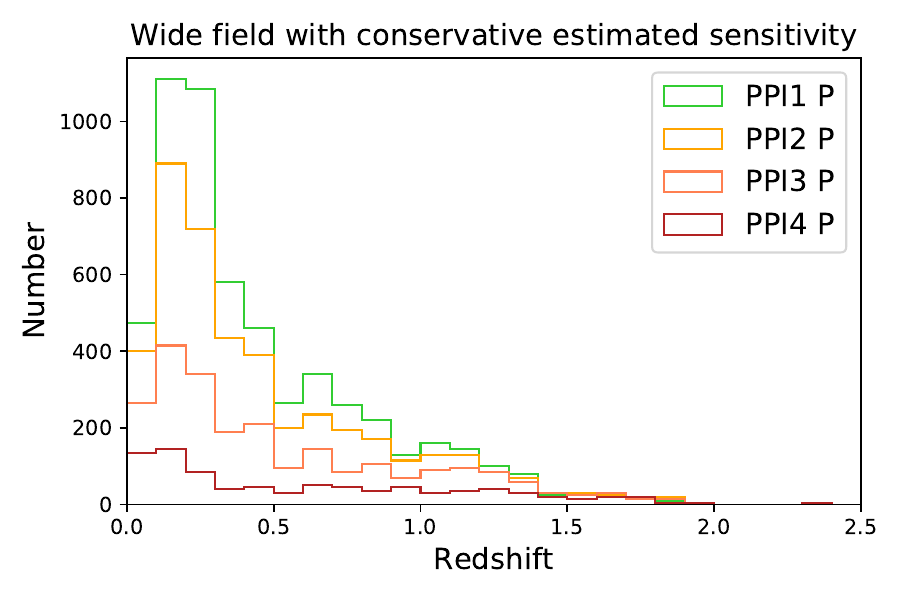}\\
\includegraphics[width=7cm]{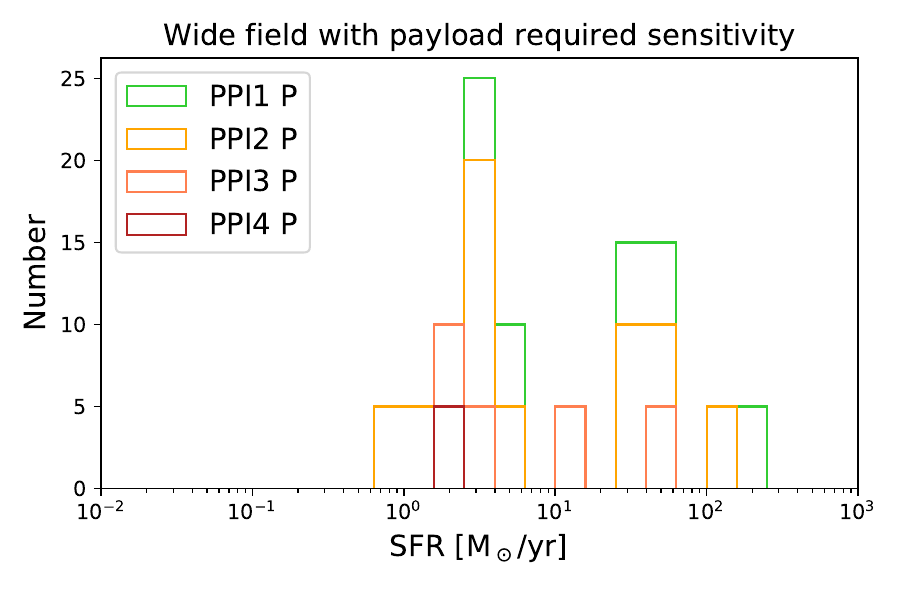} & \includegraphics[width=7cm]{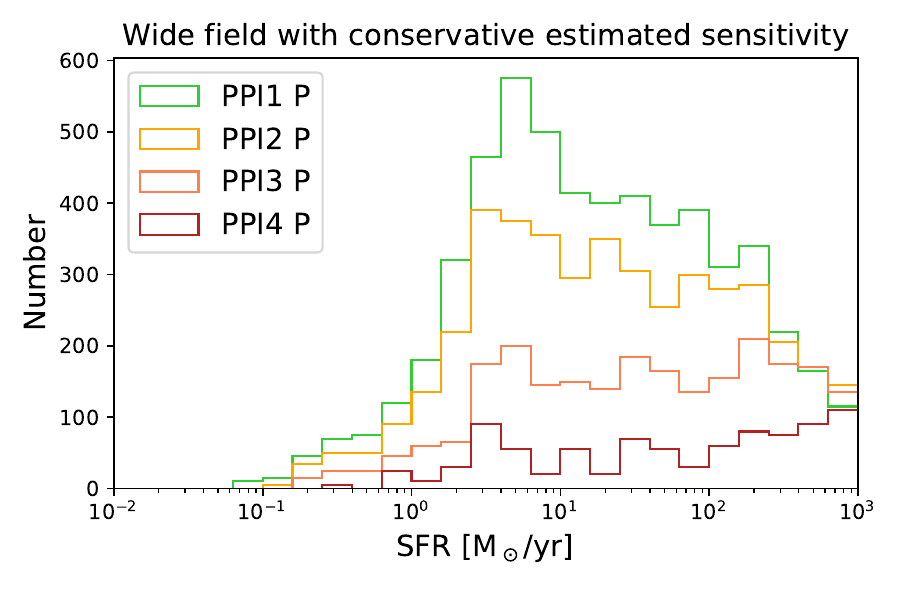}\\
\end{tabular}
\caption{\label{fig:histos_polar}Redshift and SFR distributions of the sources above the detection limit in polarized flux density $P_{\rm lim}$ (see Table\,\ref{tab:Ndet_polar} and Sect.\,\ref{sect:disc_P}). The left columns correspond to the payload required sensitivity and the right ones to the conservative estimated sensitivity. The rows are from top to bottom: redshift distribution in the deep field, SFR distribution in the deep field, redshift distribution in the wide field, SFR distribution in the wide field. The bands are color coded as indicated in the figure.}
\end{figure*}

\section{Consequences for surveys}

\label{sect:surveys}

\subsection{Expected impact of confusion in intensity}

\label{sect:disc_I}

In the previous sections, we discussed only the noiseless case corresponding to the best possible performance we could obtain for a given telescope diameter. However, it is crucial to compare the classical confusion limit with the expected instrumental performance. If the instrumental noise is much higher, the confusion can be ignored. If the instrumental noise is well below the confusion noise, advanced deblending methods will be required to make the most of the intrinsic sensitivity, but the performance may never fully match the instrumental noise.

In our analysis, we will consider two cases for the instrumental noise. The {\em payload required} sensitivity is the guaranteed performance of the instrument. It is a very conservative estimate to ensure high confidence in meeting the PI science goals. The real performance is expected to be much better, with at least 60\,\% margin and in some cases by factors of several.  We consider an intermediate estimate termed the {\em conservative estimated} sensitivity which lies between the payload requirement and the actual estimated performance. To be consistent with the confusion, we also consider a 5$\sigma$ limit. We treat the case of two fields observed 1\,500\,h each. The deep field has a 1\,deg$^2$ area, while the wide field covers 10\,deg$^2$.

In Fig.\,\ref{fig:comp_sythesis}, we compare the classical confusion limit with the instrumental sensitivity. The confusion has a steeper rise with increasing wavelength than the sensitivity for both sensitivity estimates. Consequently, the short wavelengths are noise-limited and the long wavelengths are confusion-limited. For the wide survey and a payload required sensitivity, the curves cross at 65\,$\mu$m around the center of band PHI2. As shown in Sect.\,\ref{sect:SFRz_I} and Fig.\,\ref{fig:SFRz_I}, the red end of the PHI2 and the PPI bands at the confusion limit are not probing a part of the SFR-$z$ space missed by the other sub-bands. However, PPI data are important to characterize the physics of the objects, and PHI priors will be crucial to deblend them. In the deep field with a payload required sensitivity, the classical confusion limit is reached at around 45\,$\mu$m, and the PHI2 band will thus be affected by it. At this wavelength, the 7.7\,$\mu$m PAH feature can be seen up to z$\sim$5. (see Sect.\,\ref{sect:SFRz_I}). This ensures that we will benefit from very deep priors up to this redshift before reaching the confusion, which will be essential to deblend the PHI2 and PPI bands and obtain more accurate physical constraints.

For the conservative estimated sensitivity, the confusion is reached at $\sim$55\,$\mu$m and $\sim$45\,$\mu$m in the wide and deep fields, respectively. We will thus be confusion limited in the PHI2 band, except in its bluest part for the wide fields. The  SFR-$z$ space probed will thus be similar to the pure confusion case discussed in Sect.\,\ref{sect:SFRz_I} and Fig.\,\ref{fig:SFRz_I}. We  also note that in PPI bands, the instrumental noise will be $\sim$2.5 orders of magnitude below the confusion. These data will thus be extremely close to the noiseless case discussed in this paper and will be ideal to apply modern deep learning deblending algorithm \citep[e.g.,][]{Lauritsen2021}.

\subsection{Feasibility of dust polarization surveys of distant galaxies}

\label{sect:disc_P}

Far-infrared blank-field polarization surveys are in a totally uncharted territory. With our analysis, we can now set constraints on the expected classical confusion limit and we have demonstrated that we can recover constraints on the polarized flux density and angle of a galaxy (Sect.\,\ref{sect:polar}). However, since the signal will be fainter than in intensity, it is important to check if the instrumental sensitivity will be good enough to detect a large sample of sources. In this section, we discuss only our standard simulation assuming a mean polarization fraction $\mu_p$ of 1\,\%. In Appendix\,\ref{sect:other_polfrac}, we consider the alternative cases with 3$\sigma$-lower ($\mu_p$=0.7\,\%) and 3$\sigma$-higher ($\mu_p$=1.3\,\%) values (see Sect.\,\ref{sect:simpolar}).

In Fig.\,\ref{fig:comp_sythesis}, we also compare the confusion and the instrumental limits in polarization. In PPI bands, the payload required and the conservative estimated sensitivities are very different. In the first case, the sensitivity limit in the deep fields are about an order of magnitude above the classical confusion limit (1.5\,dex above for the wide). In the second more optimistic case, the wide field is noise limited. In the deep field, the PPI1 and PPI2 band are close to the classical confusion limit, while the other bands have a sensitivity limit slightly below the classical confusion limit. This means that we should be very close to the confusion limited case in the SFR-$z$ plane discussed in Sect.\,\ref{fig:SFRz_I}.

To evaluate the impact of these two hypotheses on the sensitivity, we used the SIDES simulation to predict the number of detections expected in the various cases. We combined quadratically the 5$\sigma$ confusion and 5$\sigma$ instrumental noise to obtain a secure polarized flux density limit $P_{\rm lim}$, and used it to select the detectable sources in SIDES. Since the wide field is a factor of 5 larger than our simulation, we applied a scaling factor to the number of SIDES detections. A factor of 0.5 is applied for the deep field. The number of detections and their mean redshift and SFR are listed in Table\,\ref{tab:Ndet_polar}. We also show the redshift and SFR distributions in Fig.\,\ref{fig:histos_polar}.

For the payload required sensitivity, we expect $\sim$100\,galaxies per field in PPI1, but fewer than 10 in PPI4. Since it is a totally unexplored parameter space, it will open a new window with small but statistical significant samples. In the deep field, half of the sources detected in polarization are below $z\leq 0.2$, and only a handful are above $z\ge 0.7$ with a tail up to $z\sim1.5$. We will thus trace mainly intermediate redshifts, though the dust polarization properties of galaxies at these epochs are currently totally unexplored. In terms of SFR, we will span a large range of SFR from nearby 0.2 \,M$_\odot$/yr to high-$z$ 1000\,M$_\odot$/yr galaxies. In the wide field, objects are detected only up to $z=0.4$. As expected, the lowest SFR will not be probed. Paradoxically, we will also observe less $>$100\,M$_\odot$/yr systems than in the deep field. This is driven by the very low number density of these high-SFR objects at z$<$0.4 and the polarized flux density limit being too high to be able to detect any high-$z$ system.

With the conservative estimated sensitivity, the confusion and instrumental noise will be similar in the PPI1 and PPI2 bands. Several thousands of galaxies will be detected both in the deep and wide fields (Table\,\ref{tab:Ndet_polar}). The number of detections in the deep field is a factor of $\sim$2 smaller than in the wide field. The PPI1 band will provide the highest number of detections, but even the PPI4 band will detect several hundred sources allowing statistical studies of the polarized SEDs. Both deep and wide fields have a peak redshift distribution around z$\sim$0.4 with large tail up to $z=2.5$. In terms of SFR, we will cover 5 orders of magnitudes from 0.01 to 1000\,M$_\odot$/yr.

The payload required sensitivity would open a new window on the dust polarization of high-redshift galaxies with more than hundred detections up to z$\sim$1.5. With the conservative estimated sensitivity, the results would be totally transformational by opening this new window directly with several thousands of sources up to $z=2.5$.

\subsection{Synergies between intensity and polarization surveys}

\label{sect:synergies}

Independently of the sensitivity scenario, the PHI1 band will be dominated by the instrumental noise, while the PPI bands will always be confusion limited in intensity. However, the classical confusion limit in polarization is more than 100 times smaller than in intensity, and the confusion will only be reached in the deep field in the optimistic sensitivity scenario. This opens the opportunity for synergistic strategies between intensity and polarization.

If we undertake deep integrations that approach the classical intensity confusion limit in the PHI1 band, the PHI2 and PPI band will be limited by confusion in intensity. However, depending on the exact sensitivity ratio between bands, PPI bands may still not be affected by confusion in polarization. In addition, it will also be possible to deblend the PHI2 band using a prior-based extraction algorithm \citep{Donnellan2024}. All the bands will thus be used efficiently in such a strategy, and we will fully exploit the high PPI sensitivity through the polarization. The risk of attempting a first high-z deep polarization survey will also be mitigated, since we will get extremely deep intensity data at shorter wavelength at the same time. PRIMAger is thus a very promising instrument, which is able to open two new windows of survey parameter space with a single deep field observation.

\section{Conclusion}

\label{sect:conclusion}

%{\bf Using simulated PRIMAger maps from the SIDES simulation, we studied how confusion impacts the performance of basic blind source extractors both in intensity and polarization. We determined the classical confusion limit for all PRIMA bands. The flux density limit increases steeply with increasing wavelength. 

We produced simulated PRIMAger data (Fig.\,\ref{fig:maps}) using the SIDES simulation to study how confusion impacts the performance of basic blind source extractors, both in intensity and polarization. With this, we determined the classical confusion limit for all PRIMA bands, which increases steeply with increasing wavelength (Fig.\,\ref{fig:comp_sythesis}).

%In the wide and deep fields in intensity, the confusion limit curve crosses the sensitivity limit one around 60\,$\mu$m and 45\,$\mu$m, respectively. Above these wavelengths, advanced deblending methods benefiting from un-confused priors at shorter wavelength will be necessary. This is discussed in a companion paper \citep{Donnellan2024}. In polarization, the confusion limit is more than two orders of magnitude lower than in intensity. Surveys will thus be limited by the sensitivity instead of the confusion, except above 150\,$\mu$m in the deep field for the baseline sensitivity scenario.

For the conservative estimated sensitivities of the PRIMAger wide and deep surveys, the classical confusion limit curve crosses the sensitivity limit at approximately 60\,$\mu$m and 45\,$\mu$m, respectively. Taking advantage of the available instrument sensitivity at longer wavelengths requires using deblending methods. The PRIMAger hyperspectral architecture, which produces finely sampled $R=10$ SEDs, is particularly good at enabling these methods by providing priors for sources detected at shorter, unconfused wavelength. A companion paper \citep{Donnellan2024}, analyzes the performance of a particular deblending approach \citep[XID+,][]{Hurley2017}, showing that its application will recover fluxes out to $\lambda=100$\,$\mu$m and beyond for astrophysical SEDs. Moreover, we showed that in polarization the confusion limit is more than two orders of magnitude lower than in intensity. Surveys will thus be limited by the instrument sensitivity, except at $\lambda>150$\,$\mu$m in the deep field.

We studied the effect of galaxy clustering, showing that it has  a mild impact on confusion in intensity ($<$25\,\%), while its effect on polarization is very small (Fig.\,\ref{fig:comp_clustering}). This difference of behavior is explained by the respective scalar and vectorial nature of intensity and polarization. 

The measured flux density in intensity for $\lambda>100$\,$\mu$m is on average larger than the flux density of the brightest galaxy in the beam, because of the contamination by confused sources (Fig.\,\ref{fig:photacc_I}). In contrast, the polarized flux density and polarization angle measurements are essentially not affected (Fig.\,\ref{fig:acc_P}). The polarization fraction measurements are affected, however, because it is derived from both intensity and polarization measurements. 

We computed the probability to detect a galaxy above the classical confusion limit as a function of its position in the SFR-z plane (Fig.\,\ref{fig:SFRz_I} and \ref{fig:SFRz_P}). In intensity, galaxies at the knee of the infrared luminosity function (L$_\star$) will be above the classical confusion limit in at least one band up to $z\sim3$, while massive (10$^{11}$\,M$_\odot$) main-sequence galaxies can be recovered up to $z\sim5$. In polarization, PRIMager opens up a brand new parameter space by enabling studies for L$_\star$ and massive main-sequence galaxies which are brighter than the classical confusion limit up to $z\sim0.5$ and $z\sim1.5$, respectively. We can also observe a tail of extreme objects up to $z\sim2.5$.

%The polarized emission of high-z galaxies can also act as noise for the polarization studies of diffuse Galactic emission and resolved nearby galaxies. The 1-\,$\sigma$ noise can vary from 1.2 to 3.4\,kJy/sr depending and the band and how bright the foreground is. This noise is correlated between bands, which impact the uncertainties on foreground color measurements. We estimated that a minimum foreground polarized surface brightness of 11\,kJy/sr at 96\,$\mu$ is necessary to obtain a 10\,\% precision on the foreground P$_{96}$/P$_{235}$ color.

We estimate the effect of the background of polarized emission due to high-$z$ galaxies on measurements of polarized emission of extended low-$z$ foreground sources (Table\,\ref{tab:conf_foreground}). The 1-\,$\sigma$ noise on polarized surface brightness can vary from 1.2 to 3.4\,kJy/sr depending on the band and how bright the foreground is. This noise is correlated between bands, which has an impact on the uncertainties of foreground color measurements. We estimated that a minimum foreground polarized surface brightness of 11\,kJy/sr at 96\,$\mu$m is necessary to obtain a 10\,\% precision on the foreground $P_{96}/P_{235}$ color.

Finally, assuming the PRIMAger conservative estimated sensitivity, we expect several thousands of detections of the integrated dust polarization of high-z galaxies (up to z$\sim$2.5) in both the deep and wide fields (Fig.\,\ref{fig:histos_polar} and Table\,\ref{tab:Ndet_polar}). Considering that polarization has been reported on only two lensed high-z galaxy so far \citep{Geach2023,Chen2024}, PRIMAger has the potential to revolutionize this field of study by producing large statistical samples.

Detecting sources in polarization requires a $\sim$100 times better sensitivity than necessary to detect the same sources in intensity at the same wavelength. Opening this new window on high-z dust polarization has thus obviously a cost in observing time. However, it can be offset by the detector intrinsic sensitivity. Indeed, for a deep PRIMAger survey (1\,deg$^2$ in 1\,500\,h) and our conservative estimated sensitivities, both the short-wavelength data in intensity (PHI1) and the polarization maps at longer wavelengths (PPI) will be close to the confusion limit of the instrument. There is thus a strong synergy between pushing the instrument to this limit in intensity at short wavelengths, and performing deep polarization surveys at long wavelengths. At this depth, the long-wavelength data in intensity will be fully dominated by confusion rather than by the extremely low instrumental noise, but analysis methods using short wavelength priors are expected to allow us to deblend sources efficiently using short-wavelength data \citep{Donnellan2024}. A single survey will thus obtain well-sampled mid- to far-infrared SEDs in intensity, opening new opportunities for star formation and AGN science \citep{Bisigello2024}, together with long-wavelength far-infrared data in polarization, providing new constraints on galactic magnetic fields.

\begin{acknowledgements}
This research made use of Photutils, an Astropy package fordetection and photometry of astronomical sources \citep{photutils}. James Donnellan was supported by the Science and Technology Facilities Council, grant number ST/W006839/1, through the DISCnet Centre for Doctoral Training. Seb Oliver acknowledges support from the UK Space Agency through ST/Y006038/1. Guilaine Lagache This project has received funding from the European Research Council (ERC) under the European Union’s Horizon 2020 research and innovation programme (grant agreement No 788212). Portions of this work were carried out at the Jet Propulsion Laboratory, California Institute of Technology, under a contract with the National Aeronautics and Space Administration (80NM0018D0004).

The contributions of the authors using the Contributor Roles Taxonomy (CRediT) were as follows:
{\bf Matthieu Bethermin:} Conceptualization, Methodology, Software, Validation, Investigation, Writing - original Draft;
{\bf Alberto Bolatto, Laure Ciesla, Jason Glenn, Brandon Hensley, Guilaine Lagache, Enrique Lopez-Rodriguez, Seb Oliver, and Alexandra Pope:} Conceptualization,  Writing - Review \& Editing;
{\bf Denis Burgarella and James Donnellan:}
Writing - Review \& Editing;
{\bf Charles Bradford, François Boulanger, and Marc Sauvage:}
Conceptualization.

\end{acknowledgements}

\bibliographystyle{aa}

\bibliography{biblio}

\begin{appendix}

\section{Calibration of the polarization fraction using data from the local Universe}

\begin{figure}
\centering
\includegraphics[width=8cm]{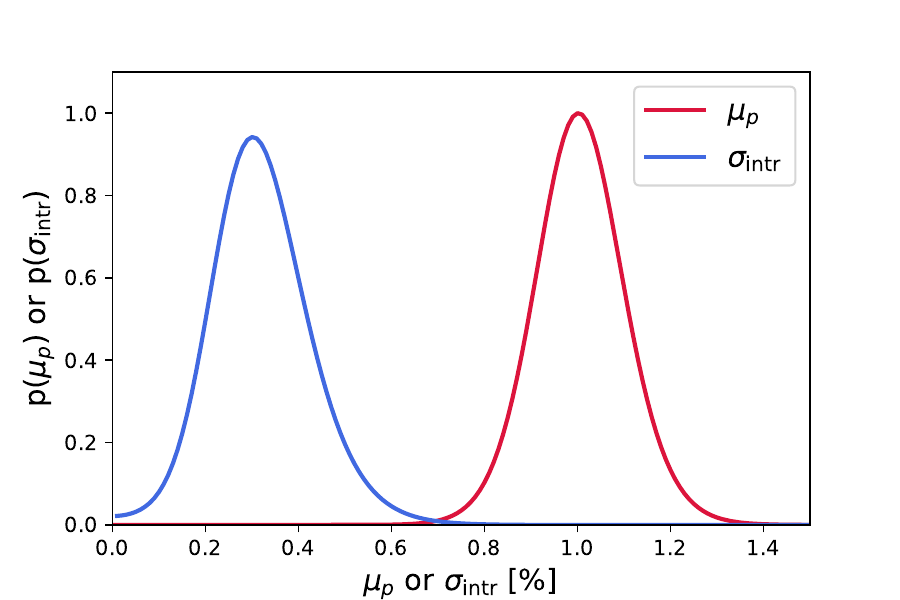}
\caption{\label{fig:polar_calib} Probability density function of the central value of the polarization fraction ($\mu_p$) and the scatter ($\sigma_{\rm intr}$) around it. These distributions were determined based on the local Universe sample of \citet[][see Sect.\,\ref{sect:simpolar}]{Lopez-Rodriguez2022}.}
\end{figure}

\begin{figure*}
\centering
\begin{tabular}{cc}
\includegraphics[width=8cm]{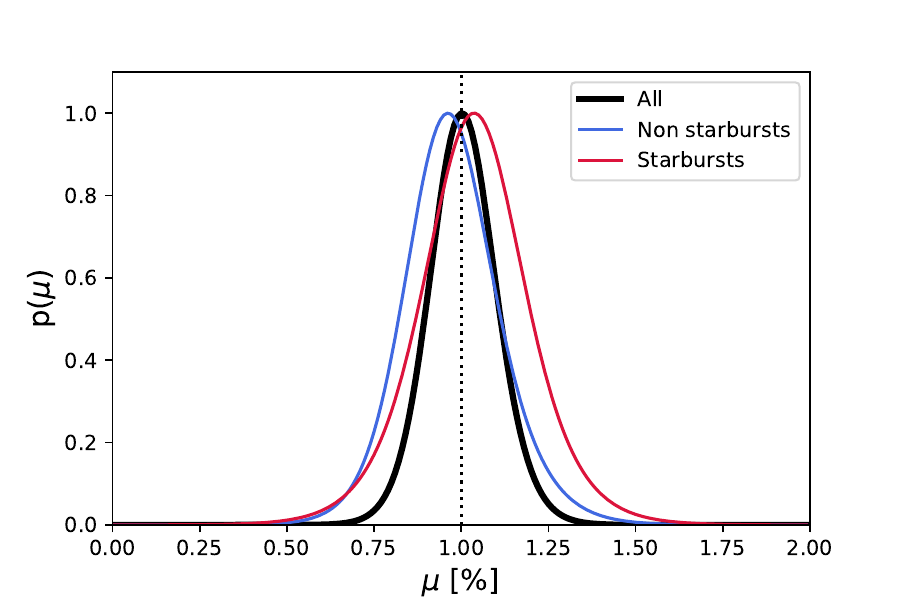} & \includegraphics[width=8cm]{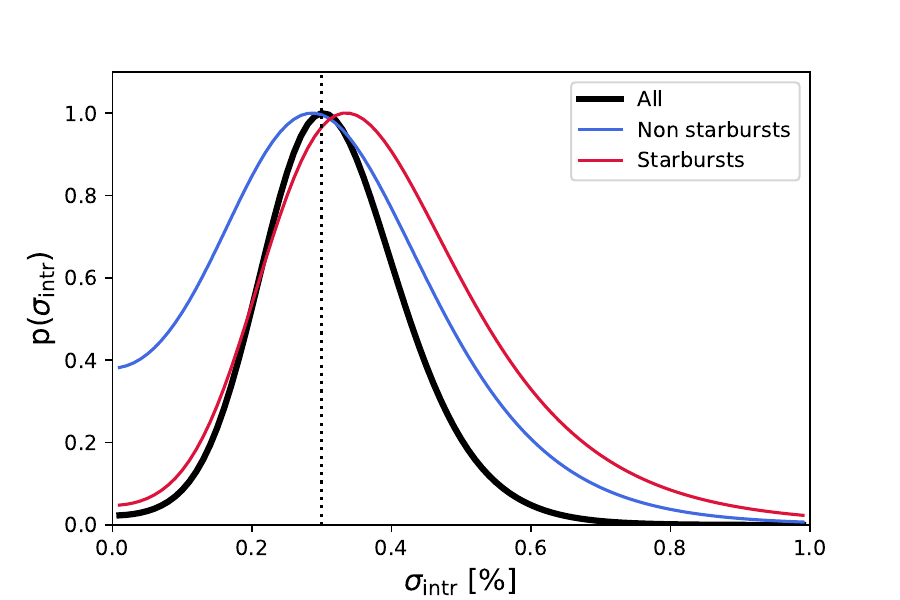} \\
\includegraphics[width=8cm]{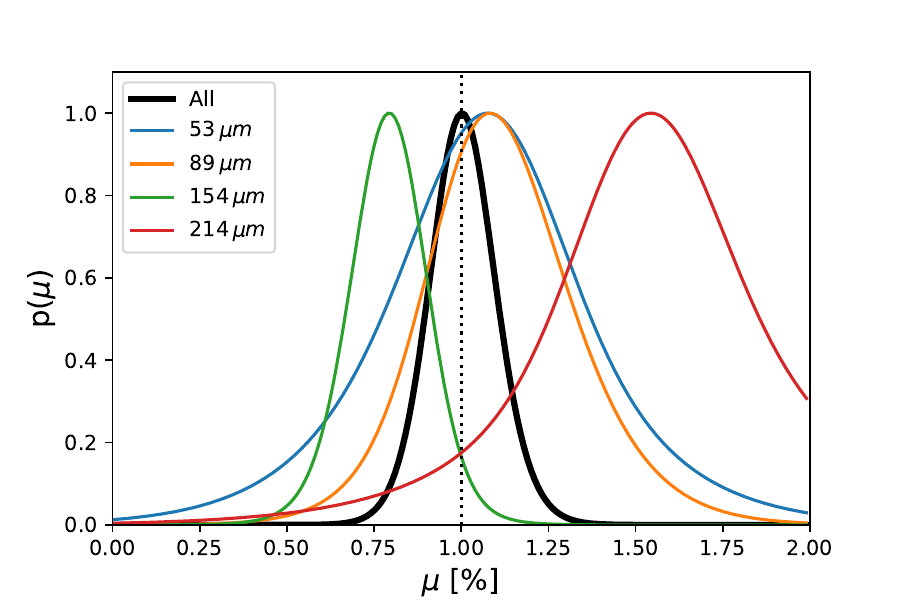} & \includegraphics[width=8cm]{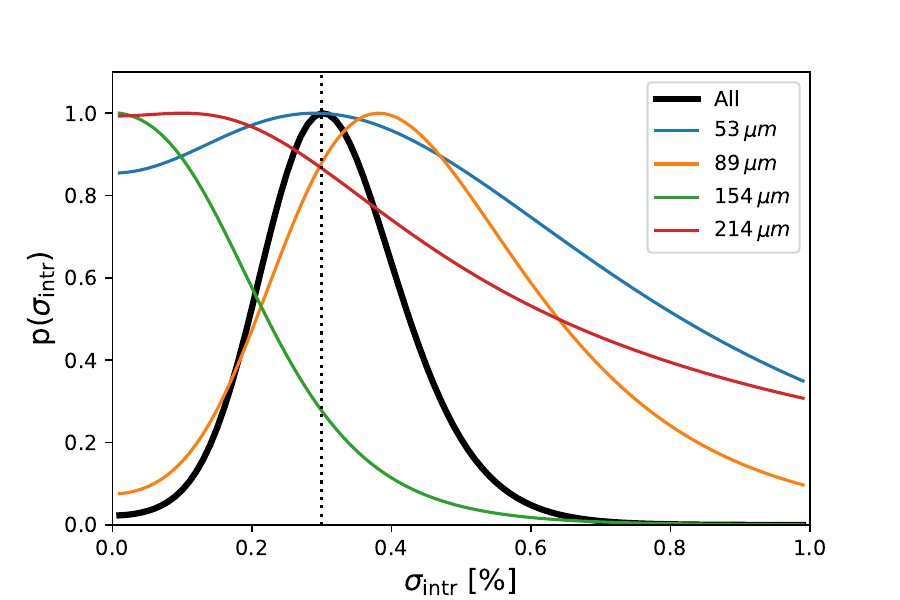} \\
\end{tabular}
\caption{\label{fig:polar_dependence} Probability density of the mean (left panels) and the standard deviation (right panels) of the polarization fraction determined using the method described in Sect.\,\ref{sect:simpolar}. The top and bottom panels show the dependence with the presence of a starburst and the wavelength, respectively. The black thick curve is the result obtained with the full sample and the dotted vertical line is the most probable value. The colored curve are obtained using subsamples.}
\end{figure*}

\label{sect:polar_dependence}

To calibrate the intrinsic distribution of the polarization fraction, we used the local-Universe SALSA sample from \citet{Lopez-Rodriguez2022}. We discarded the Circinus value at 214\,$\mu$m, which is a very strong outlier (8.4\,\%). It is a Seyfert object, which may not be representative of typical galaxy populations. It showed large polarization fractions located in the interarm regions at few kpc from the central AGN. This specific observation only had a few dozen polarization measurements due to a shallow integration time, and the significance of final polarization measurements is 3--5\,$\sigma$.

We fitted the mean polarization fraction ($\mu_p$) and the intrinsic scatter around it ($\sigma_{\rm intr}$) using their measured integrated polarization fractions (their Table\,5). The likelihood $\mathcal{L}$ is computed using:
\begin{equation}
\textrm{ln} (\mathcal{L}) = \sum\limits_{\rm i=1}^{N_{\rm sample}}-  \frac{ \textrm{ln}(2 \pi) + \textrm{ln}(\sigma_{\rm intr}^2 + \sigma_{\rm mes,i}^2)}{2} - \frac{(p_{\rm mes,i} - \mu_p)^2}{2 (\sigma_{\rm intr}^2 + \sigma_{\rm mes,i}^2)},
\end{equation}
where $p_{\rm mes,i}$ is the integrated polarized fraction of the i-th object, and $\sigma_{\rm mes,i}$ is measurement uncertainty on it. By combining all the objects and the wavelengths, we found $\mu_p = 1.0_{-0.1}^{+0.1}$\,\% and $\sigma_{\rm intr} = 0.3_{-0.1}^{+0.1}$\,\%. The marginalized probability density functions of both parameters are shown in Fig.\,\ref{fig:polar_calib}. We thus drew $p$ from a Gaussian with $\mu_p=1.0$ and $\sigma_{\rm intr}=0.3$, and replaced negative values with zero. 

In the current version of SIDES including the polarization, we use the same probability function to draw the polarization fraction $p$ for all galaxies (see Sect.\,\ref{sect:simpolar}). In Fig.\,\ref{fig:polar_dependence}, we show the probability density of the means and standard deviations obtained for various subsamples from the SALSA survey \citep{Lopez-Rodriguez2022}. The starbursts have marginally higher mean polarization fraction and standard deviation than non starbursts, but the offset between the two probability density is much smaller than the uncertainties. The probability of the values found for the full sample corresponds to 0.95 times the peak probability density of the subsamples. There is thus no significant dependence detected by our data-driven approach. 

The wavelength-dependence is more complex to interpret. Both 53 and 89\,$\mu$m have peak probability densities around 1.1\,\% of polarization, while 154\,$\mu$m peaks around 0.8\,\% and 214\,$\mu$m peaks around 1.6\,\% with large uncertainties. It could be a hint of the signature from inner outflows \citep[][, their Fig.\,10 and 11]{Lopez-Rodriguez2023}. However, the best-fit value of 1\,\% found combining all wavelengths has a probability larger than 0.2 at all wavelength. It is thus hard to conclude with such a small probability. \citet{Lopez-Rodriguez2023} used the resolved polarization data to consolidate their results, but in this paper we consider only integrated properties since high-z galaxies will not be spatially resolved.

\section{Impact of the instrument pixels}

\begin{figure}
\centering
\includegraphics[width=8cm]{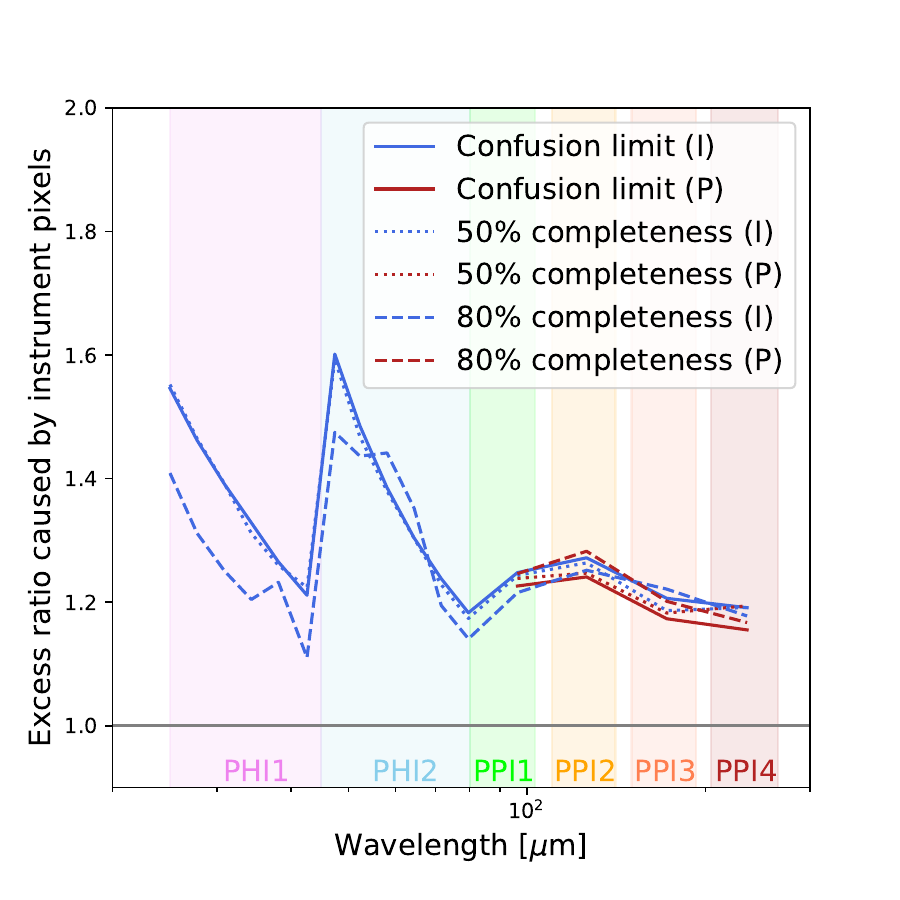}
\caption{\label{fig:pix_impact} Same figure as Fig.\,\ref{fig:comp_clustering} but for the effect of the instrument pixels on the beam FWHM in absence of thin dithers or other mitigations. The ratio corresponds to the same clustered input catalog, but with different beam FWHM.}
\end{figure}

\label{sect:instrpix_impact}

In our main analysis, we used a beam FWHM driven only the optics. However, the current design of PRIMAger has pixel sizes that undersample the PSF (particularly at short wavelengths). Pixel sizes range from $1.25\lambda/D$ to $0.7\lambda/D$ for the PHI pixels in the wavelength ranges 23 to 45 $\mu$m (PHI1) and 45 to 82 $\mu$m (PHI2). We repeated our analysis using a beam that is broadened by the response of the pixel. To compute this broaden beam, we convolve the diffraction-limited response of the system with the response of the pixel, modeled as a top-hat function with the diameter corresponding to the wavelength range. This estimate is  conservative since the  effects of pixel size can be mitigated in part by the use of sub-pixel dithers and drizzle-type map making algorithms. These calculations are thus an upper limit on the possible impact. Moreover, the instrument pixel size will likely be further optimized to reduce undersampling effects. 

In Fig.\,\ref{fig:pix_impact}, we show the ratio between the results obtained with the broader beam including the impact of the pixels and the narrower beam corresponding to the optics only. The impact on the 5$\sigma$ classical confusion limit and the 50\,\% completeness flux density is similar. There is also no significant difference between intensity and polarization in PPI bands. We find a difference by a factor of 1.6 in the blue side of both PHI bands, decreasing down to 1.2 in the red side of each band. There is thus a strong jump between band PHI1 and PHI2, which can be explained by the constant pixel size in a given band while the beam size is increasing from the blue to the red. The relative impact of the pixel size is thus stronger in the blue, where the beam is the smallest. In PPI bands, the effect is of the order of 20\,\%.  Finally, the 80\,\% completeness flux density is slightly less impacted in the PHI1 band, and it could be due to the flatter completeness curves discussed in Sect.\,\ref{sect:comp_I}.

\section{Impact of the mean polarization fraction on the expected number of detections}

\label{sect:other_polfrac}

The confusion noise scales linearly with the mean polarization fraction $\mu_p$. Since the polarized flux density is also proportional to $\mu_p$, the expected number of detections in polarization surveys affected only by confusion is independent of $\mu_p$. The situation is more complex in presence of instrumental noise. When $\mu_p$ decreases, the polarized flux density of the sources and the polarized confusion noise will both decrease, but the instrumental noise will remain constant. In the regime where the instrumental noise is dominant, the number of detections will thus decrease. In this appendix, we discuss a low ($\mu_p$=0.7\,\%) and a high ($\mu_p$=1.3\,\%) mean polarization fraction corresponding to the 3$\sigma$ bounds of the probability density distribution of $\mu_p$ determined in Appendix\,\ref{sect:polar_dependence}.

In Table\,\ref{tab:Ndet_0p7} and \ref{tab:Ndet_1p3}, we present the number of polarized detections expected for a low and high mean polarization fraction, respectively. They can can be compared with the results from our standard simulation (see Table\,\ref{tab:Ndet_polar}). The number of detections remains consistent within a factor of $\sim$2 with the standard simulation, except for the wide field in the PPI3 and PPI4 bands assuming payload required sensitivity for which the number of detection is very low in the low $\mu_p$ case. This uncertainty by a factor of $\sim$2 is much smaller than the variations by more than an order of magnitude found for our two different sensitivity scenarios.

\begin{figure*}
\centering
\begin{tabular}{cc}
\includegraphics[width=7cm]{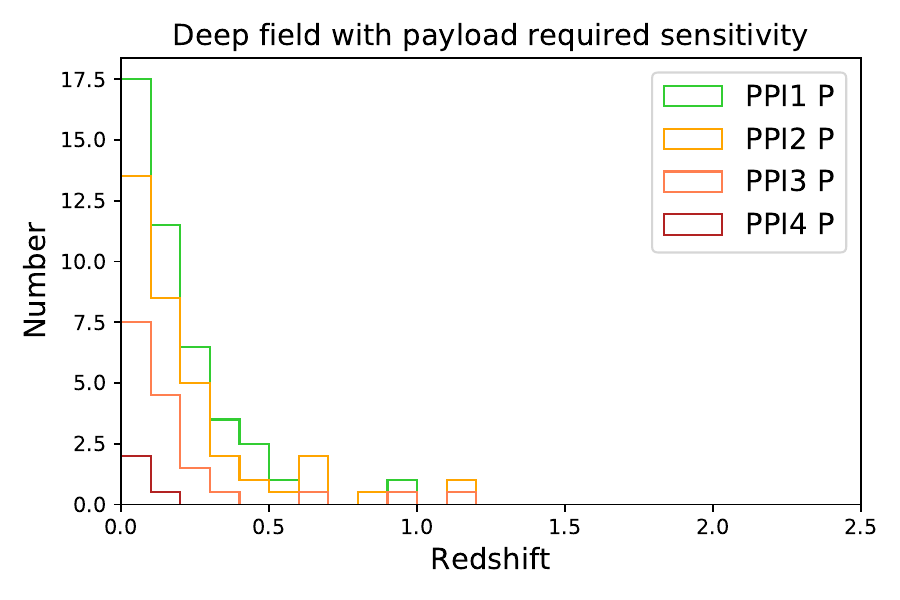} & \includegraphics[width=7cm]{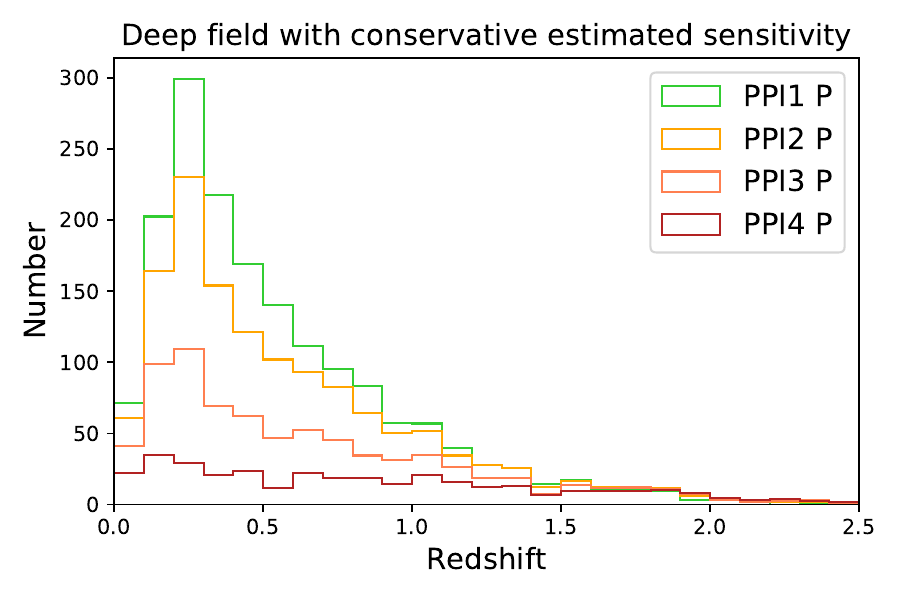}\\
\includegraphics[width=7cm]{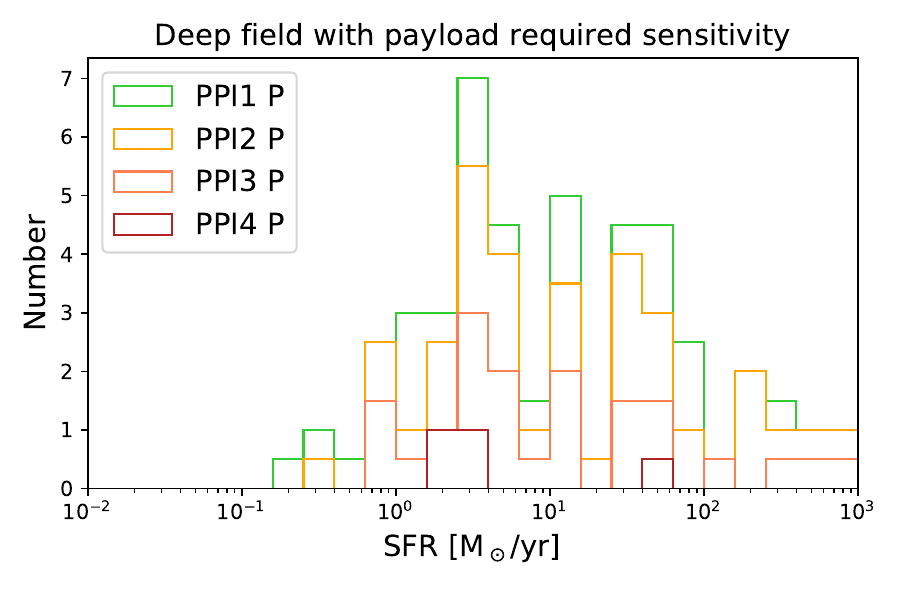} & \includegraphics[width=7cm]{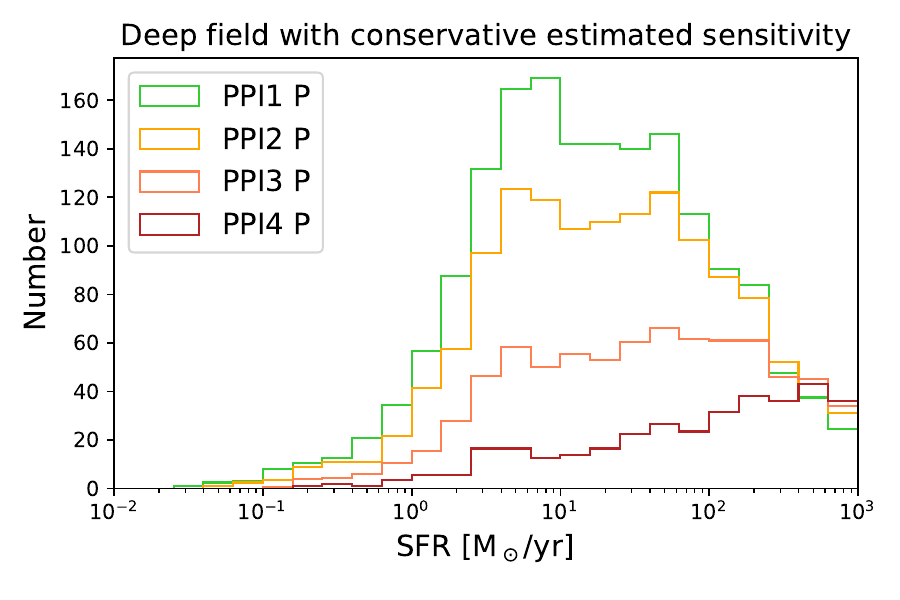}\\
\includegraphics[width=7cm]{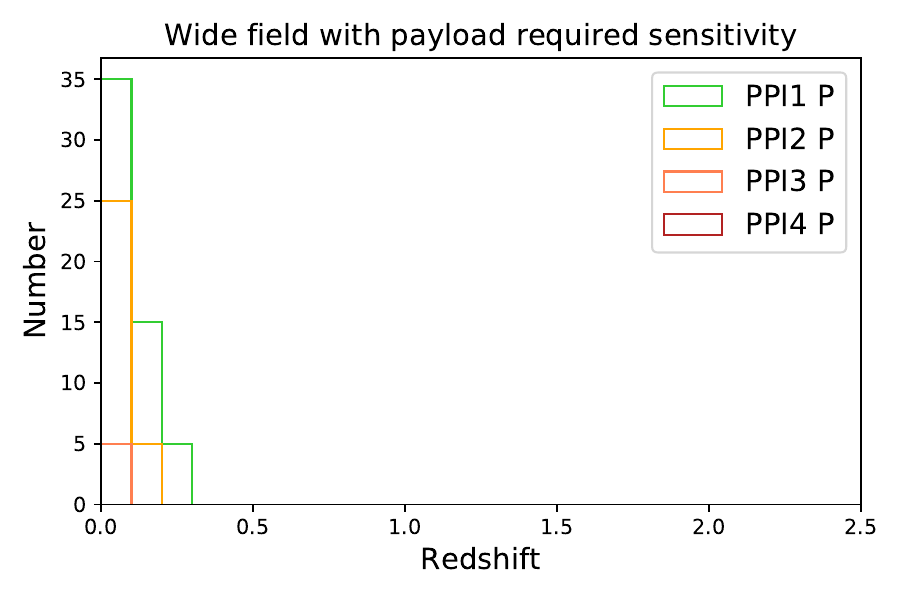} & \includegraphics[width=7cm]{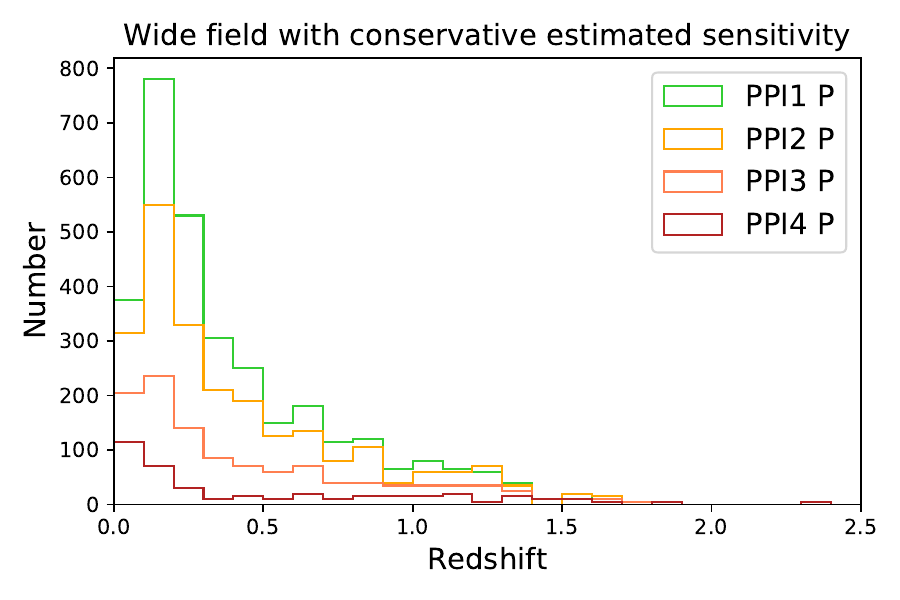}\\
\includegraphics[width=7cm]{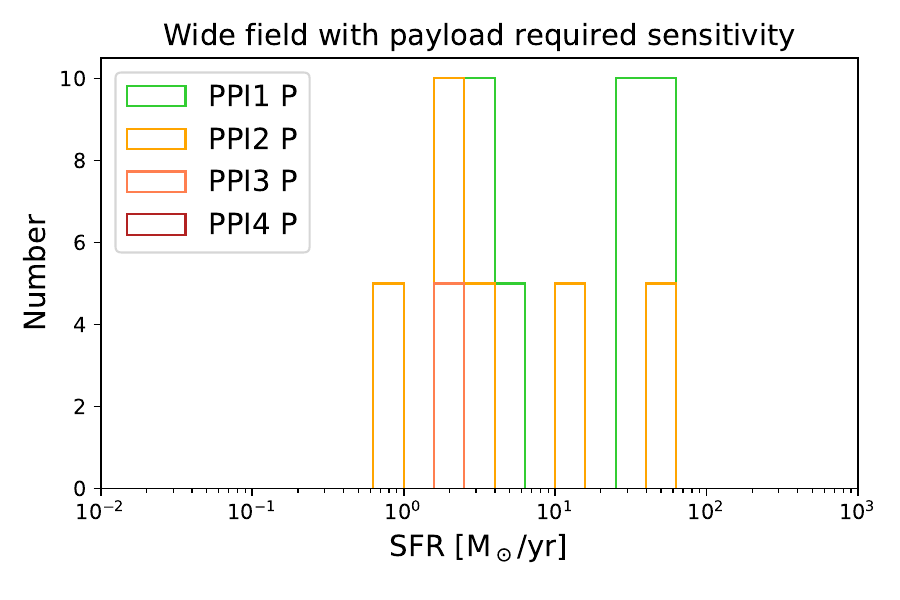} & \includegraphics[width=7cm]{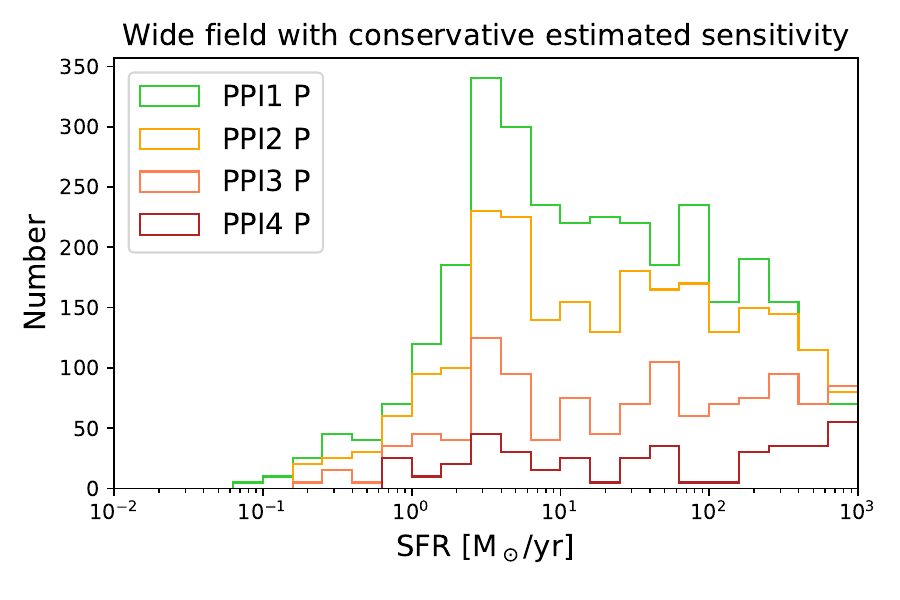}\\
\end{tabular}
\caption{\label{tab:histos_0p7} Same as Fig.\,\ref{fig:histos_polar} but assuming a mean polarized fraction of 0.7\,\%.}
\end{figure*}

\begin{figure*}
\centering
\begin{tabular}{cc}
\includegraphics[width=7cm]{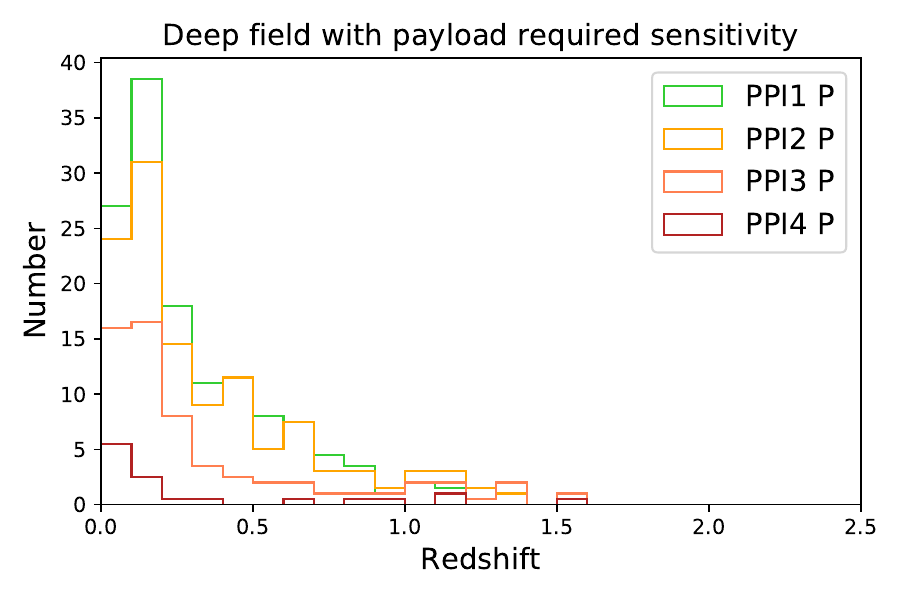} & \includegraphics[width=7cm]{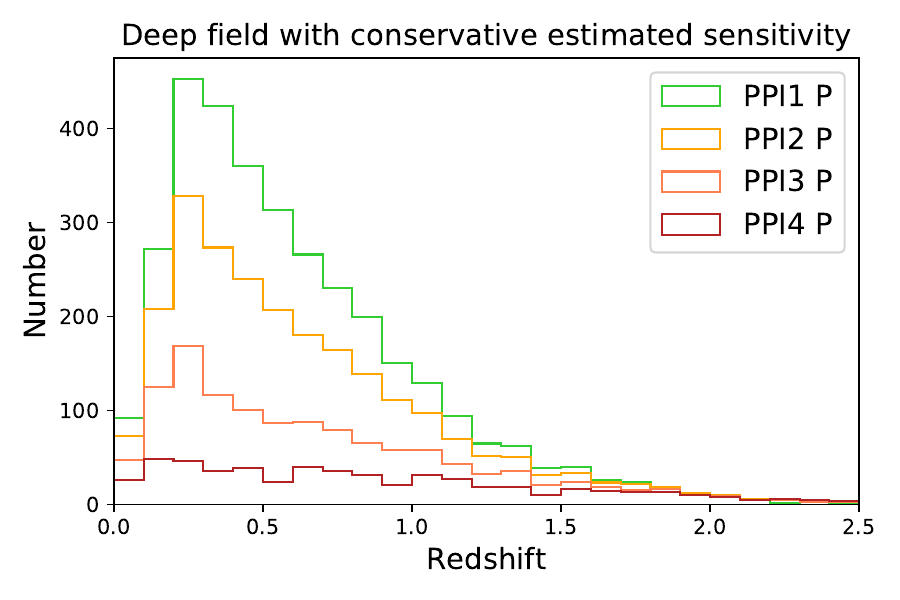}\\
\includegraphics[width=7cm]{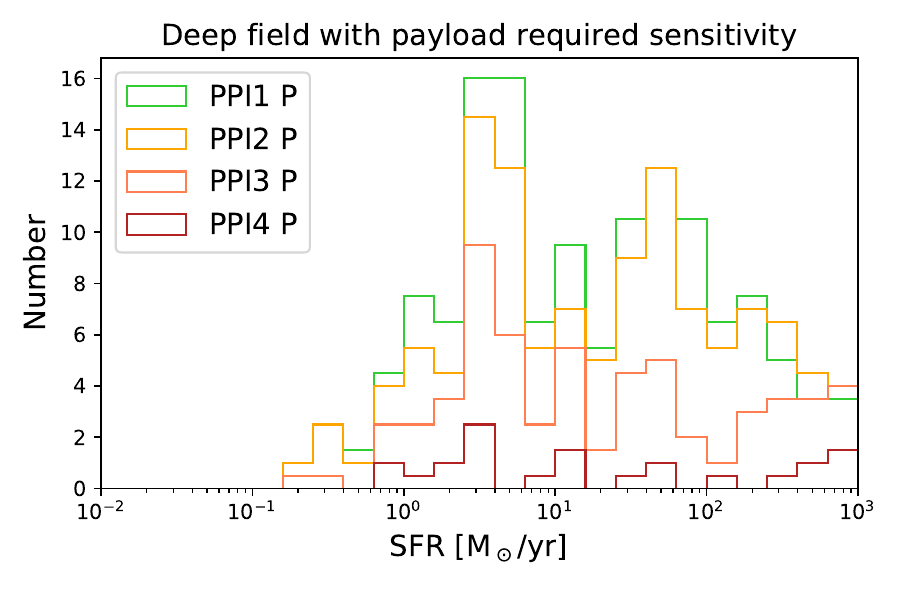} & \includegraphics[width=7cm]{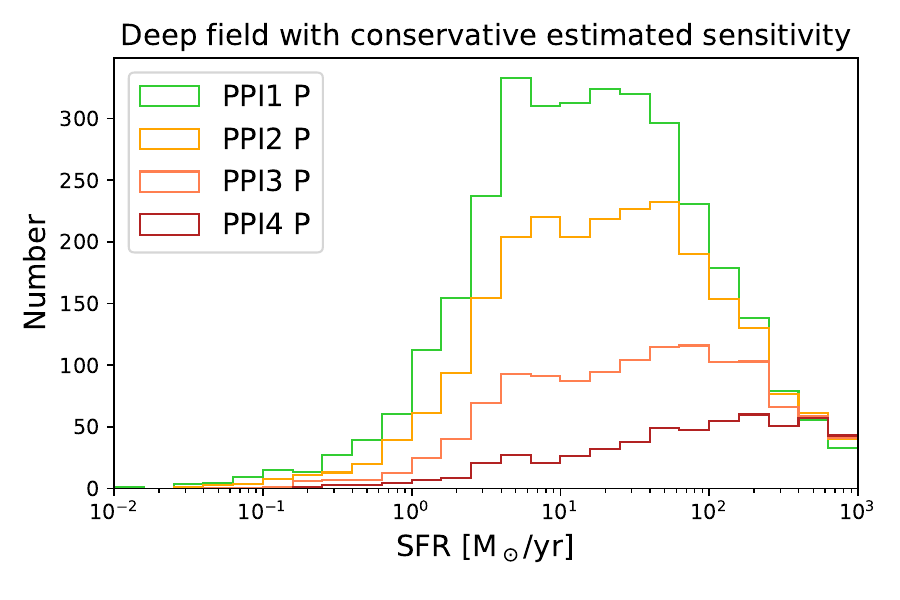}\\
\includegraphics[width=7cm]{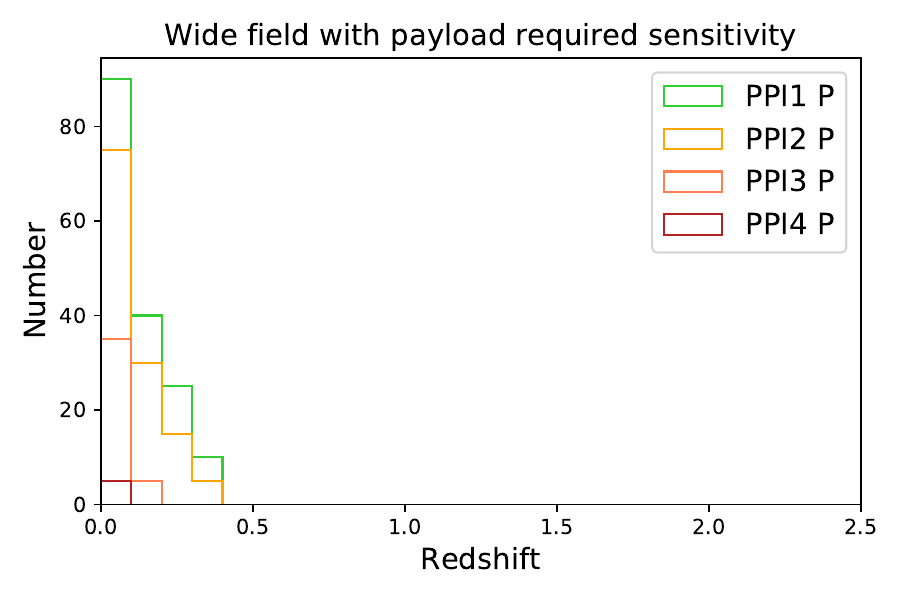} & \includegraphics[width=7cm]{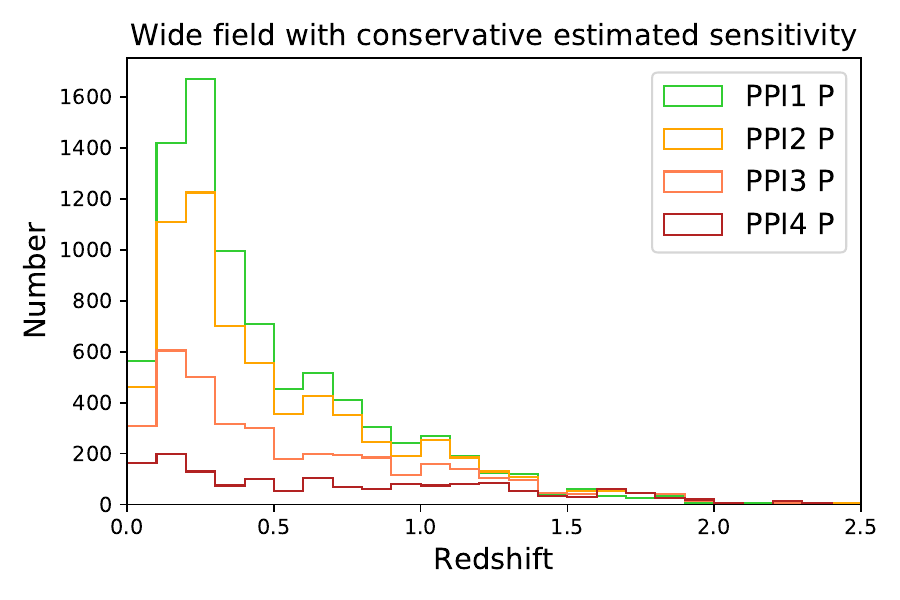}\\
\includegraphics[width=7cm]{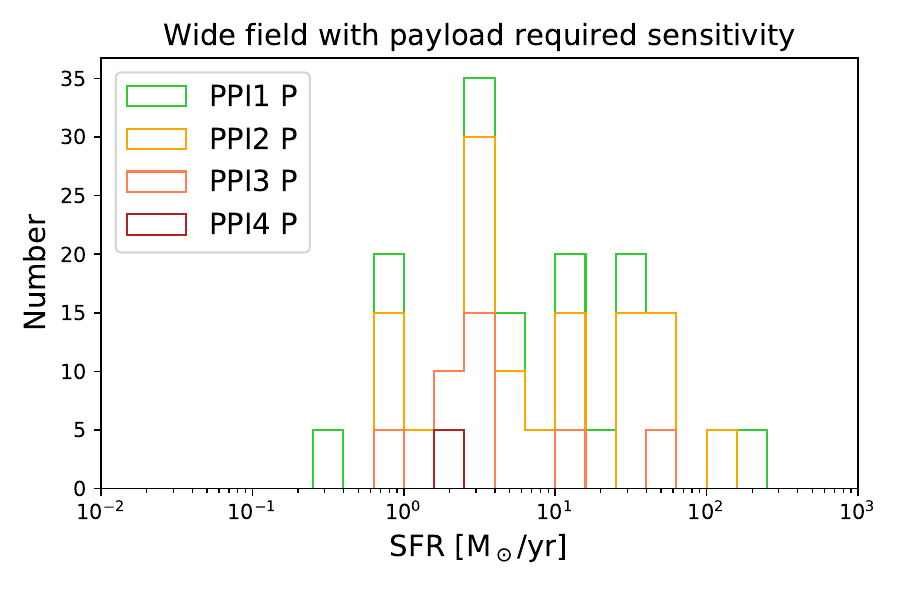} & \includegraphics[width=7cm]{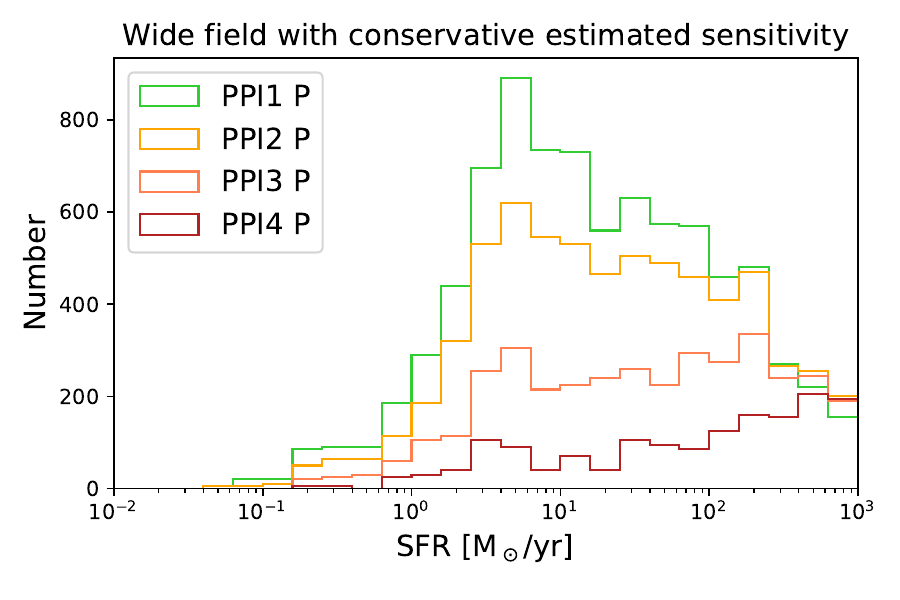}\\
\end{tabular}
\caption{\label{tab:histos_1p3} Same as Fig.\,\ref{fig:histos_polar} but assuming a mean polarized fraction of 1.3\,\%.}
\end{figure*}

Finally, in Fig.\,\ref{tab:histos_0p7} and \ref{tab:histos_1p3}, we present the redshift and SFR distributions for our various polarization and sensitivity scenarios. The impact of $\mu_p$ is mild. As shown in Table\,\ref{tab:Ndet_0p7} and \ref{tab:Ndet_1p3}, the mean redshift and SFR varies by less than a factor of $\sim$2 (after excluding the cases with $\lesssim$10 objects which are not statistically significant). For the conservative estimated sensitivity with larger numbers of detections, the variations are lower than $\sim$20\,\%.

\begin{table}
\caption{\label{tab:Ndet_0p7} Same as Table\,\ref{tab:Ndet_polar}, but assuming a mean polarized fraction of 0.7\,\%.}
\centering
\begin{tabular}{lcccc}
\hline
\hline
Band & P$_{\rm lim}$ & N$_{\rm det}$ & Mean $z$ & Mean SFR \\
& $\mu$Jy & & & M$_\odot$/yr \\
%%%copy below
\hline
\multicolumn{5}{c}{Deep field (1500h, 1deg$^2$)}\\
\multicolumn{5}{c}{with payload required sensitivity}\\
\hline
PPI1 & 552 & 46 & 0.22 & 63\\
PPI2 & 758 & 34 & 0.23 & 75\\
PPI3 & 1039 & 16 & 0.19 & 72\\
PPI4 & 1421 & 2 & 0.07 & 11\\
\hline
\multicolumn{5}{c}{Wide field (1500h, 10deg$^2$)}\\
\multicolumn{5}{c}{with payload required sensitivity}\\
\hline
PPI1 & 1718 & 55 & 0.09 & 18\\
PPI2 & 2353 & 30 & 0.07 & 11\\
PPI3 & 3213 & 5 & 0.03 & 2\\
PPI4 & 4390 & 0 & -- & --\\
\hline
\multicolumn{5}{c}{Deep field (1500h, 1deg$^2$)}\\
\multicolumn{5}{c}{with conservative estimated sensitivity}\\
\hline
PPI1 & 61 & 1669 & 0.54 & 66\\
PPI2 & 105 & 1344 & 0.58 & 84\\
PPI3 & 178 & 768 & 0.66 & 127\\
PPI4 & 238 & 352 & 0.86 & 225\\
\hline
\multicolumn{5}{c}{Wide field (1500h, 10deg$^2$)}\\
\multicolumn{5}{c}{with conservative estimated sensitivity}\\
\hline
PPI1 & 182 & 3145 & 0.40 & 85\\
PPI2 & 289 & 2345 & 0.43 & 108\\
PPI3 & 439 & 1155 & 0.47 & 150\\
PPI4 & 545 & 400 & 0.53 & 212\\
\hline
%copy above
\end{tabular}
\end{table}

\begin{table}
\caption{\label{tab:Ndet_1p3} Same as Table\,\ref{tab:Ndet_polar} but assuming a mean polarized fraction of 1.3\,\%.}
\centering
\begin{tabular}{lcccc}
\hline
\hline
Band & P$_{\rm lim}$ & N$_{\rm det}$ & Mean $z$ & Mean SFR \\
& $\mu$Jy & & & M$_\odot$/yr \\
%%%copy below
\hline
\multicolumn{5}{c}{Deep field (1500h, 1deg$^2$)}\\
\multicolumn{5}{c}{with payload required sensitivity}\\
\hline
PPI1 & 553 & 136 & 0.32 & 78\\
PPI2 & 763 & 119 & 0.35 & 96\\
PPI3 & 1054 & 61 & 0.36 & 128\\
PPI4 & 1446 & 12 & 0.34 & 177\\
\hline
\multicolumn{5}{c}{Wide field (1500h, 10deg$^2$)}\\
\multicolumn{5}{c}{with payload required sensitivity}\\
\hline
PPI1 & 1718 & 165 & 0.12 & 20\\
PPI2 & 2354 & 125 & 0.11 & 17\\
PPI3 & 3218 & 40 & 0.06 & 9\\
PPI4 & 4398 & 5 & 0.03 & 2\\
\hline
\multicolumn{5}{c}{Deep field (1500h, 1deg$^2$)}\\
\multicolumn{5}{c}{with conservative estimated sensitivity}\\
\hline
PPI1 & 69 & 3288 & 0.61 & 57\\
PPI2 & 135 & 2366 & 0.64 & 75\\
PPI3 & 255 & 1242 & 0.70 & 113\\
PPI4 & 362 & 553 & 0.87 & 194\\
\hline
\multicolumn{5}{c}{Wide field (1500h, 10deg$^2$)}\\
\multicolumn{5}{c}{with conservative estimated sensitivity}\\
\hline
PPI1 & 185 & 8195 & 0.47 & 73\\
PPI2 & 301 & 6565 & 0.51 & 94\\
PPI3 & 476 & 3660 & 0.58 & 139\\
PPI4 & 609 & 1575 & 0.75 & 239\\
\hline
%copy above
\end{tabular}
\end{table}

\end{appendix}

\end{document}